\documentclass[a4paper, 11pt, twoside]{report}

\usepackage[utf8]{inputenc}
\usepackage[T1]{fontenc}

\usepackage{amsmath}
\usepackage{amssymb}
\usepackage{graphicx}
\usepackage[dvipsnames]{xcolor}
\usepackage{float}
\usepackage{multirow}
\usepackage{ulem}
\usepackage{subfig}

\usepackage{amsfonts}
\usepackage{longtable}

\usepackage{fancyhdr}

\usepackage{tipa}

\usepackage{pdfpages}
\usepackage{enumerate}
\usepackage[shortlabels]{enumitem}

\newcommand{\clearemptydoublepage}{\newpage{\pagestyle{empty}\cleardoublepage}}

\newcommand{\dd}{\mathrm{d}}

\newcommand{\bea}{\begin{eqnarray*}}
\newcommand{\eea}{\end{eqnarray*}}

\newcommand{\beq}{\begin{equation}}
\newcommand{\eeq}{\end{equation}}

\newcommand{\bi}{\begin{itemize}}
\newcommand{\ei}{\end{itemize}}
\newcommand{\bv}{\begin{verbatim}}
\newcommand{\ev}{\end{verbatim}}

\newcommand{\beql}[1]{\beq\label{#1}}

\newcommand{\fr}{\frac}

\newcommand{\oh}{\frac{1}{2}}

\title{New topological observables in a model of Causal Dynamical Triangulations on a torus}
\author{Zbigniew Drogosz}

\begin{document}

\thispagestyle{empty}
\begin{center}
{\bf \huge New topological observables in  a model of}\\[3.5mm]
{\bf \huge Causal Dynamical Triangulations on a torus}\\[10mm]
{\bf \LARGE Zbigniew Drogosz}\\[30mm]
{
\bf 
\LARGE
Rozprawa doktorska\\
\vspace{1cm}
Promotor: prof. dr hab. Jerzy Jurkiewicz\\[2mm]
Promotor pomocniczy: dr hab. Jakub Gizbert-Studnicki\\[15mm]
}
\includegraphics[width=30mm]{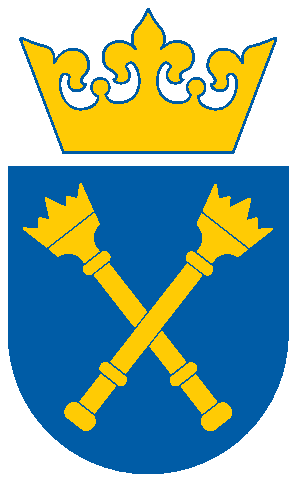}\\[15mm]

{\bf \Large Uniwersytet Jagielloński}\\[5mm]
{\bf \Large Instytut Fizyki im. Mariana Smoluchowskiego}\\[1.5mm]
{\bf \Large Zakład Teorii Układów Złożonych}\\[8mm]
{\bf \Large Kraków 2021}
\end{center}

\newpage

\tableofcontents

\clearemptydoublepage

  \addcontentsline{toc}{chapter}{Acknowledgments}

\markboth{Acknowledgments}{Acknowledgments}

{\Huge \bf Acknowledgments}

\vspace{10mm}

I would like to express my sincere gratitude to my supervisor Prof.~Jerzy Jurkiewicz for
introducing me to the fascinating world of quantum gravity. The knowledge he shared with me 
and his continual encouragement and scientific inspiration
were invaluable for my work described in this thesis.

I am also deeply grateful toward my secondary supervisor Dr.~Jakub Gizbert-Studnicki for 
the manifold support I received from him during our collaborations.

I highly appreciate the enduring collaboration and the fruitful conversations with
Prof.~Jan Ambjørn, Dr.~Andrzej Görlich and Dániel Németh.

I would also like to thank Prof.~Renate Loll and Nilas Klitgaard 
for the interesting discussions, especially about the quantum Ricci curvature.

This work was supported by the National Science Centre of Poland (NCN), grant 2019/32/T/ST2/00390.
I also acknowledge the support from project Kartezjusz.

\newpage

  \addcontentsline{toc}{chapter}{Abstract / Streszczenie}\label{chap:abstract}

\markboth{Abstract}{Abstract}

{\Huge \bf Abstract}

\vspace*{1cm}

Causal Dynamical Triangulations (CDT) is a nonperturbative formulation of a theory of quantum gravity, 
based on the Einstein-Hilbert action and Weinberg's asymptotic safety conjecture, in which quantization
is performed by calculating path integrals and using a form of the Einstein-Hilbert action proposed by Regge
for simplicial manifolds. Causality means that spacetime can be written as a Cartesian product of a space 
with a certain topology and a time interval. A local topology change would mean a creation of a singularity 
and a violation of causality. The discrete spacetime geometries are generated in computer Monte Carlo simulations.

This thesis examines the structure of simplicial manifolds in a model of Causal Dynamical Triangulations in 3+1 dimensions
with the spatial topology of a 3-torus, using new topological observables, such as loops with nonzero winding numbers and coordinates based on scalar fields with jumps at the boundaries of the
elementary cell of the torus,
to analyse non-local topological structures. It is not possible to introduce these observables in the case of CDT models with the spatial topology of a 3-sphere.
The results are given an interpretation and used in the measurements of a more 
local observable that is
the quantum Ricci curvature.
Moreover, the thesis describes the analysis of the influence of scalar matter fields on the geometry
of CDT spacetimes.

\newpage

\markboth{Streszczenie}{Streszczenie}

{\Huge \bf Streszczenie}

\vspace*{1cm}

Kauzalne Dynamiczne Triangulacje (CDT) to nieperturbacyjne sformułowanie teorii kwantowej grawitacji, 
w którym wychodzi się od działania Einsteina-Hilberta oraz od założenia asymptotycznego bezpieczeństwa
Weinberga i dokonuje kwantyzacji przez obliczenie całek po trajektoriach po możliwych geometriach Wszechświata.
Stosowana metoda regularyzacji tych całek polega na dyskretyzacji czasoprzestrzeni i wykorzystaniu formy działania
Einsteina-Hilberta zaproponowanej przez Reggego dla rozmaitości symplicjalnych. Kauzalność oznacza, że czasoprzestrzeń daje 
się zapisać jako iloczyn kartezjański przestrzeni o pewnej topologii i odcinka oznaczającego czas.
Lokalna zmiana topologii przestrzeni oznaczałaby powstanie osobliwości i pogwałcenie kauzalności.
Dyskretne geometrie czasoprzestrzeni są generowane w komputerowych symulacjach Monte-Carlo.

Niniejsza dysertacja dotyczy analizy struktury rozmaitości symplicjalnych w modelu Kauzalnych Dynamicznych
Triangulacji w 3+1 wymiarach z topologią przestrzenną 3-torusa przy wykorzystaniu nowych topologicznych
obserwabli, takich jak pętle o niezerowej liczbie nawinięć i współrzędne oparte na polach 
skalarnych o skoku na brzegu komórki elementarnej torusa, do badania nielokalnych topologicznych struktur.
Obserwable te nie są możliwe do wprowadzenia w przypadku modeli CDT z topologią przestrzenną 3-sfery.
Wyniki są poddane interpretacji, a także wykorzystane w pomiarach bardziej lokalnej obserwabli,
jaką jest kwantowa krzywizna Ricciego.
Ponadto w dysertacji przedstawione jest badanie wpływu skalarnych pól materii na geometrię 
czasoprzestrzeni w CDT.

\chapter{Introduction}\label{chap:introduction}
\markboth{Introduction}{Introduction}

\section{The search for a theory of quantum gravity}

Einstein's general relativity (GR) accurately describes macroscopic gravitational phenomena
and is one of the most precise and successful theories of physics.
However, at very high energy scales or equivalently at very short length scales,
comparable with Planck's length $l_P$, GR by itself is insufficient as it does not take into account
phenomena described by quantum mechanics.
Attempts to quantize general relativity in standard
ways used in quantum field theory, and successfully 
applied e.g.~to electrodynamics, fail  
because the gravitational constant $G$ in the 
Einstein-Hilbert action
\beq
S_{EH}[g_{\mu \nu}]=\fr{1}{16\pi G} \int_M \dd ^4 x\sqrt{-\det g} (R-2 \Lambda)
\eeq
has mass dimension $-2$, which makes the theory
perturbatively non-renormalizable.

Therefore, the unification of general relativity
with quantum mechanics in a theory that may 
describe gravity at the shortest lengths
and in scenarios with extreme energy densities,
such as the early Universe, must proceed in a different
fashion. And so, for example, string theory \cite{polchinski}, which
is certainly a bold attempt at creating not only a theory 
of quantum gravity but also at providing an overarching
framework for describing all interactions, removes some 
singularities by replacing at the Planck length scale
point particles with one-dimensional objects. It depends, however, upon the existence of
supersymmetry and compactified dimensions,
no traces of which have yet been detected \cite{super1,super2, dimensions1, dimensions2}.
Another candidate theory, loop quantum gravity, is founded on
a representation of solutions to the Wheeler-DeWitt equation 
in terms of functionals of sets of loops.
In that picture, conventional spacetime is absent, and the physical space is
a quantum superposition of aggregates of quanta of the gravitational field,
represented by the nodes of a spin network (network of loops)
\cite{rsm, rovelli}.

In lattice approaches to quantum gravity (see below),
it is hoped that the continuum limit is recovered
when the lattice spacing goes to zero.
In loop quantum gravity, on the other hand, there is no such regulating parameter, and the
discretization is introduced 
by the very definition of the scalar product in the Hilbert space of spin network states,
which makes obtaining the semiclassical limit a difficult task
\cite{lqg}.
Among other unsolved problems of that theory is to
establish and write down the precise form of the quantum Hamiltonian \cite{rovelli}.

Perhaps there exists another, simpler solution to the quantization of gravity, though,
if Weinberg's asymptotic safety conjecture \cite{weinberg}, which suggests that general relativity
may be renormalizable but in a non-perturbative way, is true.
An asymptotically safe theory is one in which 
the essential coupling constants
approach a fixed point as the energy scale of their renormalization group flow goes to infinity.
The number of free parameters in such a theory is 
equal to the number of dimensions of the critical
surface of the ultraviolet fixed point (UVFP);
one of these parameters is dimensional
and connected to the energy scale, and the others are dimensionless.
Therefore, if the ultraviolet critical surface
is finite dimensional, the condition of asymptotic
safety fixes all but a finite number of parameters of the theory.
Weinberg showed that
for fixed points associated with second-order phase transitions
the ultraviolet critical surface is finite dimensional. Then, Weinberg proceeded to consider GR in $2+\epsilon$ dimensions, $0 < \epsilon \ll 1$, 
and concluded that in that case 
there exists an asymptotically safe theory of pure gravity 
with a one-dimensional critical surface.
There remained, however, the problem of extending the results to 
four dimensions, where the parameter $\epsilon = 2$ is no longer small.

The search for a nontrivial ($\mu \neq 0$) ultraviolet fixed point
is one of the aims of the analysis of the phase diagram of Causal Dynamical Triangulations (CDT).
As mentioned, the UVFP should be associated with a second- or higher-order phase transition point.
 Some evidence of such an UVFP is provided by calculations in $2+\epsilon$ dimensions \cite{kawai1, kawai2, kawai3, kawai4, kawai5} and from the use of the
exact renormalization group \cite{reuter1, reuteretc1, reuteretc2, niedermaier, reuteretc4}, although there
are still no conclusive results \cite{via}. 
The phase structure observed in CDT 
may also possibly be an indication of the existence of a 
point described by the asymptotic safety theory.

 In addition, it should be possible to define the renormalization group flow lines in the lattice coupling constant space leading from an infrared limit to the UVFP. This in general requires finding a region in the lattice coupling constant space where the semiclassical limit (consistent with the classical GR) can be defined, 
together with some physical observables.
These physical observables should be such that keeping 
their values fixed defines a path in the lattice coupling constant space that allows the interpretation of a decreasing lattice spacing when moving away from the semiclassical region. If the lattice spacing goes to zero at the endpoint of the path, this endpoint will be an UVFP. 

Another aim, independent of the search for an UVFP,
is the investigation of non-perturbative aspects of quantum gravity treated as an effective theory valid up to some finite energy scale.

\section{Earlier lattice models of quantum gravity}\label{subsubs:dt}

In a quantum theory of gravity, a goal is to calculate the quantum amplitude of the transition between two physical states. 
The amplitude is defined as a path integral over field configurations, which in the case of gravity
are equivalence classes of metrics with respect to diffeomorphisms, defining various spacetime geometries,
\beq
\mathcal{Z} = \int \mathcal{D}[g_{\mu \nu}] e^{iS_{EH}[g_{\mu \nu}]}.
\label{eq:z}
\eeq
$S_{EH}$ is the Einstein-Hilbert action
\beq
S_{EH} [g_{\mu \nu}] = \frac {1}{16 \pi G} \int_M \dd{}^4 x \sqrt{- \det \ g} \, (R-2 \Lambda ),
\eeq
where $R$ is the scalar curvature and $\Lambda$ is the cosmological constant. At this point, that expression is
still only formal and requires regularization and 
a choice of the integration measure
over $g_{\mu\nu}$ and the domain of integration over spacetimes.

In lattice models of quantum gravity, the integral (\ref{eq:z}) is regularized as a sum over a suitable ensemble of piecewise linear manifolds.
General relativity on such 
discretized spacetimes
was analyzed by Regge \cite{regge},
who showed that the continuously differentiable $d$-dimensional
manifolds of GR can be approximated 
with simplicial manifolds to arbitrary precision
and without introducing any coordinates.
In that setting,
curvature is determined by deficit angles around $d$-2-dimensional simplices (called hinges),
and the scalar curvature term
of the Einstein-Hilbert action is given by the sum of
products of volumes of the hinges
and the deficit angles at them.
Ultimately, the Regge action $S_R$,
which is the piecewise linear analogue of the Einstein-Hilbert action
$S_{EH}$ can be expressed in terms of the number of $d$-dimensional simplices, their edge lengths, 
and the way they are glued together along their 
$d$-1-dimensional faces.

A naive approach to apply Regge calculus to general relativity would be to consider
a fixed lattice, i.e., one in which the number of simplices and the connections
between them are kept constant, while the lengths of the edges are variable \cite{hamber}.
That method, however, is ill-suited for quantum gravity, where the goal is to calculate
a path integral over all geometries, since generally many such
lattices would correspond
to the same physical geometry, 
and taking into account all the combinatorial factors 
would likely present an intractrable problem. 

That is the reason why lattices of different construction, named Dynamical Triangulations (DT),
were invented. Dynamical Triangulations are shape-shifting lattices, where
all the simplices are equilateral and of the same 
and constant size, with the edge length $a$
acting as the ultraviolet cutoff scale, 
whereas it is the number of simplices and the way they are glued together 
that are variable
\cite{bosonic-string1, bosonic-string2, aaa2}. 
Crucially two different triangulations correspond 
to two different physical geometries, and all the combinatorial factors
can be eliminated by labeling the vertices and considering labeled triangulations.
It is assumed that in the limit $a \to 0$ the ensemble of triangulations 
becomes dense in a suitable way
in the set of continuous geometries that appears in the continuum path integral,
and that the bare constants can be adjusted
in such a way that it is possible to take 
that limit while keeping the physics unchanged.

The only existing analytical solutions of DT are in two dimensions.
They use conformal symmetries and tools of the Random Matrix Theory in large N limit
\cite{bbb1, bbb2, bbb3, bbb4, bbb5}. 
In that low-dimensional case, the continuum limit
$a \to 0$ was successfully taken, and the results agreed with 
quantum Liuville theory, a continuum two-dimensional Euclidean quantum gravity theory, which had 
already been solved analytically in all cases where these
solutions exist (central charge $c \leq 1$). In higher dimensions no analytical solutions
of DT have been found, and the theory is
studied using Monte Carlo simulations.
In the literature, there exist extensive
studies in three and four dimensions
of the theory without added matter
\cite{3dEDT2, 3dEDT1, 3dEDT3, 3dEDT4, 3dEDT5, 4dEDT1, 4dEDT2, 4dEDT3}, 
together with generalizations where matter fields were added to the action \cite{matterEDT1, matterEDT2, matterEDT3, matterEDT4, matterEDT5}.

\section{Causal Dynamical Triangulations (CDT)}\label{subsubs:cdt}

Models of (Euclidean) Dynamical Triangulations, while constituting an interesting early
attempt at creating a theory of quantum gravity, failed
to produce physically viable
spacetimes of more than two dimensions
because of the formation of ``baby universes'', whereby space was invariably
made fractal and divided into disconnected regions, regardless of 
the chosen definition of time. 
No suitable UVFP was found \cite{firstorderEDT1, firstorderEDT2}
(although there are some newer attempts revising this theory, see e.g. \cite{jack-us1, jack-us2, Laiho:2016nlp}).

Therefore, a new class of models was created
as an attempt to curtail the excessive formation of baby universes
\cite{ddd}.
The modified theory was named Causal Dynamical Triangulations (CDT), after  
its key component, causality, by which is meant the preservation of the spatial topology
in the time evolution of the system.
The notion is taken from general relativity, where a change of the topology of spacetime
would entail a creation of a singularity and a causality violation.
Accordingly, a CDT spacetime is topologically a Cartesian product of a 
$d$-dimensional
spatial manifold  $\Sigma$ (it is additionally assumed that this manifold is closed)
and the interval $I$ representing proper time: $M = \Sigma \times I$.
This global time foliation defines in a natural way a time coordinate,
ranging from $t=1$ to $t=T$, which is the total number of leaves 
(often called slices) of the foliation.
As will be described below, to avoid choosing the initial and final states, 
as well as to facilitate computer simulations, the standard procedure is to
impose periodic boundary condition on the time interval and to
identify the spatial geometry at time $t$ with that at time $T+t$.
It is uncertain if the requirement of constant spatial topology  
must hold in every theory of quantum gravity, and if 
the introduction of a preferred foliation is compatible 
with general 4D spacetime diffeomorphism invariance.
At any rate, the lack of classification of topologies of manifolds in dimension three and higher 
makes any calculation and interpretation of a path integral over all topologies
difficult to imagine.

In CDT, as in EDT, spacetime is a triangulation built by joining together fixed-size 
simplices in a way that satisfies certain topological requirements. 
In the 3+1-dimensional case, the simplicial building blocks are four-dimensional.
Each simplex is the convex hull of five vertices that lie on two neighboring slices $\Sigma$.
There are thus two types of four-simplices: $\{4,1\}$-simplices with four vertices on a slice $t$
and one vertex on a slice $t\pm 1$, and $\{3,2\}$-simplices with three vertices on a slice $t$
and two vertices on a slice $t\pm 1$. 
Each simplex abuts along its 
three-dimensional faces on five other simplices, called its neighbors.
All spacelike links, i.e., line segments that connect two vertices
on the same time slice, are of length $a_s$, and all timelike links, i.e., line segments
that connect two vertices on neighboring time slices, are of length $a_t$.
Those lengths are constant, serve as the ultraviolet cutoff,
and their ratio squared is the asymmetry factor:
$\alpha = - a_t^2/ a_s^2$. Note the minus sign.
All four-simplices are flat fragments of Minkowski spacetime, $\alpha>0$, and the curvature,
as mentioned, is encoded at $d-2$-dimensional simplices.

Through the discretization of the geometry 
the continuum transition amplitude between
the spatial geometries at the initial and final global times, eq. (\ref{eq:z}),
is regularized by the sum over triangulations $\mathcal{T}$:
\beq
\mathcal{Z} = \sum_{\mathcal{T}} \fr{1}{C_{\mathcal{T}}} e^{iS_R[\mathcal{T}]},
\eeq
where $S_R$ is
the Regge action with a cosmological term, and the combinatorial factor
$C_{\mathcal{T}}$ is the order of the automorphism group of the triangulation.
No simple method to calculate it for a general triangulation exists, but
the necessity for it can be avoided by assigning indices to vertices
and henceforward considering labeled triangulations $\mathcal{T}_L$.

If the number of vertices in a triangulation $\mathcal{T}$ is denoted by $N_0[\mathcal{T}]$, then
\beq
C_{\mathcal{T}} = \fr{N_0[\mathcal{T}]!}{N(\mathcal{T})},
\eeq
where $N(\mathcal{T})$ is the number of labeled triangulations isomorphic with
an unlabeled triangulation $\mathcal{T}$.
The partition function takes the form
\beq
\mathcal{Z} = \sum_{\mathcal{T}_L} \fr{1}{N_0[\mathcal{T}_L]!} e^{iS_R[\mathcal{T}_L]},
\eeq
with $\mathcal{T}_L$ denoting a labeled triangulation.
To obtain real probability distributions, it is necessary to perform a Wick rotation
of each simplex to Euclidean signature (this is made possible by the time foliation and performed by
analytically continuing $\alpha$ to the negative real axis in the lower complex plane \cite{physrep}) with the Regge action of the triangulation changed accordingly \cite{ddd}.
The exponent in the sum becomes real, and the complex amplitudes become real probabilities, which makes the model well-suited for studying by means of Monte-Carlo simulations \cite{ajl1}:
\begin{equation}
{\cal{P}}({\cal T}) \propto e^{-S_R({\mathcal{T}_E})}, 
\label{eq:prob}
\end{equation}
where $\mathcal{T}_E$ is now a labeled Euclidean triangulation whose four-simplices
have an Euclidean metric.
The Regge action $S_R$ takes the following very simple form \cite{physrep}:
\beq
S_R[\mathcal{T}_E]=-\left(K_0+6\Delta\right)N_{0}[\mathcal{T}_E]+K_4\left(N_{4,1}[\mathcal{T}_E]+N_{3,2}[\mathcal{T}_E]\right)+\Delta N_{4,1}[\mathcal{T}_E],
\eeq
\noindent where $N_{4,1}$ and $N_{3,2}$ denote the number of $(4,1)$- and $(3,2)$-simplices,
while $\kappa_{0}$, $\Delta$ and $\kappa_{4}$ are bare dimensionless coupling constants related to the inverse of Newton's constant $G^{-1}$, the cosmological constant $\Lambda$, and the asymmetry parameter $\alpha$ respectively.

Numerical evidence indicates that the number of triangulations grows up to the leading
order exponentially as $e^{K_4^c N_4}$,
where $N_4=N_{4,1}+ N_{3,2}$ is the total number of four-simplices.
Taking the limit $N_4 \to \infty$ is replaced in simulations by analysing
a series of spacetimes with increasing $N_4$ and inferring from that the asymptotic behavior.
If the asymptotic safety hypothesis holds,
then, after calculating the partition function, it should in principle be possible to
take in a certain way the limit in which
the edge length $a$ of the simplices goes to $0$, and with the knowledge of the partition  
function 
to calculate expectation values and correlation functions
of all observables:
\beq
\langle O_1 \dots O_n \rangle = \fr{1}{Z} \int \mathcal{D}[g_{\mu \nu}] O_1(g_{\mu \nu}) \dots O_n (g_{\mu \nu}) e^{iS_{EH}[g_{\mu \nu}]}.
\eeq

\section{An overview of previous results in four-dimensional CDT}

For many years since the beginning of the study of four-dimensional CDT,
the only topology of the spatial slices that was considered was the 
spherical topology $S^3$, 
which seemed to be the obvious choice because of its simplicity.
The study led to several important discoveries, however
later, as will be described below, the interest largely shifted to the toroidal
topology $T^3=S^1 \times S^1 \times S^1$, and most of the previous analysis  
was repeated in this new topology.
Therefore, most of this section will describe both topologies simultaneously.
 
Among the most important CDT results is the discovery of the phase diagram
with four phases, which is quite non-trivial considering the simplicity of the 
Regge action. Although only
two topologies have been studied, the phase diagram looks similar in
both cases and a plausible conjecture is that it
might be independent of the spatial topology chosen 
\cite{phasestorus}.
As described in the previous section, four-dimensional CDT has three coupling constants:
$K_4$, $K_0$ and $\Delta$. The limit where
the number of four-simplices $N_4$ grows to infinity is unmanageable in computer
simulations, so the method used is to set $K_4$ in such a way that
the volume of a simulated configuration oscillates around a certain value of $N_4$
(the volume is controlled by adding a volume-fixing potential
term $S_{VF}$ to the action).
This is repeated for several $K_4$ to estimate the asymptotic behavior
for $N_4 \rightarrow \infty$, corresponding to the critical value of $K_4$. In that limit,
the only coupling constants left are
$K_0$ and $\Delta$, which makes the phase space two-dimensional.

\begin{figure}
\centering
\includegraphics[width=\textwidth]{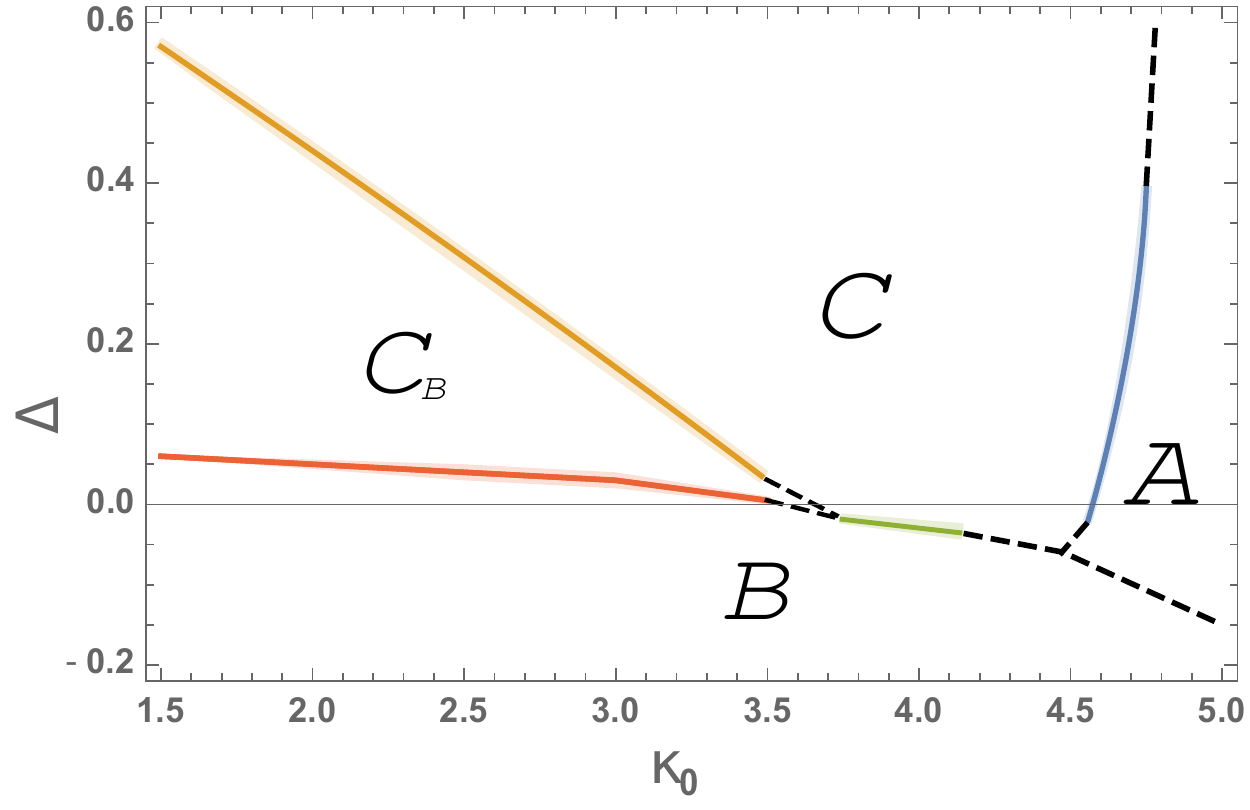}\\
\caption{
The phase diagram of four-dimensional CDT with the toroidal spatial topology. 
}\label{fig:phases}
\end{figure}

The phase diagram for four-dimensional CDT with toroidal spatial topology is 
shown in Figure \ref{fig:phases}. 
A diagram for the spherical case looks very similar.
For technical reasons connected with probabilities of moves (see the next chapter)
some regions are easier studied in the toroidal topology.
As mentioned, there exist four phases labeled with 
$A$, $B$, $C$ and $C_b$
\cite{signature, phasetransition24}, out of which the most interesting one is the $C$ phase, where semiclassical limit can be obtained.
Some of the boundaries of the $C$ phase may possibly be
higher order phase transition lines
\cite{phasetransition24,phasetransition21, phasetransition22,phasetransition23,   phasetransition25, scalar}, and the two triple points in which phase transition lines intersect are candidates for the ultraviolet fixed point \cite{phasetransition11,phasetransition12, phasetransition13},
although at present these are still conjectures.

\begin{figure}
\centering
\includegraphics[width=0.8\textwidth]{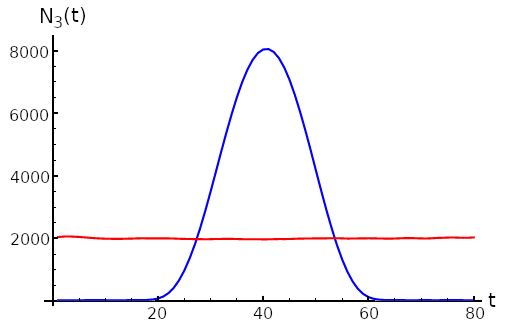}
\caption{The spatial volume profile $N_3(t)$ at the point $K_0=2.2$, $\Delta=0.6$ in the C phase in the spherical (blue line) and the toroidal (red line) spatial topology.}\label{fig:volume}
\end{figure}

Only the $C$ phase seems to 
correspond to physically relevant semiclassical universes \cite{agjl}, and
for this reason it has always been the one most extensively studied.
The simplest observable measured was the spatial volume profile $N_3(t)$, defined as the number
of spatial tetrahedra on a time slice $t$ (see Fig.~\ref{fig:volume}).
In the case of spherical spatial topology, the spatial volume plot of
a typical system with a sufficiently large number of slices $T$
consisted of a wide peak or blob, which contained most of the tetrahedra,
and a thin stalk necessary to satisfy the periodic time boundary 
conditions. 
After making use of the invariance with respect to discrete translations in time
to shift the position of the peak to the same place 
in the data from all simulations, it was shown that both the average volume
$\langle N_3(t)\rangle$ and its fluctuations could be derived from the discretized version of the
effective Hartle-Hawking mini-superspace action for the isotropic and homogeneous four-dimensional universe \cite{mini, hh}. The peak was well 
described by an appropriately scaled and translated $cos^3$ function \cite{agjl}.
The classical solution in this
case corresponded to a 4D de Sitter sphere, which
is the origin of the other name of the $C$ phase in CDT, the de Sitter phase.
In the toroidal case, the volume profile was flat, which could also be derived
from the mini-superspace action with a certain correction.
Four-dimensional CDT with toroidal topology was first analyzed in the paper \cite{torusz}.
The volume profile for the point with the same parameters as in the spherical case
was constant (see again Fig. \ref{fig:volume}):
\beq
N_{41}(t)=\fr{\bar{V}_4}{T},
\eeq
which could also be described by an appropriately modified mini-superspace model.

Thus, in phase $C$, the scale factor behaves semiclassically \cite{agjl,Ambjorn:2005qt,Ambjorn:2004qm, Ambjorn:2007jv}, despite the fact that 
a typical universe studied in CDT simulations has
a size of only about 10-20 Planck's lengths, as determined by 
relating the measurements of certain observables to a continuous theory \cite{physrep}.
The lattice spacing varies across the phase space \cite{LatSpacing}.

There are important differences in the assumptions of the mini-superspace model 
and CDT. In that model in four-dimensional general relativity, isotropy and homogeneity
of the universe are put in by hand, and the scale factor $a(t)$ is the only degree of freedom.
In CDT, however, the typical geometries are very complicated and far from 
classical solutions of general relativity. The approximate agreement with the classical mini-superspace solution is obtained from an average of an ensemble of fluctuating geometric states and is caused by a nontrivial interplay between the physical action and the entropy of configurations. 
Another difference is the sign of the effective action, opposite to 
the one found
in general relativity. In CDT the solution of classical equations of motion gives a 
stable classical vacuum state, where at each $t$ there are all possible geometric realizations
with a particular value of $N_3(t)$. The existence of this nontrivial classical general 
relativity limit of the model was one of the most important results in the early studies of 
CDT.
For reviews of further results see \cite{physrep, reviews2}. Specifically for the most recent
results in the toroidal topology, see also the overview article \cite{toroidalreview}.

Although the initial reason to consider systems with the toroidal spatial topology 
$T^3 = S^1 \times S^1 \times S^1$ was mostly
to test how much the earlier findings depend on the topology choice, 
a little later one more, deeper reason for considering toroidal spacetimes was understood, namely,
the simplicity of the $S^3$ topology was found to be unhelpful in the analysis of the geometry. 
No simple way for
introducing spatial coordinates in that case was discovered,
whereas in the case of $T^3$ the existence of
noncontractible loops of nonzero winding numbers and
surfaces dual to them led to several new methods of
investigating the spatial structure of the universes, presented in this thesis.

\section{The demand for good observables}

The CDT model has a proper time, but in the spatial directions it is genuinely coordinate free.
There is no background geometry in the definition of the path integral. Geometric information provided by the model is local, in the form of the neighborhood relations between the elements of the geometry.
On the one hand, the absence of coordinates is commendable and fits in the spirit
of general relativity with its diffeomorphism invariance.
On the other hand, well-chosen coordinates are often very useful 
for describing physical phenomena. 
Physics should of course be independent of the choice of coordinate system, but
the complete lack of coordinates may also be considered problematic, 
as it makes it difficult to define physical observables possible to relate to
more analytic approaches to quantum gravity.
The existence of the proper time coordinate was 
crucial for determining the effective mini-superspace action,
where the spatial geometries were integrated over, yielding a description 
of a semiclassical mini-superspace geometry
and its quantum fluctuations
\cite{ajl1, agjl, Ambjorn:2005qt, Ambjorn:2004qm, Ambjorn:2007jv}. 
If there existed good coordinates in spatial directions, it might be feasible to 
measure an effective action of CDT taking into account 
not only the scale factor but also the spatial degrees of freedom.
 
In this thesis, several ideas to introduce coordinates and observables 
probing the geometry of triangulations are presented.
Capturing the global properties of the system without a good choice of coordinates is difficult. It was
not a priori clear if such a choice is at all possible for a highly fluctuating geometry. 
In the following chapters it is shown that such coordinates are not only possible but 
also provide a much better understanding of the structures present in quantum geometric configurations.
All the methods utilize the properties of the toroidal topology of the spatial geometry, in which 
there exist loops of nonzero winding numbers and hypersurfaces orthogonal to them.

It is true that a single configuration in the path integral
of a quantum theory is not physical and cannot be measured in the real world.
At most, its discretized finite version can be generated in a computer simulation.
A value of a physical observable is one that is averaged over all configurations of the path integral. 
Nevertheless, a single CDT configuration from the path integral is
not without interest. If it is adequately large and ``typical'' to be representative
of the whole ensemble at the given point of the phase space,
then in some cases it might be sufficient for calculating the value
of an observable and obtaining the correct answer (up to finite-size corrections).
In principle, most of the results mentioned 
in the previous section could have been determined that way. Thus, it would be advantageous to understand the nature of an individual configuration in the path integral, which might be used to calculate certain observables
even if it is not an observable itself.
Perhaps in the near future these methods will be used to
formulate an effective action that would include all the spacetime directions.

\section{The author's contributions to the field}

Four-dimensional CDT with the toroidal spatial topology was first studied by a
team whose part was the present author \cite{torusz}, 
before the work described in this thesis was started.
Subsequently, the author pioneered the approach of using fictitious boundaries of the 
elementary cell and shortest winding loops to study the structure of quantum geometries
of CDT.

The first part of the work involved adding and testing new procedures to the basic program for 
simulations of CDT, 
written and maintained by A. Görlich and 
based on an earlier Fortran program written by J. Ambjørn and J. Jurkiewicz \cite{ajl1}.
Some of the procedures written were used for encoding the boundaries of the elementary cell, 
some for updating the position of the boundaries during the moves, while 
making the boundaries possibly small and regular (see the following chapter), and some for measuring each of the observables described.
The procedures written for finding shortest loops had several versions with various treatment of the boundaries.

Another part of the work was designing and maintaining simulations to generate appropriate
well-thermalized configurations. Generating each of the simulations with scalar fields took several months of computer time. During that time, they needed some fine-tuning of the $K_4$ parameter, archiving, and restarting after power outages and computer failures.

When the configurations appropriate for purposes described in each of the following 
chapters were generated, the next step was to perform the measurements.
The first stage involved specially written procedures in C.
The configurations were copied to a larger number of computers belonging to the Jagiellonian University
Theory of Complex Systems Department, including the cluster Shiva, as, contrarily to the
simulations themselves, these calculations could be split into smaller parts
and parallelized. Especially resource-consuming was the search for loops of high winding numbers starting from each simplex of a configuration, which
lasted for several months, often using more than a hundred processors at a time.
The second stage was the analysis and visualization of obtained data.
This was done using procedures written in Mathematica.
The final part of the work was 
presenting the results and editing the articles themselves.

The author also implemented the measurement
of the quantum Ricci curvature in four-dimensional CDT
on the torus. The results can be compared with the similar measurements performed by Nilas Klitgaard and Prof.~Renate Loll 
in the case of the spherical spatial topology. The results confirm the richer structure of geometry in the toroidal case.
They serve at the same time as a link between non-local, topological observables such as the length of the shortest non-contractible
loop passing through a simplex and observables that are more local, such as the quantum Ricci curvature and the Hausdorff dimension.

Gradually, the promising results obtained by the author
and the increased understanding of the spatial geometry of triangulations
inspired more members of the team working on CDT in the 
Theory of Complex Systems Department at the
Jagiellonian University to use configurations with boundaries and some
of the same procedures.
The author collaborated with the rest of the team in the research on CDT with dynamical scalar fields,
personally generating and analyzing a number of configurations with various values of 
scalar field jumps. The results of that analysis are described in Chapter 6.

\section{Outline of the thesis}

The thesis is based on six articles:

\begin{enumerate}[start=1,label=\lbrack \arabic*\rbrack]
\item J.~Ambjørn, Z.~Drogosz, J.~Gizbert-Studnicki, A.~Görlich, J.~Jurkiewicz,
\textit{Pseudo-Cartesian coordinates in a model of Causal Dynamical Triangulations},
 Nucl.\ Phys.\ B \textbf{943} (2019) 114626
\item J.~Ambjorn, Z.~Drogosz, A.~Görlich, J.~Jurkiewicz,
\textit{Properties of dynamical fractal geometries in the model of Causal Dynamical Triangulations},
Phys.\ Rev.\ D \textbf{103} (2021) 086022
\item J.~Ambjorn J, Z.~Drogosz, J.~Gizbert-Studnicki, A.~Görlich, J.~Jurkiewicz, D.~Németh, 
\textit{CDT Quantum Toroidal Spacetimes: An Overview}, Universe  \textbf{7}(4) (2021) 79
\item J.~Ambjorn, Z.~Drogosz, J.~Gizbert-Studnicki, A.~Görlich, J.~Jurkiewicz, D.~Németh, 
\textit{Cosmic Voids and Filaments from Quantum Gravity}, arXiv:2101.08617 
\item J.~Ambjorn, Z.~Drogosz, J.~Gizbert-Studnicki, A.~Görlich, J.~Jurkiewicz, D.~Németh,
\textit{Matter-driven phase transition in lattice quantum gravity}, arXiv:2103.00198
\item J.~Ambjorn, Z.~Drogosz, J.~Gizbert-Studnicki, A.~Görlich, J.~Jurkiewicz, D.~Németh, 
\textit{Scalar Fields in Causal Dynamical Triangulations}, arXiv:2105.10086.
\end{enumerate}

Chapter 2 reviews the most salient aspects of the computer implementation of CDT Monte Carlo 
simulations. Chapter 3 describes the first attempt
at introducing coordinates to the spatial leaves of CDT, the so-called pseudo-Cartesian coordinates.
Chapter 4 describes observables based on non-contractible loops with nonzero winding numbers, which 
behave in many ways better than the pseudo-Cartesian
coordinates and have allowed for a quite precise analysis of the
shape of a generic CDT configuration. Chapter 5 compares the results of analysis based on
topological observables with a more local observable, the recently invented
quantum Ricci curvature, and shows that the four-dimensional model of CDT with toroidal
spatial topology has a richer structure than one with spherical spatial topology. Chapter 6 describes
the latest attempt at introducing coordinates to CDT via scalar fields with nontrivial boundary conditions. The same chapter also presents some results in 
a model of CDT with scalar fields dynamically coupled to geometric degrees of freedom. Finally, Chapter 7 concludes the thesis with an interpretation of the results and an outlook on future research.

\chapter{Implementation}\label{chap:implementation}
\markboth{Implementation}{Implementation}

In the absence of analytic solutions of Causal Dynamical Triangulations in four dimensions,
the main way to proceed is to analyze results of numerical Monte Carlo simulations.
This chapter discusses some of the technical details of those computer simulations, including the 
description of the encoding of the configuration and the choice and execution of moves
by means of which the triangulation is evolved.
Of special note are the boundaries defining an elementary cell, which are the key tool
in this work.

\section{The encoding and storage of configurations}

The program used for Monte Carlo simulations was written in C in 
years 2006-2015 by A. Görlich \cite{phdatg} on the basis of the Fortran program used earlier 
by J. Ambjorn and J. Jurkiewicz \cite{ajl1}. 
As of the time of writing this thesis, A. Görlich continues to update
the base of the program, improving its efficiency.
In order to measure new observables, suitable procedures are added to the program,
according to the needs.
The present author has added procedures for the encoding and updating of boundaries
and for the measurement of quantities described in this thesis.

The ensemble of triangulations appearing under the path integral of CDT is probed by 
performing a random walk, starting from a simple configuration and evolving
it gradually by a long series of moves, each of which operates locally,
modifying a region containing a few simplices.
The moves preserve the topology and causality of the triangulation and 
adhere to the so-called detailed balance condition, which ensures that the 
probabilities of obtaining any configuration are correct and consistent with the action.
Moreover, the set of all the moves is ergodic, which means that any configuration can be 
reached from any other configuration with the same topology through a finite series of moves.

The configurations model the labeled triangulations appearing under the path integral of Causal Dynamical Triangulations. 
The data that are sufficient unambiguously to encode a configuration are: 
(i) a list of labels of all vertices together with their corresponding
time parameters; (ii) a list of labels of all four-simplices together with their
vertices and the neighboring simplices positioned opposite to each of the vertices.
If there were no periodic boundary conditions imposed on the time interval,
then it would moreover be necessary to indicate the initial and final state.
Both during the execution of the program and during the analysis of the stored configurations
it is advantageous to have an immediate access to other data as well.
Therefore, the program also stores information about the
links (one-dimensional simplices), and triangles (two-dimensional simplices).

It is worth mentioning that any CDT configuration has also an equivalent description,
called the \textit{dual lattice}, which is a graph whose vertices (nodes) correspond to the 
four-simplices of the triangulation and the edges (links) to the common three-dimensional faces
of simplices and their neighbors. As each four-simplex has exactly five neighbors, so 
also each vertex in the dual lattice picture is connected by edges to exactly five 
other vertices. The dual lattice will be referred to in the following chapters.

All Monte Carlo simulations in four-dimensional Causal Dynamical Triangulations
with toroidal spatial topology start from essentially the same simple configuration.
Each time slice (each leaf of the foliation) is constructed in an identical way,
and the only choice left is the number of time slices.
In the initial
phase of the simulations, the configuration 
expands rapidly to reach a number of simplices close to the number chosen beforehand; during this phase and the subsequent
thermalization, the number of moves performed 
is so large that any influence of the precise shape of the initial configuration
is erased, apart from only the choice of the spatial topology and the number of time slices. 
Therefore, a logical choice for the initial
configuration is one that is easy to create, picture and encode.
Such exactly is the initial configuration used,
in which each time slice is built from a regular Cartesian grid of
$4^3=64$ four-dimensional hypercubes, each of which in turn is constructed from $16$ simplices.
The way to triangulate a four-dimensional cube was described in \cite{cube}.
In most of the simulations whose results are described in this thesis,
the number of time slices $T$ was chosen to be $4$, so the initial configurations contained $4^4 \cdot 16=4096$ simplices. 
To make the topology toroidal, the faces of the outermost hypercubes were
pairwise identified in the appropriate way.
As mentioned in the previous chapter, for technical reasons
(to avoid choosing the initial and final states and to facilitate
the simulations), the time interval is made periodic. Thus, the  
configuration is topologically a four torus, i.e., a Cartesian product of four circles.

\begin{figure}
\centering
\includegraphics[scale=0.5]{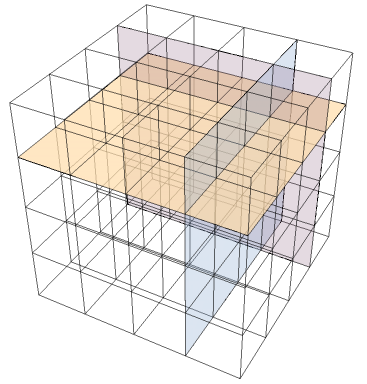}
\caption{A picture of a time slice in the initial configuration, showing the $4^3$ hypercubes (each divided into 16 simplices) and one of the possible initial positions of the boundaries.}\label{fig:cube}
\end{figure}

A four-torus can be regarded not only as a finite manifold, but also, equivalently, 
as an infinite periodic object.
All the $N_4$ four-simplices of the torus are contained in an elementary cell, 
repeated infinitely many times in four basic directions.
In the initial configuration, the aforementioned outermost faces
might serve as boundaries of the elementary cell. However, there are also
many other equivalent choices of the boundaries, such as the one pictured in Fig. \ref{fig:cube}.
The initial position of the boundaries is unimportant; it only matters that they should have
the correct topological properties (see Section \ref{sec:simulations}).

In order for the boundaries (labeled in four directions by $\sigma=1,2,3,4$) 
to delimit an elementary cell 
of the evolving configuration at every moment of the simulation, 
their position has to be kept track of and updated during each move.
The boundaries are encoded by five quadruples of numbers for each four-simplex:
one quadruple for each face of the simplex, and one number within 
a quadruple for encoding each of the four boundaries.
Each of these numbers is equal either to $0$ if the face does not belong to 
the given boundary or $\pm 1$ if the face belongs to the given boundary;
the sign defines the orientation of the boundary.
Perhaps an easier way to visualize those numbers and to use them in calculations is to
imagine them as four matrices $\mathbf{B}^\sigma$ with
elements $B_{ij}^\sigma$ equal to $\pm 1$ if the face between simplices $i$ and $j$ exists and belongs to the boundary and to $0$ otherwise. 
The boundaries are oriented,
$B_{ij}^\sigma=-B_{ji}^\sigma$, and the sum of $B^\sigma$ along a closed loop 
is equal to the winding number in direction $\sigma$.

\section{The simulations}\label{sec:simulations}

During the simulations, the triangulations are updated step by step
through the execution of moves. In order to obtain a well-thermalized configuration,
typically billions of moves are performed, which may take several months of processor time. 
The number of moves required for thermalization depends mostly on the total number
of simplices in the configuration and on the point in the phase space
at which the simulation is performed.
This large consumption of computing time 
is the main obstacle against increasing the volume of the configurations further,
especially since no method of parallelization of these computations
has been invented.

The moves used in CDT in four dimensions are of seven types.
Let us omit a detailed description of each of them 
(available, e.g., in \cite{phdatg}).
It is enough to mention that each move is local and transforms, according to fixed rules,
a cluster of a given number of connected simplices (from 2 up to 8, depending on the move)
into a different cluster of simplices. Each move is either self-dual or has an inverse move.
In each step, first the type of the move to be performed and the place of its execution
are randomly chosen. Then, if a procedure confirms that the move executed at that place 
would not cause a topological defect, the move is either performed or rejected,
with a probability consistent with the action.

In order for the position of the boundaries to be updated correctly, the procedures
responsible for the execution of each move were expanded by the present author.
Essentially, if a boundary passes through the simplices that are to be affected by 
the move, the boundary is deformed to sidestep that region. Then, after the execution
of the move, the boundary may reenter that region, if a checking procedure determines that it would
lead to a decrease of the boundary's size without introducing topological defects.

While planning to run a new simulation, one has to decide on the choice of
parameters $K_0$ and $\Delta$, the number of time slices $T$, the average total number of $(4,1)$-simplices and, of course, on the spatial topology.
From a practical point of view, it is convenient to keep $N_4$ or $N_{4,1}$ fixed 
by fine-tuning $K_4$.
Most of the simulations whose results are described
in this thesis were performed at the canonical point in the phase space
of toroidal CDT, i.e., in the $C$ phase, for the parameters $K_0 = 2.2$ 
and $\Delta = 0.6$ and for $N_{4,1} = 160000$ 
(customarily denoted $N_{4,1}=$160k or even simply volume=160k).

\chapter{Pseudo-Cartesian coordinates}\label{chap:cart}
\markboth{Pseudo-Cartesian coordinates}{Pseudo-Cartesian coordinates}

This chapter is based on the following publication:
\begin{enumerate}[start=1,label=\lbrack \arabic*\rbrack]
\item J.~Ambjørn, Z.~Drogosz, J.~Gizbert-Studnicki, A.~Görlich, J.~Jurkiewicz,
 Pseudo-Cartesian coordinates in a model of Causal Dynamical Triangulations,
 Nucl.\ Phys.\ B \textbf{943} (2019) 114626.
\end{enumerate}

As explained in the previous chapter, the quantum geometries studied in this work
have the topology of the four-torus $T^4$, i.e., the Cartesian product of four circles.
Each geometry is a piecewise linear manifold, which can equivalently
be thought of as an infinite periodic structure consisting of
an elementary cell delimited by non-contractible three-dimensional boundaries
that is periodically repeated in four directions.

This chapter describes the use of these boundaries 
as a reference frame for a set of coordinates defined by the distance 
(on the dual lattice) from a simplex to 
each of the boundaries (the set of coordinates is termed ``pseudo-Cartesian'' because 
of the similarity
to a Cartesian coordinate system in the case of Euclidean space; some intuition
from that case can be used here despite the very irregular structure of the triangulations).
The precise position and shape of the boundaries are inconsequential for the
possibility of this
construction, as long as they define the elementary cell correctly
and have no topological defects; nonetheless, clearly, it is convenient and beneficial for
the properties of the coordinates if the boundaries are ``regular'' in some sense, 
e.g., contain as few 3-simplices as possible rather than fill a greater part of the configuration.
This is why a procedure trying to minimize the boundary was introduced,
as described in the previous chapter.
Although boundaries are fictitious and do not influence the action and the physics, they are useful,
as evinced by this and the following chapters, and the 
possibility of their introduction is one of the reasons for which
the toroidal CDT can be considered a more interesting object of study than spherical CDT.

For each boundary and each simplex two values of pseudo-Cartesian coordinates can be defined,
depending on the direction (one could say clockwise or counter-clockwise)
in which the path connecting the simplex to the boundary 
goes around the torus (along the loop orthogonal to the given boundary).
Hence in the four-dimensional case, with the dimensions labelled
in a standard way by the letters $x$, $y$, $z$ and $t$,
each simplex can be assigned eight numbers, which can be denoted
by the letters $\mathbf{x}$, $\mathbf{y}$, $\mathbf{z}$, $\mathbf{t}$, and
$\mathbf{x'}$, $\mathbf{y'}$, $\mathbf{z'}$, $\mathbf{t'}$, for the clockwise and the 
counter-clockwise directions, respectively.

\begin{figure}
\begin{center}
{\includegraphics[width=\textwidth]{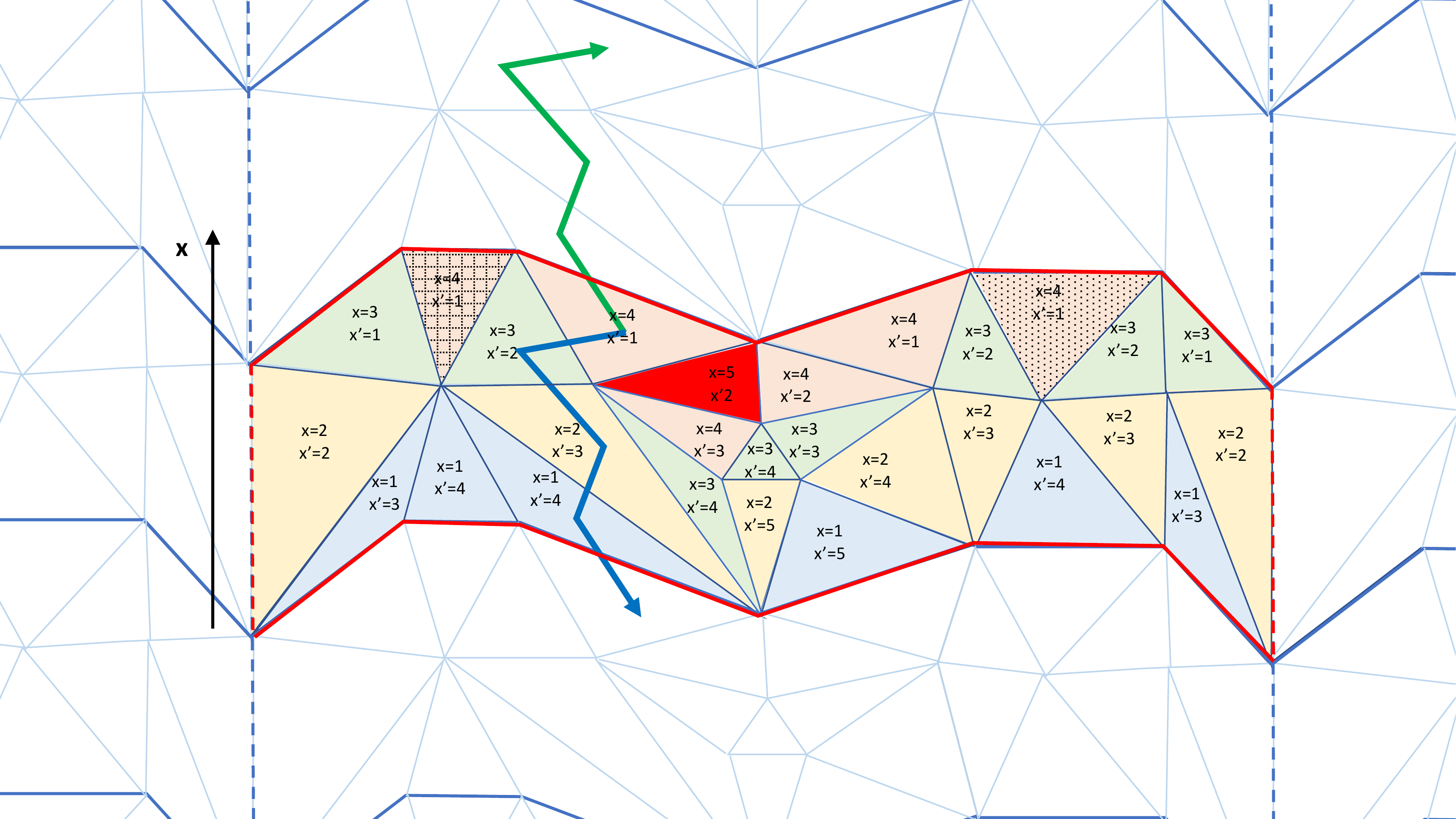}} 
\end{center}
\caption{The assignment of the pseudo-Cartesian coordinates to simplices in a 1+1-dimensional example. The boundaries of the elementary cell are marked as red lines. The green and blue arrows show the minimal loop from one chosen simplex to its copy in a neighboring elementary cell.}
\label{fig:visualize}
\end{figure}

An advantage of the pseudo-Cartesian coordinates (as compared with the more precise observables
to be defined in the following chapters)
is that they can be assigned very quickly to all simplices in a configuration. 
The algorithm used works in the following way:
first, the simplices touching the boundary from one of the two sides are
marked with the number one; then from these simplices a diffusion wave, for which
the given boundary is impermeable, is started. The number of steps the diffusion wave requires
to reach a given simplex, plus one, is the distance (on the dual lattice) to the boundary
and therefore the value of the coordinate of the simplex.
Figure \ref{fig:visualize} shows a visualization of the procedure in the 1+1-dimensional case.

\begin{figure}
\begin{center}
{\includegraphics[width=12cm]{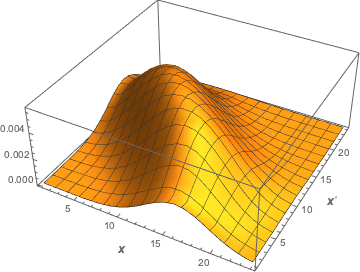}} 
\end{center}
\caption{The  distribution  $P({\bf x},{\bf x'})$ for a system with $N_{4,1}=$160k.}
\label{xxprim}
\end{figure}

The pseudo-Cartesian coordinates in the time direction are trivially equivalent to the built-in coordinates defined by the time foliation. Also trivially, ${\bf t}+{\bf t'}=4T+1$.
The coordinates in the spatial directions have more interesting properties.
Figure \ref{xxprim} presents the distribution
of coordinates ${\bf x}$ and ${\bf x'}$ in a system with $N_{4,1}=$160k.
The data were obtained by averaging over all simplices in 800 statistically independent configurations.

\begin{figure}
\begin{center}
{\includegraphics[width=7cm]{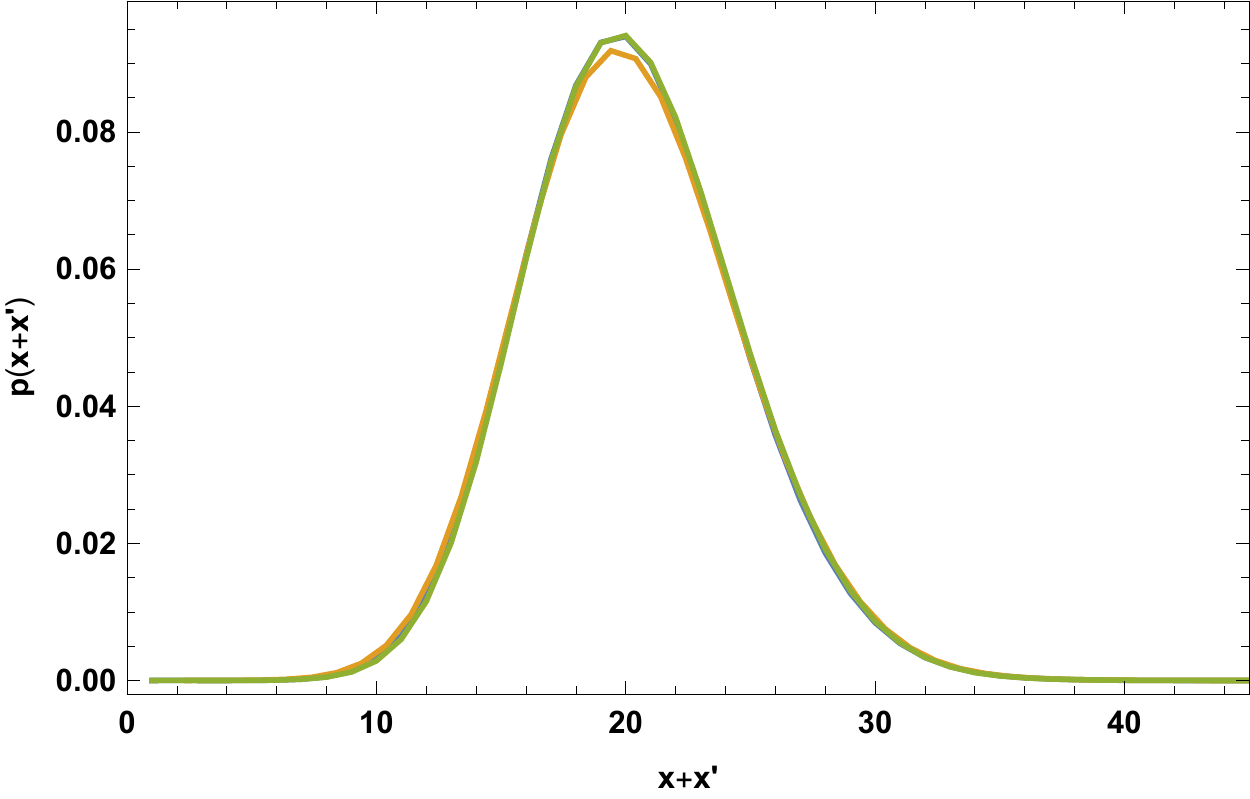}} 
{\includegraphics[width=7cm]{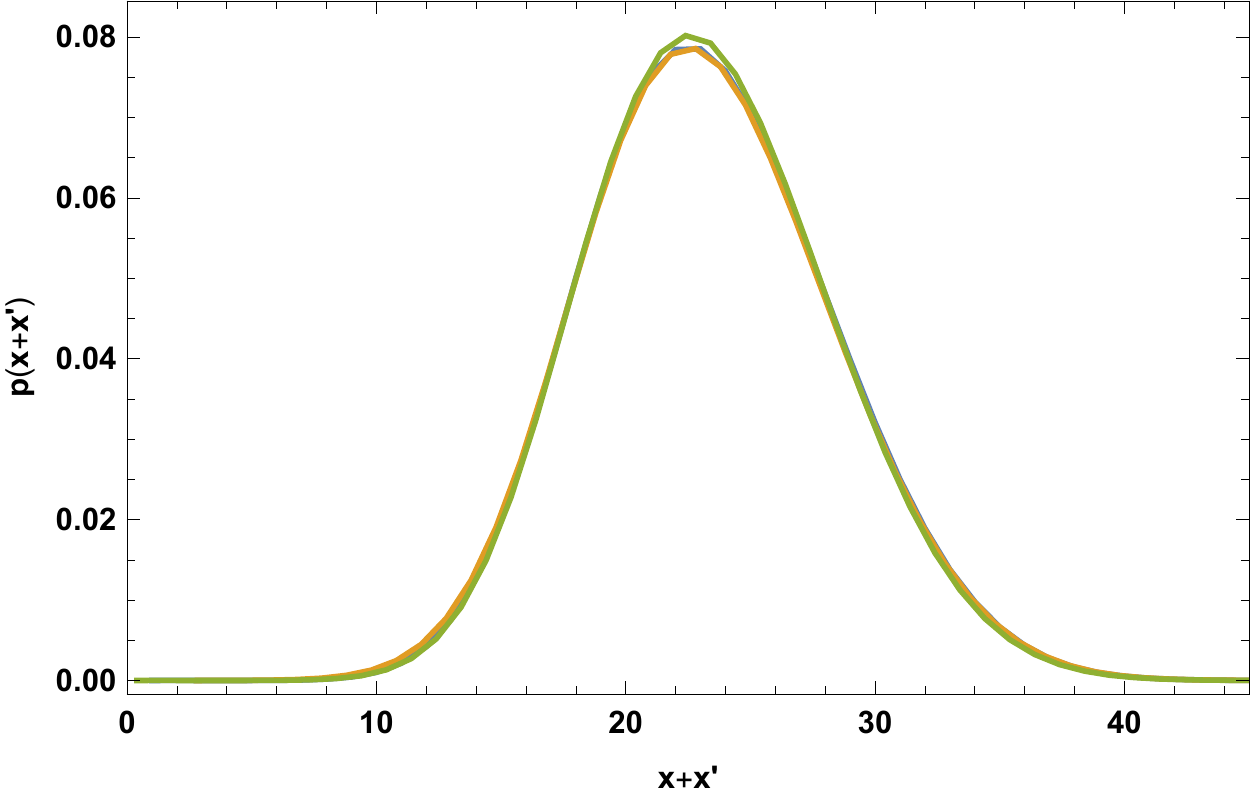}} 
\end{center}
\caption{Distributions of $L_x$(blue),  $L_y$ (green) and $L_z$ 
(orange) for systems with $N_{4,1}=80k$ (left) and $N_{4,1}=$160k (right).}
\label{shifts}
\end{figure}

\begin{figure}
\begin{center}
{\includegraphics[width=10cm]{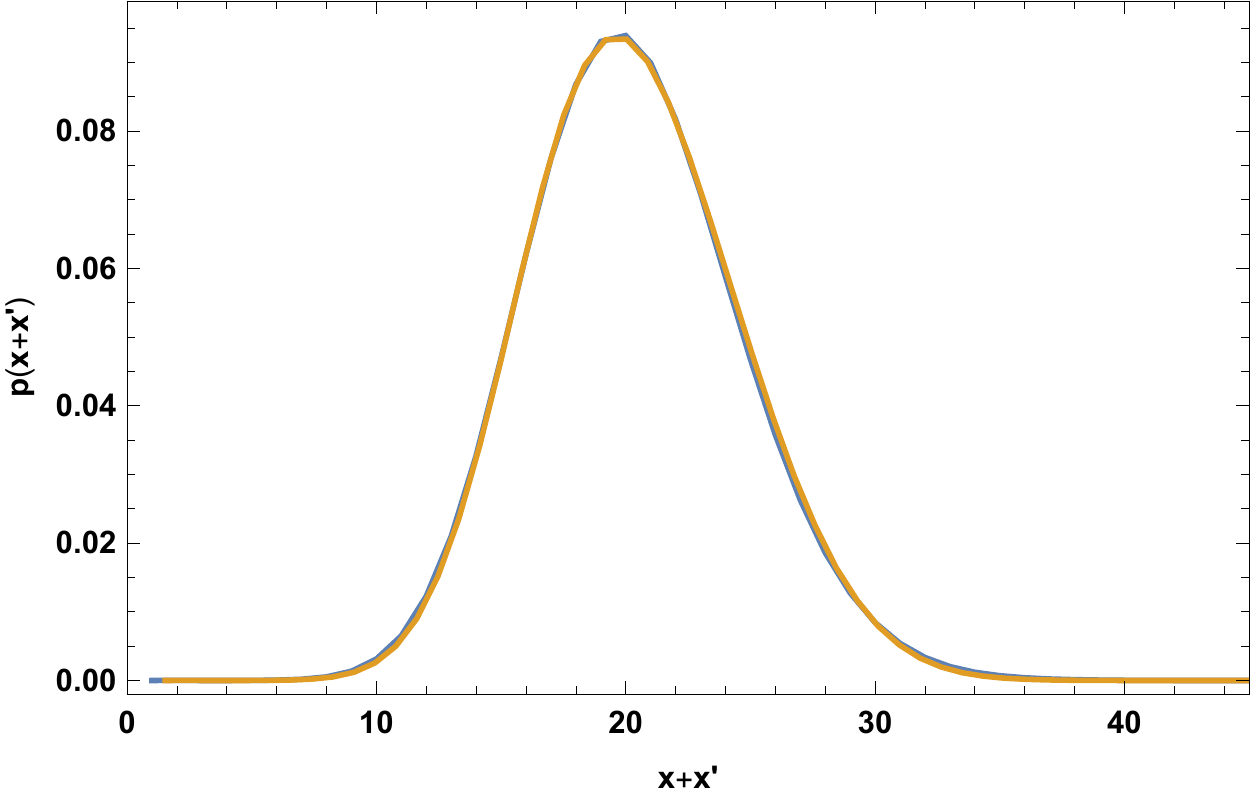}} 
\end{center}
\caption{The scaling of a distribution $p({\bf x}+{\bf x'})$ for two system sizes ($N_{4,1}=$80k and 160k,
colored  blue and orange, respectively).
For the larger system the distribution is rescaled by a factor $1/2^{1/4}$ }
\label{scaling}
\end{figure}

\begin{figure}
{\includegraphics[width= 0.5 \textwidth]{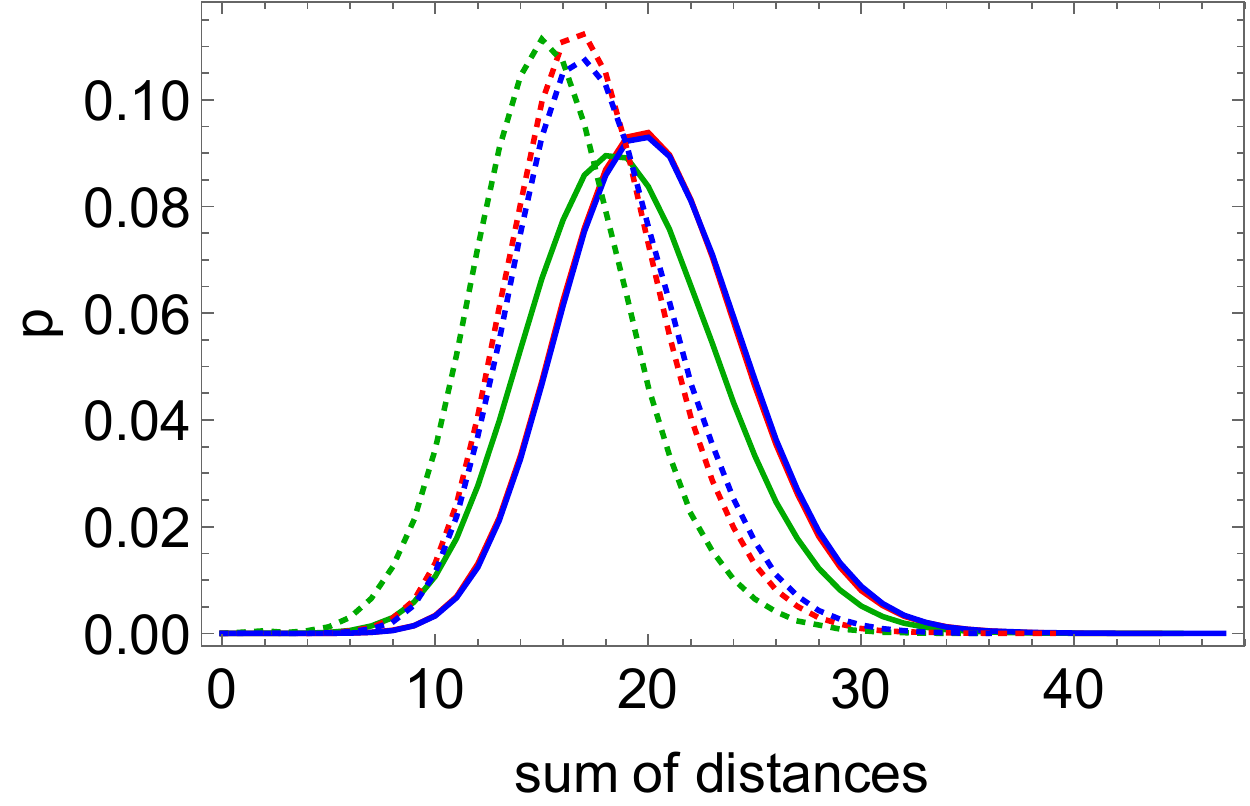}}
{\includegraphics[width= 0.5 \textwidth]{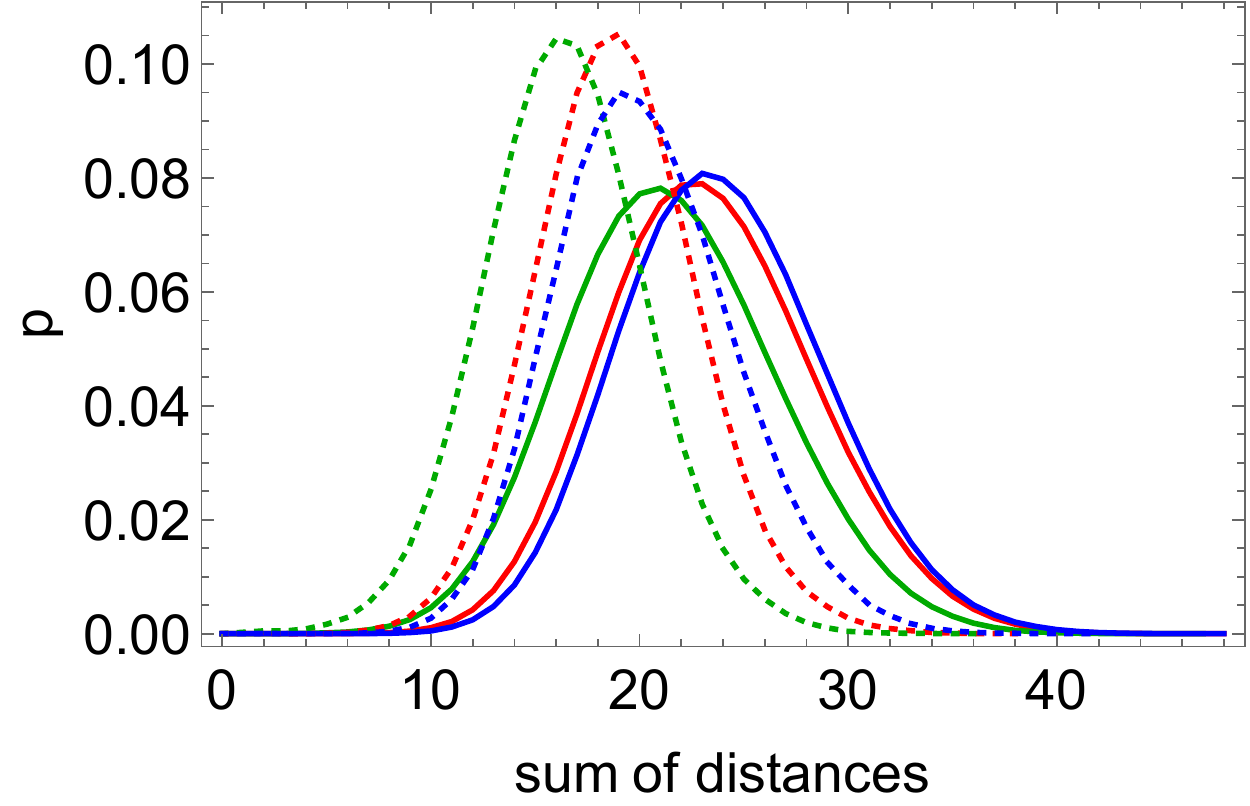}} 
\caption{Distributions of a sum of distances from simplices to the two opposite boundaries
in the $x$ direction (red), the $y$ direction (green) and $z$ direction
(blue) for systems with $N_{4,1}=80\mathrm{k}$ (left) and $N_{4,1}=160\mathrm{k}$ (right).
The distributions scale consistently with the Hausdorff dimension $d_H=4$.
The dotted lines refer to simplices adjacent to the boundary,
$x = 1$ or $x' = 1$ respectively for the two sides of the boundary.}
\label{fig:shifts}
\end{figure}

The quantity $L_x = {\bf x}+{\bf x'}$ (and analogous quantities $L_y$ and $L_z$)
is closely correlated (but not equivalent)
to the length of a minimal non-contractible loop passing through the given simplex,
to be discussed in the next chapter.
In a regular hypercubic lattice
a sum of distances from any simplex to the two opposite boundaries is a constant,
equal to the geodesic distance between the boundaries.
In a generic CDT configuration this is, however, not the case.
The distribution of these values 
is non-trivial (Fig. 3.3),
which may indicate either that the shape of the elementary cell is far from being rectangular,
or that quantum fluctuations of the geometry can be viewed as ''mountains`` and in effect simplices close to the top of the mountains have a larger geodesic
distance to the boundaries than those lying in the valleys between the mountains.
Results indicate that both effects may be important.
The latter effect is supported by the difference visible in Fig.~\ref{fig:shifts} between
the distributions $P(x + x')$ for all simplices (solid lines) and simplices adjacent to a boundary (dotted line, $x = 1$ or $x' = 1$).
Boundaries are chosen to locally minimize their area,
thus they prefer the central region of a torus (valleys) and omit outgrowths (mountains).
Therefore, simplices adjacent to one side of a boundary are closer to the 
other side of the boundary than an average simplex.
Figure \ref{shifts} shows those distributions for systems with $N_{4,1}=80k$
and $N_{4,1}=160k$. In both cases an overlap is achieved by applying shifts of order one.
The shape of distributions seems to be the same up to a scaling, 
depending on the total volume. Figure \ref{scaling} compares the two volumes ($N_{4,1}=$ 80k and 160k), after a scaling factor $1/2^{1/4}$ was applied to the 
distribution in the larger system. This agrees with the expected scaling, if the
Hausdorff dimension in the C phase is indeed $d_H=4$.

Another interesting quantity is the difference between $x$ and $x'$, which indicates
the position of a simplex relative to the boundaries in a given direction.
The distribution of this quantity for each spatial direction might be considered 
an analogue of the volume profile in the time direction. As the plot of the latter
quantity is approximately constant in CDT with toroidal spatial topology,
\cite{torusz, torus22}, a similar shape could also be expected here.
One difference is that the period in the spatial directions is not constant as in the
time direction, where it is equal to the number of time slices.
Even more importantly, the boundary in the time direction is simply built out of 
a set of faces of simplices lying at one of the time slices, and its size is not minimized,
whereas the boundaries in the spatial directions do not lie at a generic section
of the torus but are minimized by an appropriate procedure, as mentioned
in the previous chapter, and, so to speak, find a much narrower section.
Therefore, after averaging over many configurations
the number of simplices near a boundary in a spatial direction will likely remain 
lower than the average for a random section of the torus. Thus the spatial 
translational symmetry of the plot is broken by the positioning of the boundary.

\begin{figure}
\begin{center}
{\includegraphics[width=10cm]{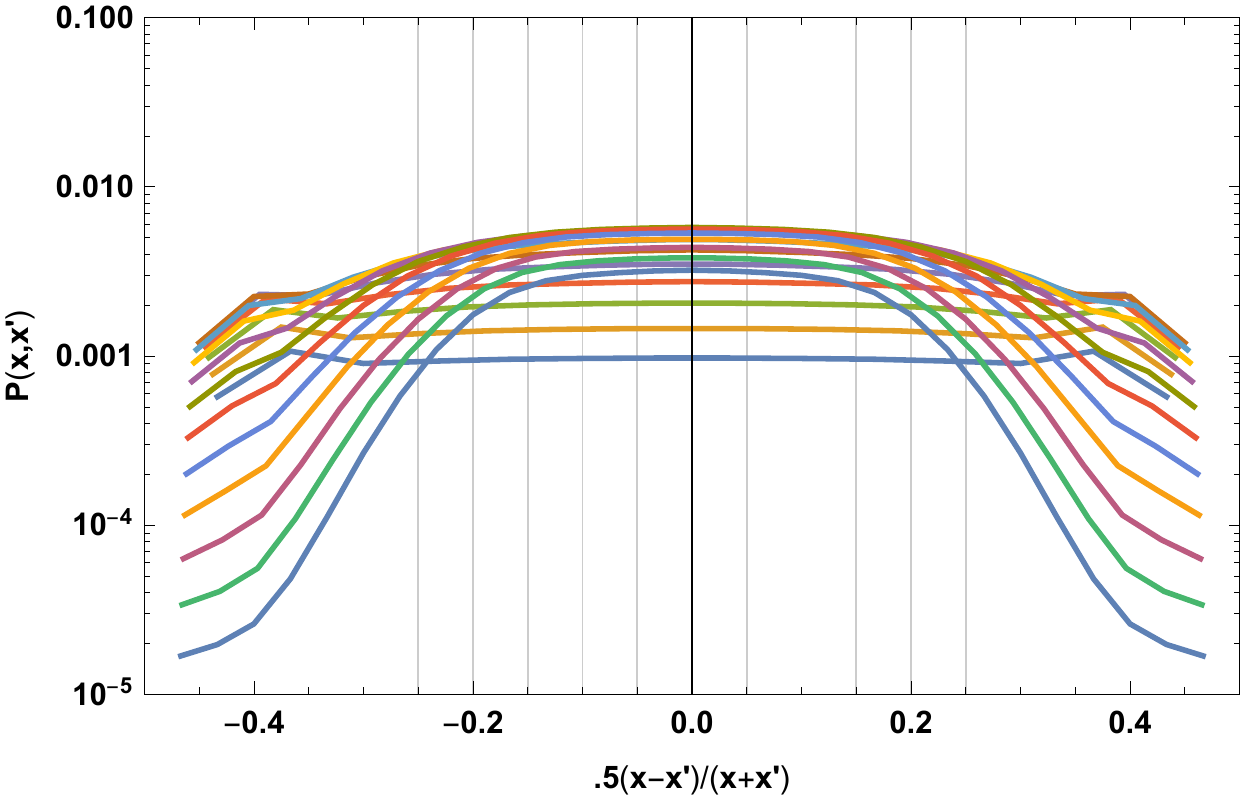}} 
\end{center}
\caption{The central part of the distribution $P({\bf x},{\bf x'})$ as a function
of the rescaled variable $\oh({\bf x}-{\bf x'})/({\bf x}+{\bf x'})$. The curves correspond to increasing values
of ${\bf x}+{\bf x'}$ in a sequence of colors: blue, orange, green etc.}
\label{middle}
\end{figure}

\begin{figure}
\begin{center}
{\includegraphics[width=10cm]{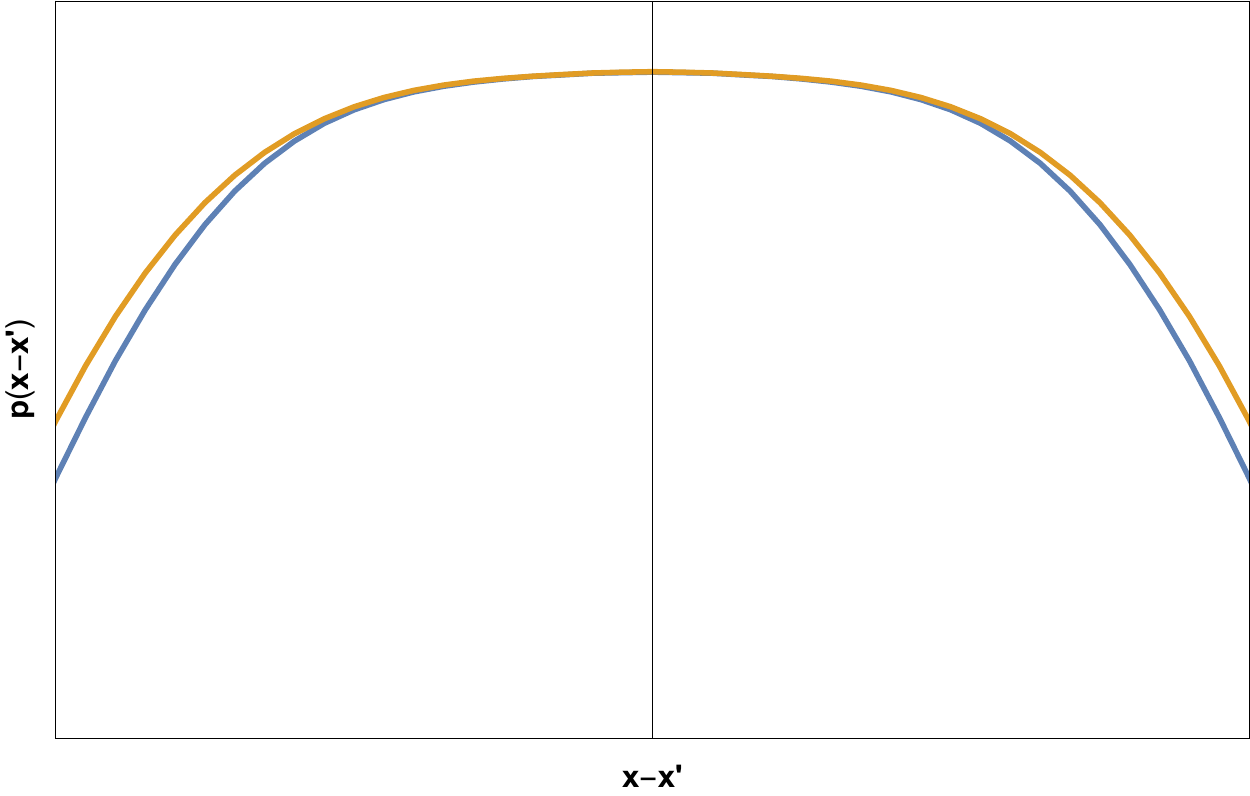}} 
\end{center}
\caption{The distribution $p({\bf x}-{\bf x'})$, integrated over ${\bf x}+{\bf x'}$, as a function
of ${\bf x}-{\bf x'}$ for volumes $N_{4,1}=$80k (blue) and 160k (orange).}
\label{middlescale}
\end{figure}

Still, inside and near the central area, where ${\bf x}\approx {\bf x'}$,
the influence of the position of the boundary should be low, 
and the distributions are expected to be approximately flat in the infinite volume limit.
In Figure \ref{middle}, the dependence of the volume distribution $P({\bf x},{\bf x'})$ 
as a function of the rescaled variable $\oh({\bf x}-{\bf x'})/({\bf x}+{\bf x'})$ for a range of values of 
$15\leq {\bf x}+{\bf x'}\leq 30$ is shown for a system with $N_{4,1}=$160k.
The distributions are indeed approximately flat in the central range, and 
the plateau appears to increase and flatten with increasing size of the configurations,
as seen by comparing distribution with volume 80k and 160k (Figure \ref{middlescale}).

The pseudo-Cartesian coordinates 
were the first observable to show that the spatial structure of a toroidal
quantum geometry could be imagined as a bulk of connected four-simplices
decorated by quantum outgrowths consisting mainly of simplices with large $L_x$, $L_y$, $L_z$,
whose existence is evinced by the plots.
This suggested a semiclassical interpretation of the four-dimensional
universe, 
which is not so different from the situation in two-dimensional Euclidean quantum gravity, where
a typical configuration, when sliced appropriately (choosing the appropriate coordinates for the configuration 
in question), can be viewed as a main universe dressed with quantum outgrowths (baby universes) \cite{kawai6}. 
In the case of two-dimensional Euclidean quantum gravity, this slicing and its associated fractal structure determines most aspects of the theory.
The analysis of pseudo-Cartesian coordinates suggested that the structure is 
less fractal (i.e., more semiclassical) than in the two-dimensional case, 
in the sense that there are fewer outgrowths, and they carry less volume than in two-dimensional quantum gravity, 
although the outgrowths still contain the greatest part of all simplices.
 
The advantage of the pseudo-Cartesian coordinates is their simplicity and 
quickness of calculation.
However, their properties are not ideal.
Although each three-dimensional boundary is simply connected, and, by construction,
the hypersurface of x=1 is connected as well, this is often not the case
for hypersurfaces of constant pseudo-Cartesian coordinates of higher value.
The values of coordinates and of the quantities of the type $L_x$,
moreover, depend on the position of boundaries.
Therefore, the search for better observables continued.

\chapter{Non-contractible loops}\label{chap:loops}
\markboth{Non-contractible loops}{Non-contractible loops}

This chapter is based on the following publications:
\begin{enumerate}[start=1,label=\lbrack \arabic*\rbrack]
\item J.~Ambjørn, Z.~Drogosz, J.~Gizbert-Studnicki, A.~Görlich, J.~Jurkiewicz,
 \textit{Pseudo-Cartesian coordinates in a model of Causal Dynamical Triangulations},
 Nucl.\ Phys.\ B \textbf{943} (2019) 114626
\item J.~Ambjorn, Z.~Drogosz, A.~Görlich, J.~Jurkiewicz,
\textit{Properties of dynamical fractal geometries in the model of Causal Dynamical Triangulations},
Phys.\ Rev.\ D \textbf{103} (2021) 086022.
\end{enumerate}

\section{Loops with unit winding numbers}

The pseudo-Cartesian coordinates described in the previous chapter shed some light on the structure of typical geometries in the C-phase of four-dimensional toroidal CDT. They 
suggested a picture of the geometry in which there is a small,
seemingly semiclassical, toroidal center
and  outgrowths of almost spherical topology and  
complicated, fractal-like structure, 
which can be interpreted as quantum fluctuations.
However, the pseudo-Cartesian coordinates
are not optimal, 
as they depend on the position and shape of the boundaries. Moreover,
a hypersurface of constant coordinates is typically not connected. 
Although the boundaries tend to be minimized by the appropriate procedures, as mentioned
in previous chapters, they
still usually have complicated structure,
which impairs the properties of the coordinates.
Therefore, another, more precise way of analyzing 
and mapping the configurations was sought.

\begin{figure}
\centering
\includegraphics[height=55mm]{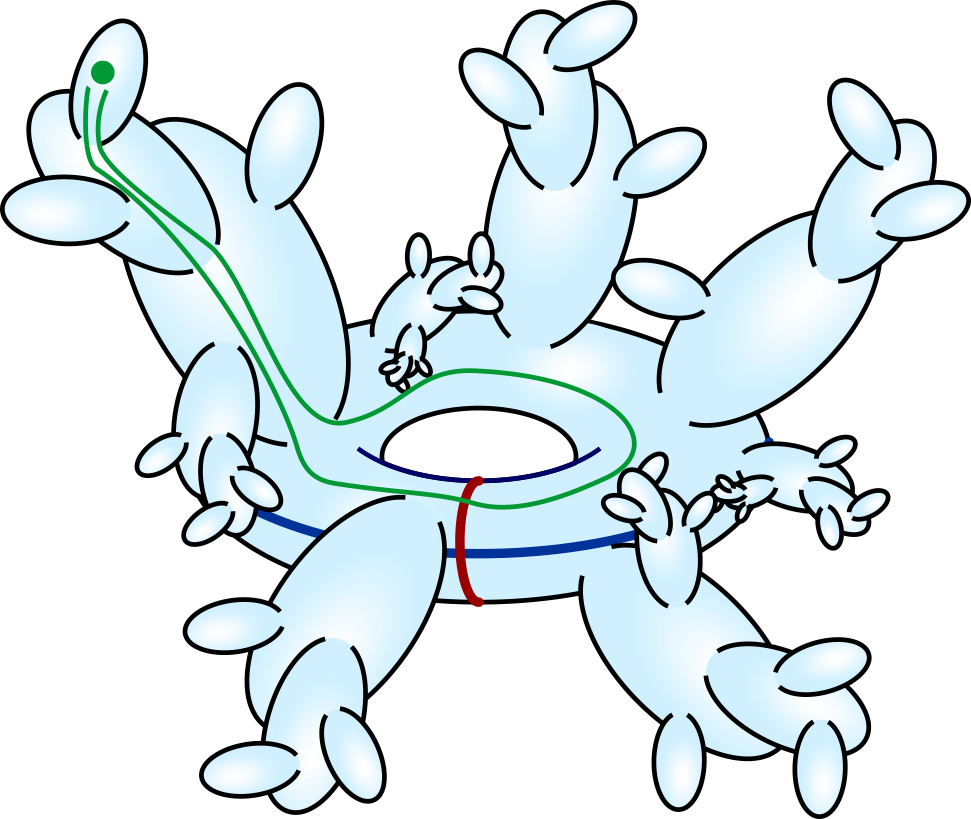}
\caption{
Illustration of a torus with outgrowths. 
The blue and red lines represent two non-equivalent and non-contractible loops.
The green loop is the shortest loop passing through the green point with the same winding number as the blue line.}
\label{fignew}
\end{figure}

The non-contractible geodesic loops, described in this chapter, are topological observables
independent of the position and shape of the boundaries.
As suggested by the interpretation of the results obtained via the pseudo-Cartesian coordinates, the triangulation can be imagined
to consist of a semiclassical toroidal structure and many outgrowths which contain most of the simplices.
The shortest possible non-contractible loop passing through a given simplex in an outgrowth would be significantly longer than 
the shortest non-contractible
loop starting at a simplex laying in
the semiclassical part.
Fig.~\ref{fignew} illustrates schematically 
the idea of this approach.
An example point in an outgrowth is marked with a green dot, and the green line is a non-contractible geodesic loop starting at that point. To form a loop, a curve starting in an
outgrowth must first exit the outgrowth,
then circle once around the torus, and 
then reenter the outgrowth and return
to the starting simplex.

In four-dimensional CDT with
the spatial topology of a three-torus $T^3$,
each configuration is topologically a Cartesian product of four circles, since
time interval also has periodic boundaries
imposed, as explained in Chapter \ref{chap:implementation}. 
Each closed curve within the configuration is homotopically equivalent to
a combination of those circles, and the coefficients of that linear combination are the four winding numbers of the loop,
which can be called the winding numbers in the $x$, $y$, $z$ and $t$ directions.

As described in a previous chapter,
an equivalent way of picturing a toroidal 
configuration is one in which the
configuration is an infinite system
periodic in four directions.
A loop corresponds in this picture to a path 
joining the same simplices in two different copies of the elementary cell.
It is natural then to assign a set of four numbers to each copy of the elementary cell
in such a way that the differences between them for any two copies 
are identical to the four winding numbers of the corresponding loop. 

Arguably it is the most convenient to look at loops in the dual picture (see Chapter 2), and so
henceforth the word ``loop'' will usually mean not a 
spacetime curve but an ordered set of connected simplices whose image in the dual lattice is a non-contractible directed cycle.
The length of a loop is the number of links in the cycle. (For simplicity it is assumed that all links
have the same length.)
Similarly, a geodesic between two simplices will mean a line connecting them whose 
image in the dual lattice has minimal length.
 This is not necessarily the shortest path 
if we consider our triangulation as a piecewise linear manifold. However, this difference
should be unimportant for generic fractal properties, 
such as the Hausdorff dimension,
in the limit of infinitely large triangulations.

In order to find the shortest loop of a given winding number
passing through a simplex, we treat the four-torus
as an infinite periodic system and follow step by step the front of a diffusion wave beginning at the chosen simplex
(using a diffusion wave in a system infinite in four directions is applicable to the case of low winding numbers;
otherwise this method becomes computationally inefficient and should be modified). 
The number of loops with a given winding number that pass through a simplex grows (eventually) exponentially fast with the loop length.
Thus, while it is feasible to list all the shortest loops of a given type,
in the case of longer loops we usually have to pick one sample loop, representing their
universal properties.

\begin{figure}
\begin{center}
{\includegraphics[width=10cm]{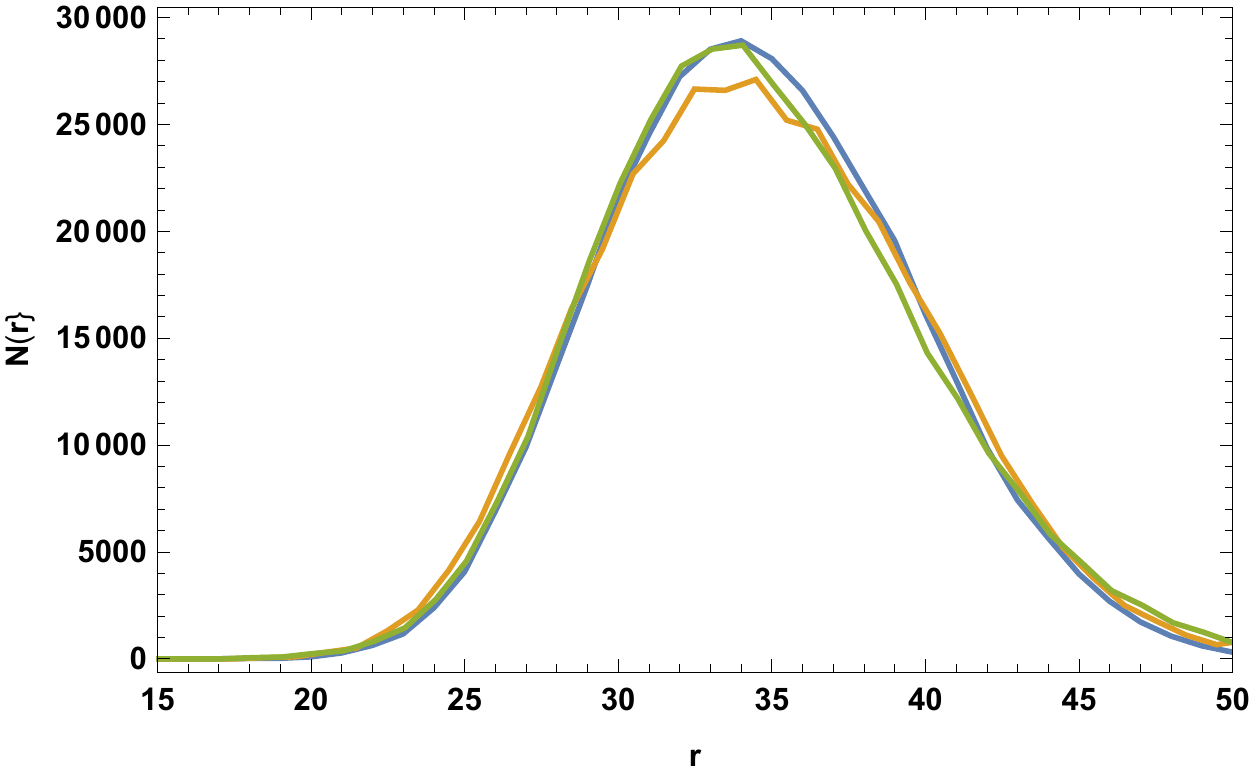}} 
\end{center}
\caption{Distributions of loop distances for loops with winding numbers $\{1,~0,~0,~0\}$ (blue), $\{0,~1,~0,~0\}$ (orange) and $\{0,~0,~1,~0\}$ (green) shifted in $r$ by a shift of order 1.}
\label{wind1}
\end{figure}

The configuration used for measurements described in
this chapter was a well-thermalized configuration 
with the total number of simplices equal to $N_4 =N_{4,1}+N_{3,2} = 370724$.
The distributions of lengths of the shortest
loops connecting the neighboring cells in $x$, $y$ and $z$ directions were observed to be
of approximately the same shape,
up to a possible small shift in the length $r$, as expected. Fig.~\ref{wind1} 
shows the length distributions of the x, y and z 
loops of winding number one centered using a shift of order smaller than 1. 
This shift signalizes a change from the regular symmetric shape of the initial hypercube. 

\begin{figure}
\begin{center}
{\includegraphics[width=10cm]{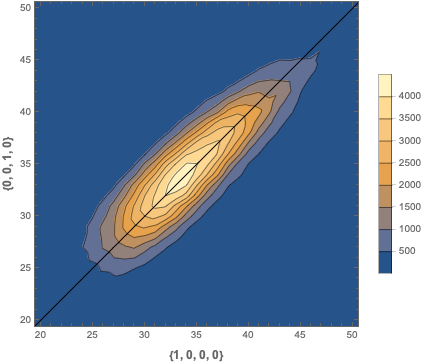}} 
\end{center}
\caption{Correlation between loop distances for loops with winding numbers $\{1,~0,~0,~0\}$  and $\{0,~0,~1,~0\}$.}
\label{loop13}
\end{figure}

\begin{figure}
\begin{center}
{\includegraphics[width=10cm]{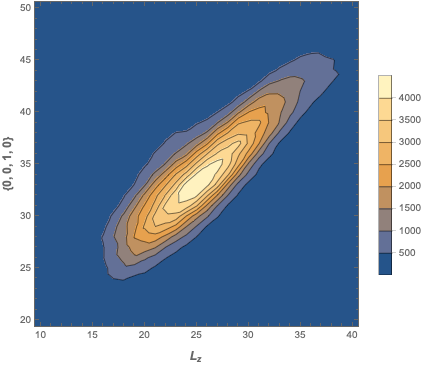}} 
\end{center}
\caption{Correlation of a distribution of $L_z$ (horizontal axis) with a loop distance in $\{0,~0,~1,~0\}$ direction.}
\label{zvs001}
\end{figure}

\begin{figure}
\begin{center}
{\includegraphics[width=12cm]{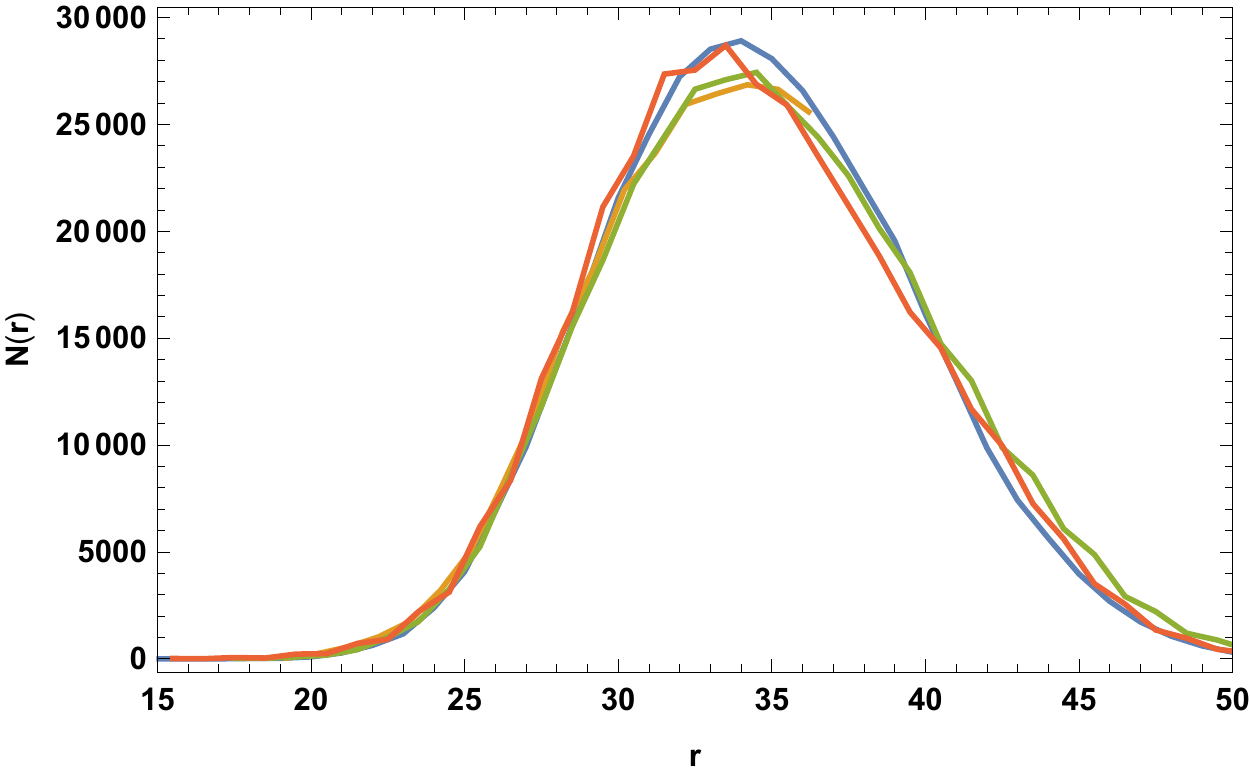}} 
\end{center}
\caption{Distributions of (shifted) loop distances  in directions $\{1,~0,~0,~0\}$ (blue), $\{1,~-1,~0,~0\}$ (red), $\{1,~-1,~1,~0\}$
(green) and $\{1,~1,~1,~0\}$ (orange) showing  the universal blob structure.}
\label{collected}
\end{figure}

\begin{figure}
\begin{center}
{\includegraphics[width=12cm]{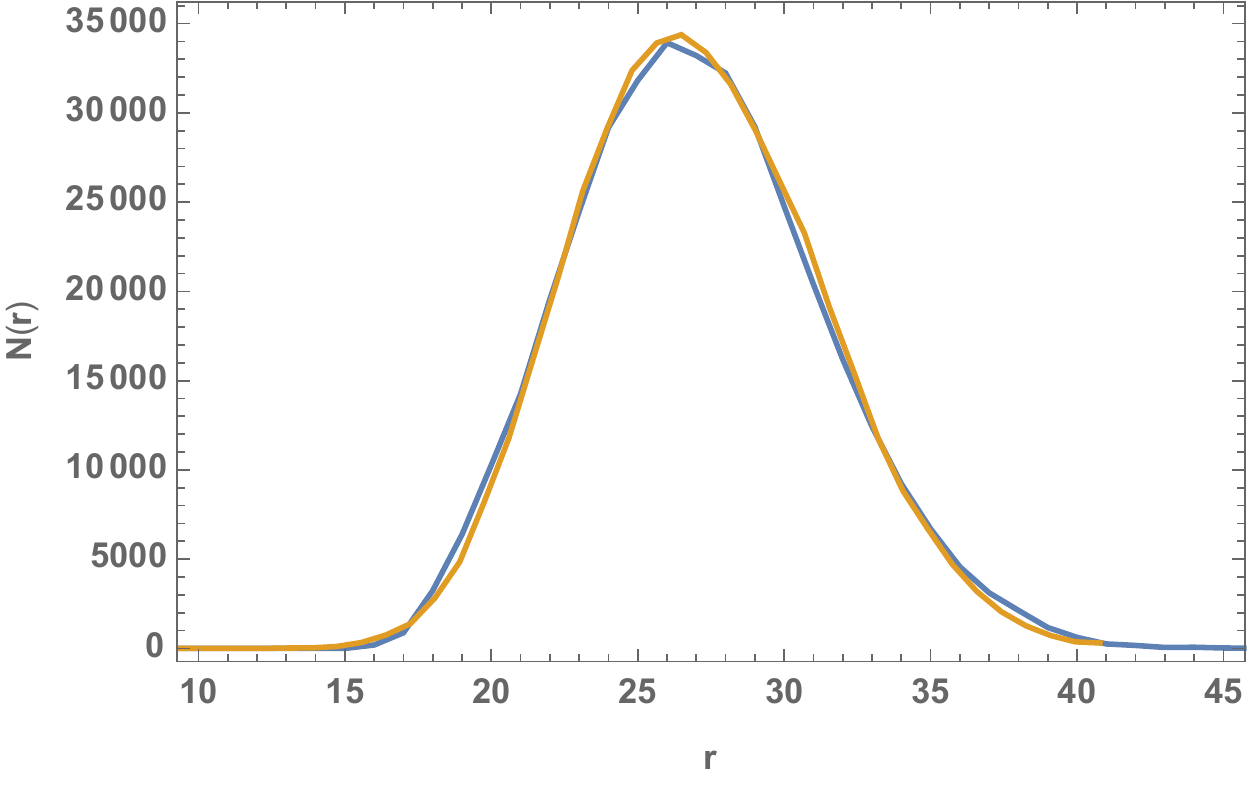}} 
\end{center}
\caption{The loop distance distribution in the time direction, compared to that in the $x$-direction. { Distribution in the time direction was shifted and rescaled by a factor equal to 1.19.}}
\label{loopt}
\end{figure}

Shortest loop distances in different directions are highly 
correlated, as can be seen in Fig.~\ref{loop13} displaying the correlation between the loop 
distance in the $\{1,~0,~0,~0\}$ and $\{0,~0,~1,~0\}$ directions.
There is also a strong correlation between the lengths of the shortest loops and the sum of pseudo-Cartesian coordinates in a given direction, e.g., $L_z=z+z'$, 
as shown in Fig.~\ref{zvs001}.
Fig.~\ref{collected} shows the comparison of distributions for
loop distances in directions $\{1,~0,~0,~0\}$ (blue), $\{1,~-1,~0,~0\}$ (red), $\{1,~-1,~1,~0\}$
(green) and $\{1,~1,~1,~0\}$ (orange). 
Despite the foliation in the time direction, the distribution of loop lengths in time has a shape similar as in the spatial directions, as can be seen in
Fig.~\ref{loopt} comparing the distribution of loop lengths in the time direction (blue)
to the distribution of loop lengths in the $x$-direction (orange; shifted and rescaled by a factor equal to 1.19).
The information encoded in the loop distance
in spatial and time directions is very similar. The average loop length in the time direction 
will obviously depend on the time period $T$.
All these plots show that 
the valleys-and-mountains (bulk-and-outgrowths)
structure is independent of the directions in which the loops are measured.
The minimal and average loop lengths in the plots may differ,
because the shape of the elementary cell may be elongated in some
directions, but in general simplices in the outgrowth have large loop lengths in all directions.

The distribution of lengths of loops with low winding numbers was described in the first of the articles on which
this chapter is based.
The universality of its shape and the strong correlations between lengths of minimal loops
in different directions were noted.
Very short loops are rare: in a configuration containing $N_4 = 370724$ simplices there are only 20 simplices belonging to loops of length 18. Moreover, loops of length
from 19 to 21 often differ from each other only by a few simplices;
the number of separate short loops of the same length – the number of distinct deepest valleys – is very small.
The results in the other spatial directions are similar.

Geometry of a random configuration generated in Monte Carlo simulation is very complex. 
The distribution of the lengths of minimal loops with a unit winding number in spatial or time directions, having the maximum at a length above 30, much above the length of the shortest loops 
in the configuration,
seems to be a signal that in most cases the starting simplex is located inside one of the fractal-like structures
(mountains, outgrowths),
whereas the (rarer) simplices belonging to the loops with a length 
that is minimal or close to minimal
correspond to the (relatively simple) basic structure of valleys in the configuration. With this 
interpretation, the length of a loop starting from a particular simplex reflects the 
position of the simplex relative to the 
valleys. 

Each simplex can be assigned a set of four numbers: the lengths of shortest loops 
(among those passing through the given simplex)
with a unit winding number in a particular direction and winding number zero in the other three directions. These numbers can be called the \textit{heights} of the simplex,
as they reflect its position above the basic structure (the valleys built of simplices
belonging to the shortest non-contractible loops in the configuration).
For the sake of brevity, those four numbers can be called $x$-height, $y$-height,
$z$-height and $t$-height.
It was checked that, as expected, the height values of the five neighbors of 
any simplex differ from its own height by $\pm 2$, $\pm 1$ or 0.
In general, any loop starting deep inside an outgrowth (at a simplex of large height)
is expected to move closer to the base 
(passing through simplices of decreasing height)
and then eventually climb back to simplices 
in the same outgrowth. 
There are only few loops whose simplices are all of equal height. This property is possessed 
by the shortest loop in the configuration and a few dozen other short loops,
which are, so to speak, the ``locally deepest valleys''.

To summarize,
the \textit{heights} of a simplex are defined as the lengths of the shortest loops passing through it
in each of the four basic directions.
Often more than one loop of the same length
and with the same set of winding numbers pass through a given simplex.
In such a case only one of them was selected for further analysis described in the following sections.
If a simplex is not located at a far end of an outgrowth, then 
usually many loops that are minimal for other simplices (see the discussion near the end of Sec.~\ref{alternativeBC}) pass through it as well.

\section{Loops with higher winding numbers}\label{sec:higher}

The shortest loops with nonzero winding
numbers contain important information about the underlying structure of the 
piecewise linear manifold and about the
distribution of valleys and outgrowths. 
The analysis of minimal loop length distributions 
can be extended to include simplices in cells with an arbitrary set of winding numbers $\{n^\mu\}$. 
The results show that among the minimal loops 
there are not only loops in the four basic
directions, as discussed above, but also loops with other winding numbers,
which, perhaps surprisingly, are sometimes even shorter.
Since the winding numbers of a loop
do not depend on a particular choice of boundaries delimiting the elementary cell,
this implies that the shape of the torus is twisted in some way and far from regular.

Using a four-dimensional diffusion wave on the dual lattice
in a system treated as infinite in four directions 
is a simple method to ensure that the shortest loops with a given set of winding
numbers are found, as the diffusion wave cannot ``miss'' any shorter path. 
The diffusion could be continued to find loops with any higher winding numbers.
However, the number of visited simplices at a distance $R$ from the initial simplex grows as $R^4$,
which means that eventually, for large $R$, the procedure would become too time- and memory-consuming,
and computationally inefficient. Therefore, the boundary conditions should be modified
in such a way that the number of simplices visited by the diffusion wave grows more slowly.
One example is to consider the four-torus as a system infinite in only one direction –
for example the $x$-direction –
and periodic in all other directions. A diffusion wave in such a system
can be used to find loops with higher winding numbers of the form 
$\{n,0,0,0\}$, $n=1,2,\dots$. A small modification of this idea is to assume that 
cell boundaries in all directions except $x$ are impenetrable for 
the diffusion wave, and the system is again infinite in only one direction.
In both of the methods the number of visited simplices 
in the $R^{\mathrm{th}}$ shell, for $R$ large enough, stabilizes and becomes independent of the distance $R$.
The growth is faster only up to the range where the 
diffusion wave reaches the boundaries of the system in the finite directions. 
A similar result can be obtained if the impenetrable walls are put
in all directions except $x$ at boundaries between cells number $\pm 1$ and $\pm 2$
rather than between $0$ and $\pm 1$.
This choice improves the behavior of the diffusion process in case the initial simplex is
near the boundary of the $\{0,0,0,0\}$ elementary cell.

\begin{figure}
\centering
\includegraphics[width= 0.5 \textwidth]{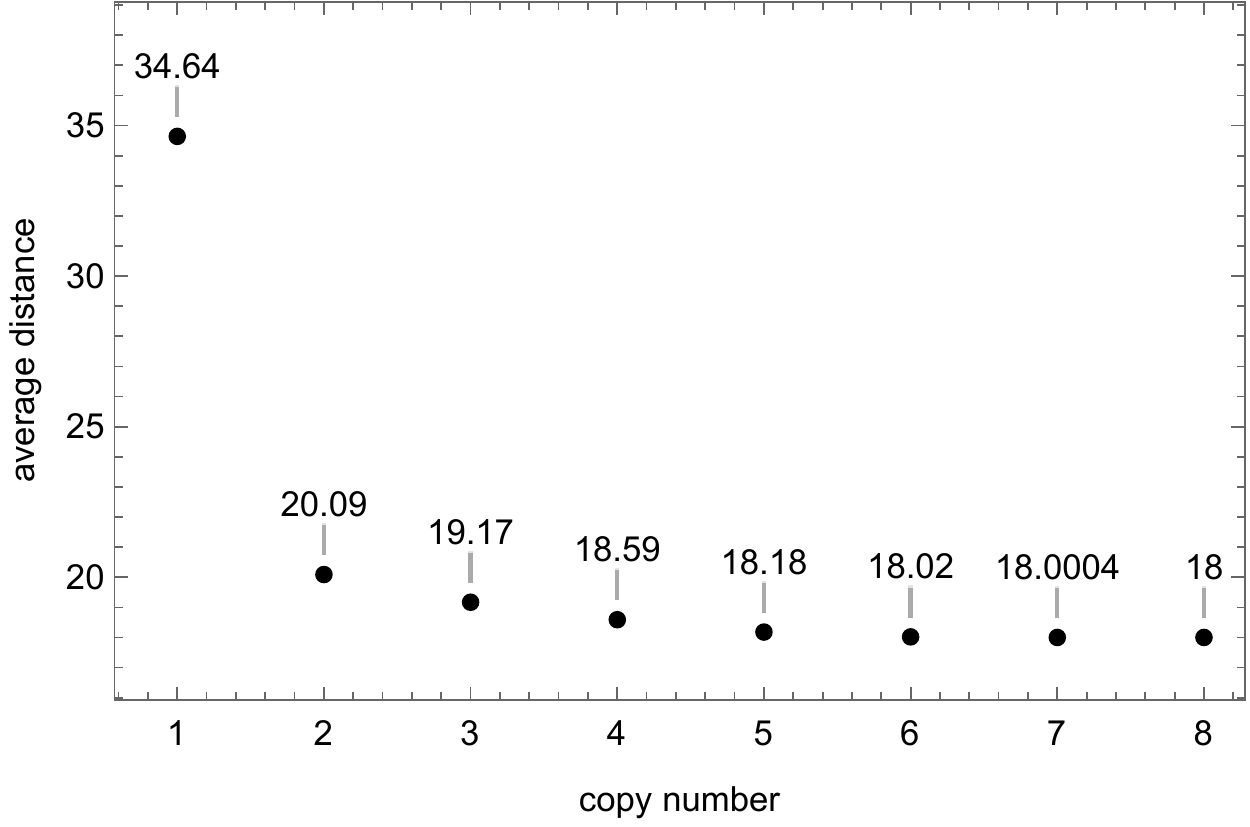}
\caption{Distance from a starting simplex to its copy in cell $\{n,0,0,0\}$ minus distance from the same simplex to its copy in cell $\{n-1,0,0,0\}$, averaged over all the simplices of the configuration.
Already for $n=8$ the minimal value of 18 is reached.}
\label{fig:av}
\end{figure}

\begin{figure}

\includegraphics[width= 0.5 \textwidth]{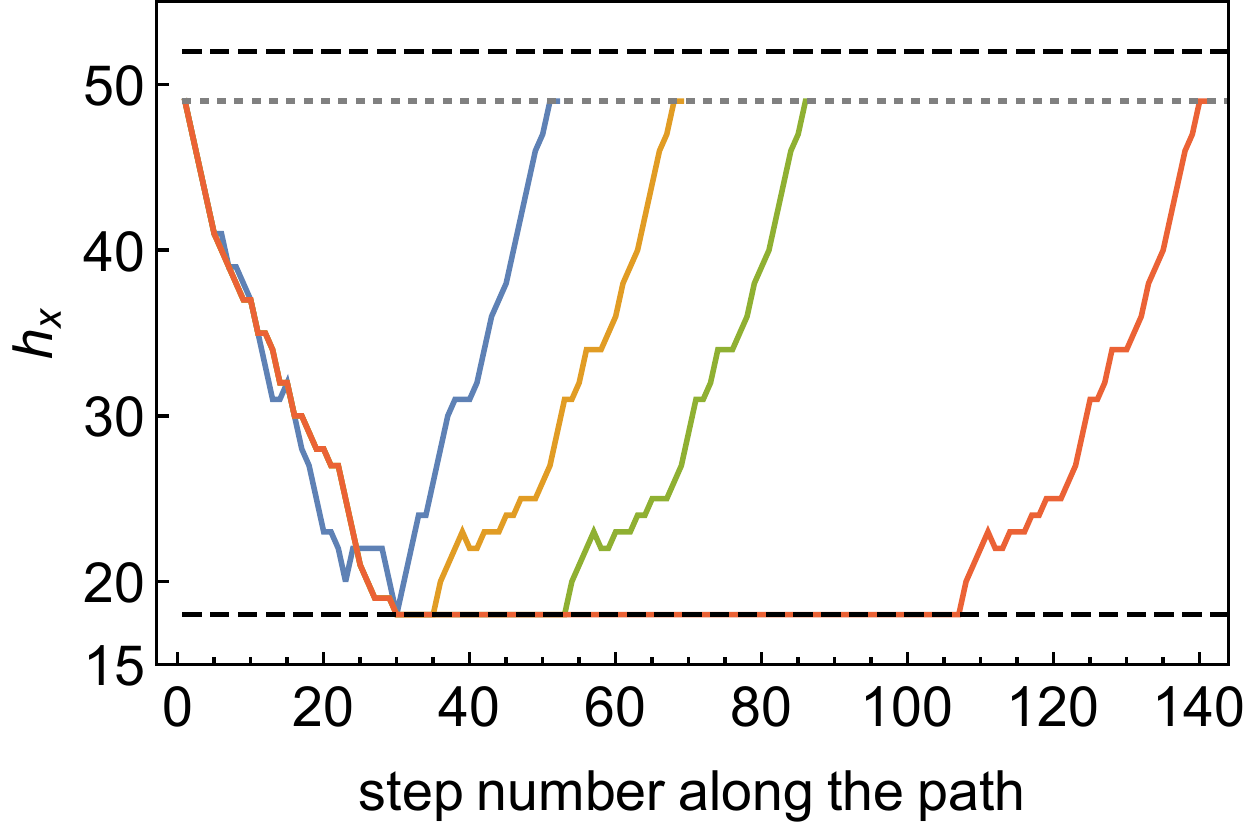} 
\includegraphics[width= 0.5 \textwidth]{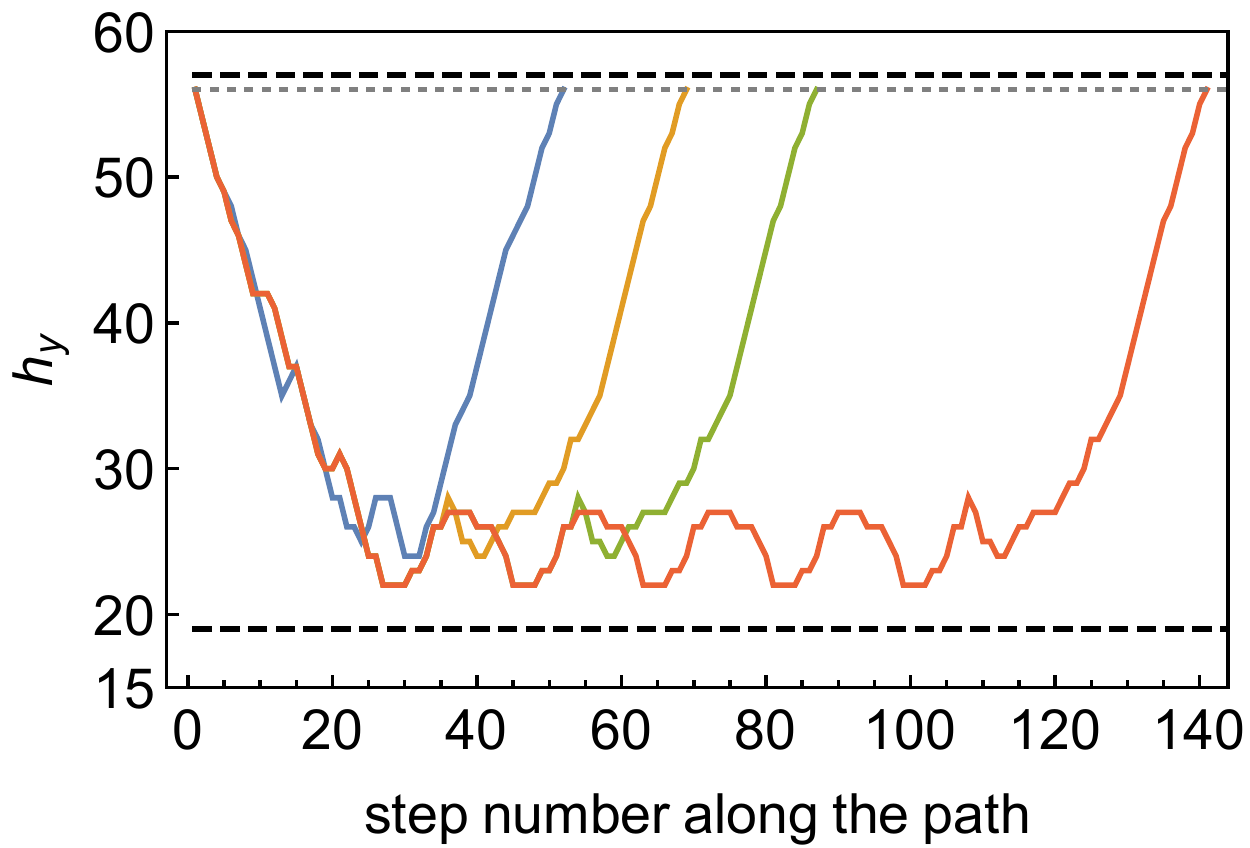}
\includegraphics[width= 0.5 \textwidth]{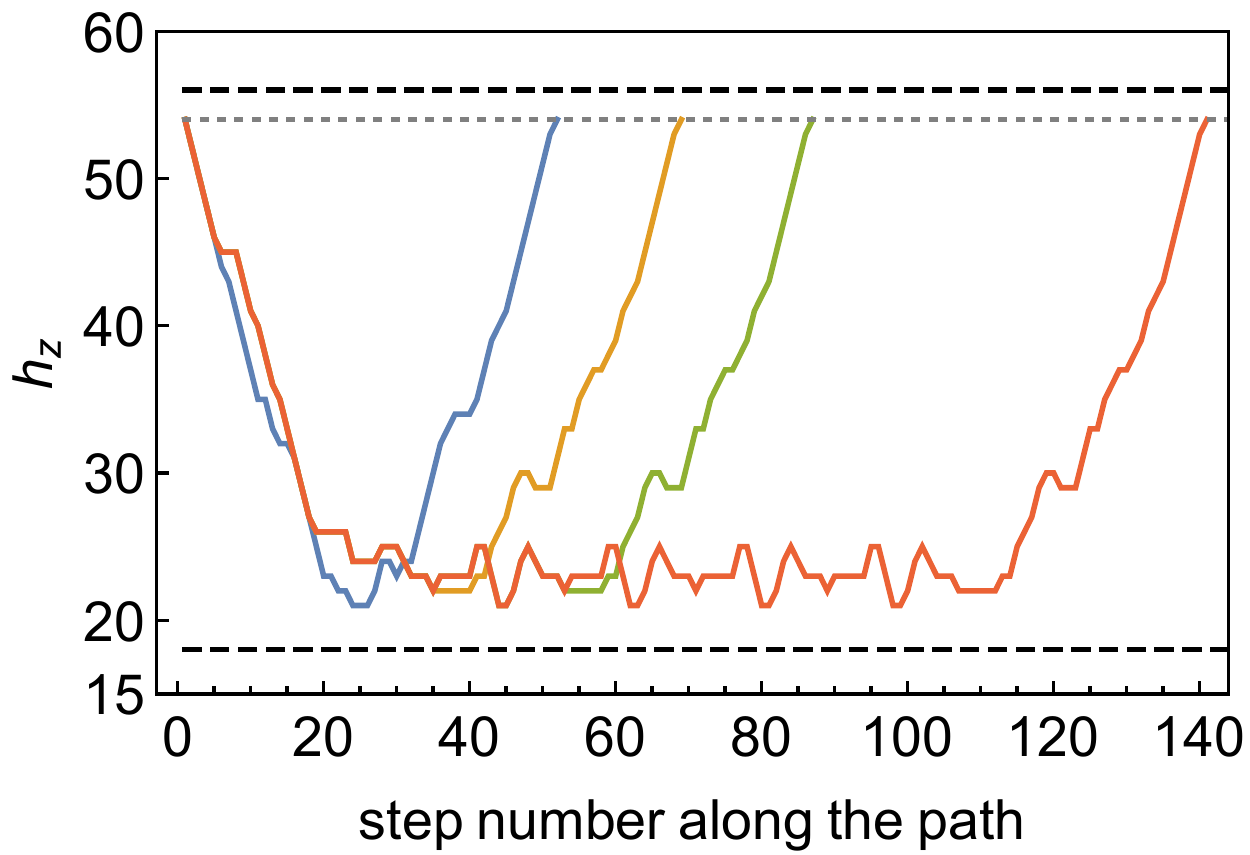} 
\includegraphics[width= 0.5 \textwidth]{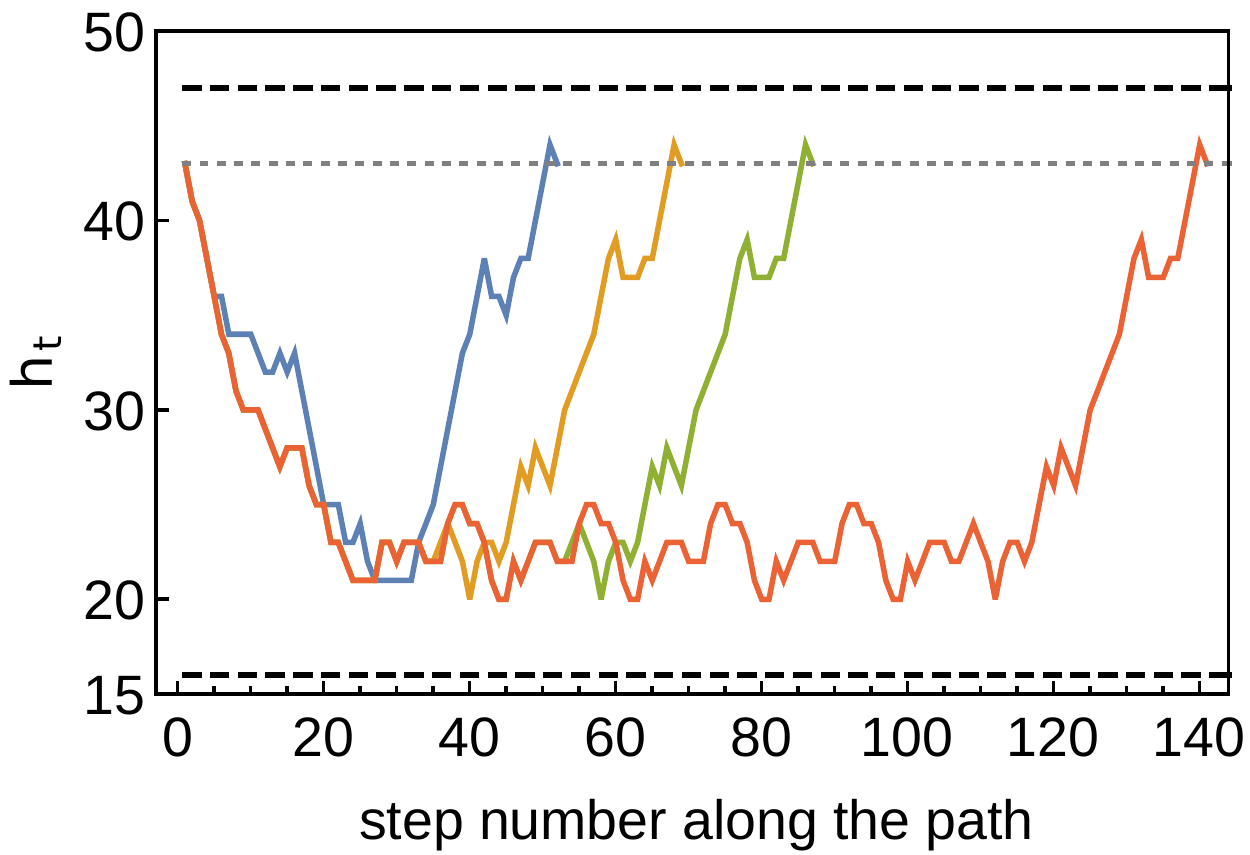}

\caption{Heights in the 4 basic directions of consecutive simplices along loops of winding
numbers $\{1,0,0,0\}$ (blue), $\{2,0,0,0\}$ (orange), $\{3,0,0,0\}$ (green), $\{6,0,0,0\}$ (red) starting from a simplex
in an outgrowth. The dashed lines indicate the minimal and maximal heights in the configuration,
and the dotted line indicates the height of the initial simplex.}
\label{fig:sim2}
\end{figure}

Fig.~\ref{fig:av} presents results obtained with the method using an impenetrable wall
between cells number $+1$ and $+2$ and an impenetrable wall between cells $-1$ and $-2$
in all directions except $x$ to measure the average lengths of loops with 
winding numbers $\{n,0,0,0\}$, $n=1,2,\dots,8$. Similar measurements were performed in
the directions $y$ and $z$.
It was found that the difference between the distance to the copy number $n$ and the distance 
to the copy number $n-1$ of a given simplex decreases rapidly with increasing $n$, down to
the length of the minimal shortest loop (in the given direction) of the whole configuration 
(equal to the minimal height among all simplices), and then it remains constant.
The explanation is simple: in order to minimize the length of a loop with a high winding number $n$ in a given direction,
it becomes advantageous for the loop to connect the initial simplex to a simplex of the 
lowest possible height,
then trace the shortest loop of unit winding number $n$ times, and finally to return to the initial simplex.
As can be seen in Fig.~\ref{fig:av}, the shortest loop of the configuration
is a part of all the loops of winding number $n \geq 8$. 
Although the average $x$-height in the analysed configuration is 34.64,
which shows that most simplices belong to the outgrowths, 
the length of the shortest loop of winding numbers $\{2,0,0,0\}$
is greater than that by only 20.09, which shows that already the second winding of the loop
passes mostly through the valleys (the central part of the configuration).

Fig.~\ref{fig:sim2} shows the heights of consecutive 
simplices along loops with an increasing sequence of winding numbers in 
the $x$-direction, starting from a simplex lying far within an outgrowth.
One can readily see once again that as the winding number of the loop increases, 
usually the minimal $x$-height of the simplices belonging to the loop decreases,
ultimately down to 18, which is the length of the shortest
$\{1,0,0,0\}$ loop in the configuration.
As there are no two completely separate $x$-loops of length 18,
this means that all the loops of a high winding number pass many times mostly through the same set
of simplices.
The graphs showing the heights in the $y$, $z$ and $t$ directions for 
the same set of loops
demonstrate that although there is a correlation between height in all directions,
the correlation is not perfect, especially after the loop leaves the outgrowth.
The repeating saw-like pattern is a loop of a high winding number tracing one of the shortest loops
of unit winding number in direction $x$ several times. The heights of simplices belonging to a loop that is
shortest in the $x$-direction are not minimal in the other three directions.
However, even though in the other directions the height fluctuates,
it still remains close to the minimal height, because the simplices in the semiclassical region
generally have low heights in all directions.

\section{Alternative boundary conditions}\label{alternativeBC}

The other method for finding loops of high winding number,
mentioned in Sec.~\ref{sec:higher}, is to search for a shortest path 
connecting a simplex to its copy in another elementary cell in a
system that is infinite in one direction and has periodic boundary conditions imposed in the other three directions.
The values of the winding numbers in the other directions are ignored by this method, i.e., it finds loops with winding numbers of the form
$\{n,a,b,c\}$, $a,b,c$ being any integers, whichever are shortest, instead of only $\{n,0,0,0\}$. 

\begin{table}
\begin{center}
\begin{tabular}{ |c|c|c|c|c|  }
\hline
\multicolumn{4}{|c|}{Winding numbers}&\multirow{2}{*}{Length}\\
 \cline{1-4}
 x&y &z&t&\\
 \hline
 \multicolumn{5}{|c|}{Shortest loops in the $x$-direction}\\
 \hline
1   & 0    &0&   0 & 18\\
1 & -1 & 1 & 0 & 16\\
2 & -1 & 1&0 & 27\\ 
\hline
 \multicolumn{5}{|c|}{Shortest loops in the $y$-direction}\\
 \hline
 0 & 1 & 0 & 0 & 19 \\
 0 & 1 & 0 & 1 & 16 \\
 -1 & 1 & -1 & 0 & 16\\
 -2 & 3 & -1 & 1 & 43 \\
 \hline
  \multicolumn{5}{|c|}{Shortest loops in the $z$-direction}\\
 \hline
 1 & 0& 0 & 0 & 18\\
 1 &-1& 1 &0 &16 \\
 2 & -1 & 3 & 0 & 47 \\
 \hline
\end{tabular}
\end{center}
\caption{Lengths of the shortest loops in the three basic spatial directions.}\label{tab:loops}
\end{table}

It turns out that also for these boundary conditions paths starting at various simplices
tend to converge and follow a handful of very short loops. However, those loops are almost never
the shortest loops with unit winding numbers, e.g. $\{1,0,0,0\}$. Rather than that, 
there exist loops with
winding numbers of the form $\{n,a,b,c\}$ that are shorter than $n$ times the length of the 
shortest $\{1,0,0,0\}$ loop in the configuration (see Table \ref{tab:loops}, where 
data from a diffusion in a
four-dimensional infinite system were used to find the precise winding numbers).
One could conjecture
that as loops of higher winding numbers are probed, loops with even smaller ratio 
of total length to the winding number should be found, but in fact
it turns out that even loops of winding numbers of order 50 utilize the loops
described in Table \ref{tab:loops}, so it appears that those loops are minimal.

\begin{figure}

\includegraphics[width= 0.5 \textwidth]{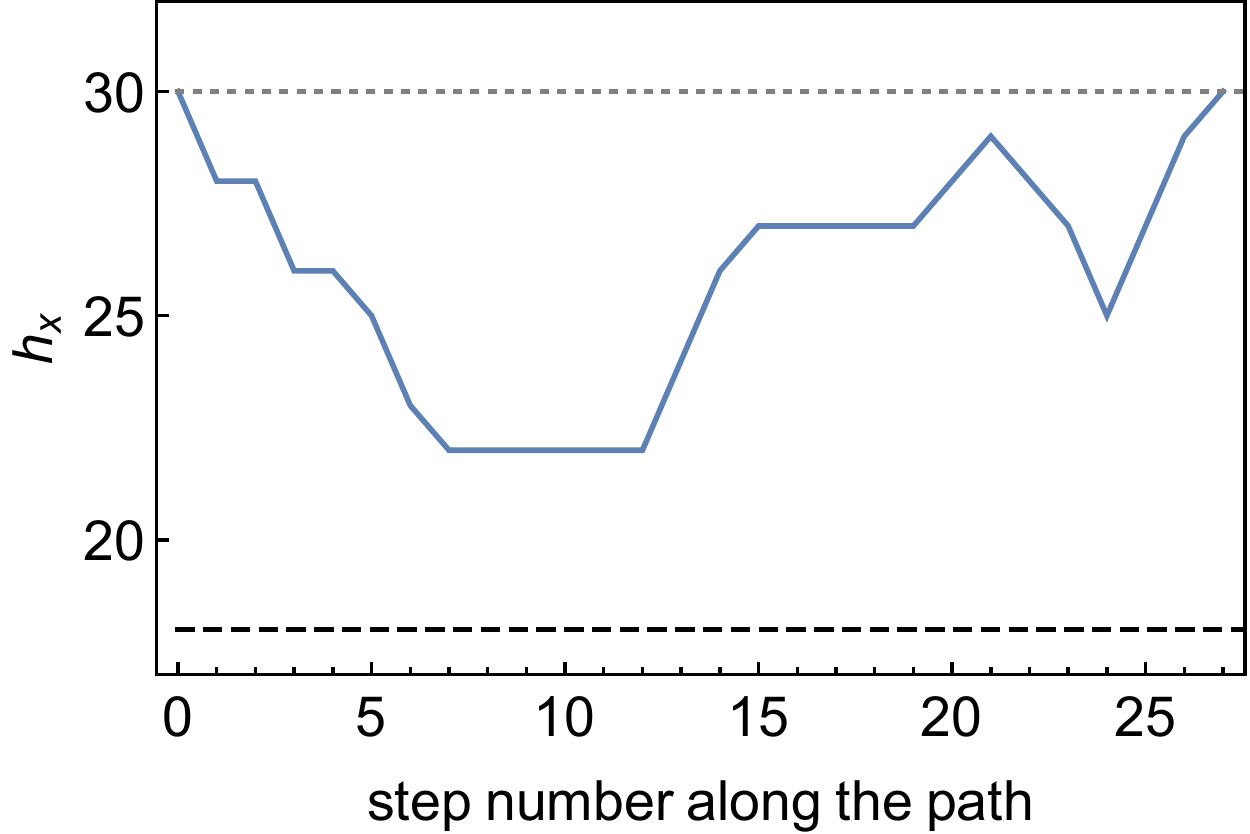} 
\includegraphics[width= 0.5 \textwidth]{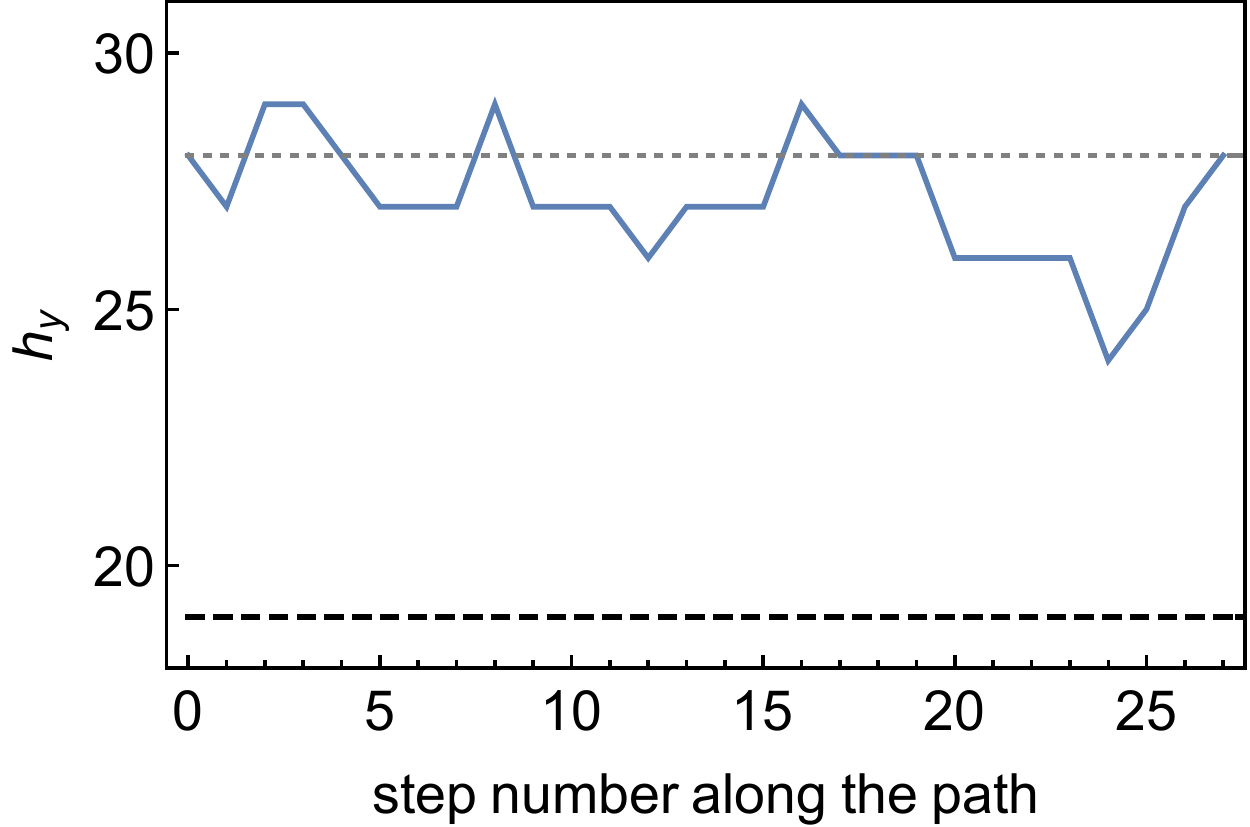}
\includegraphics[width= 0.5 \textwidth]{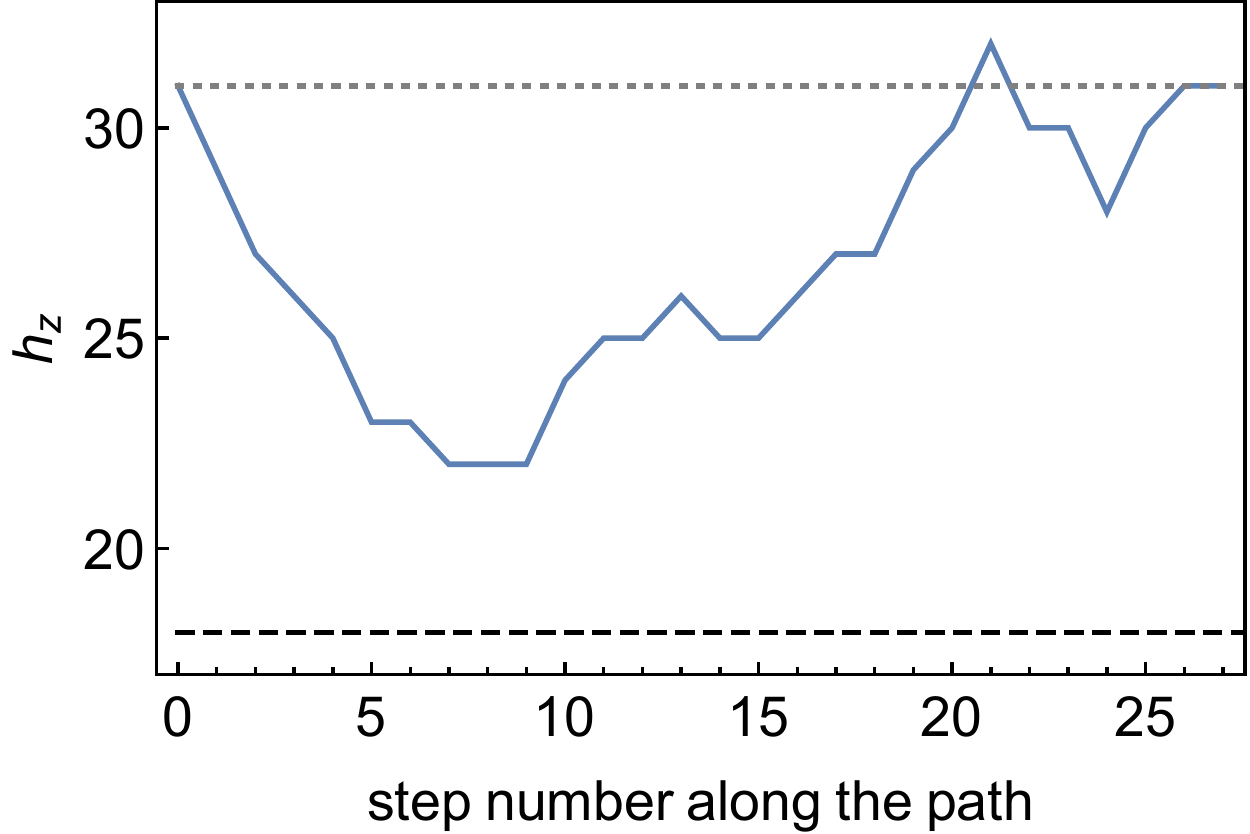} 
\includegraphics[width= 0.5 \textwidth]{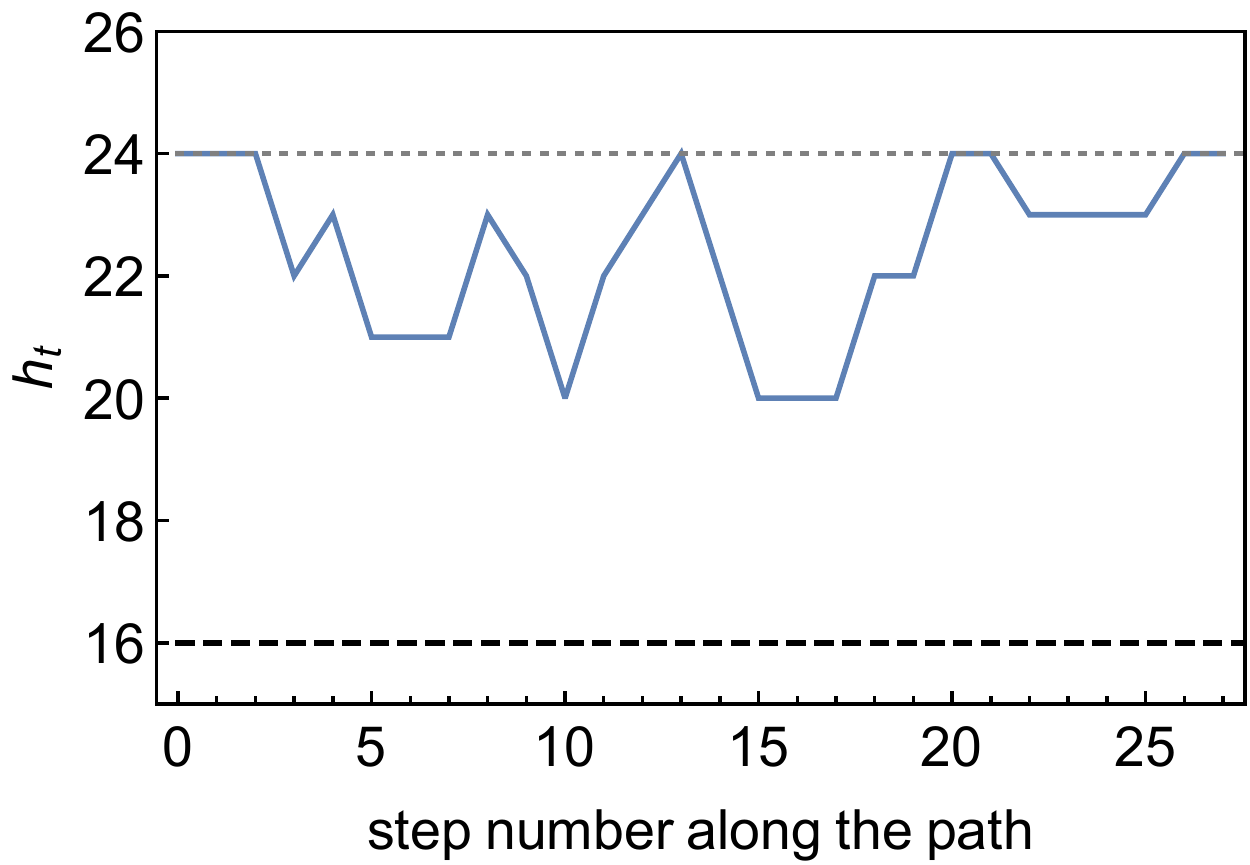}

\caption{The four basic heights of simplices belonging to 
one of the loops from Table \ref{tab:loops}: a loop of length 27 with winding numbers  $\{2,-1,1,0\}$.}
\label{fig:a}
\end{figure}

\begin{figure}

\includegraphics[width= 0.5 \textwidth]{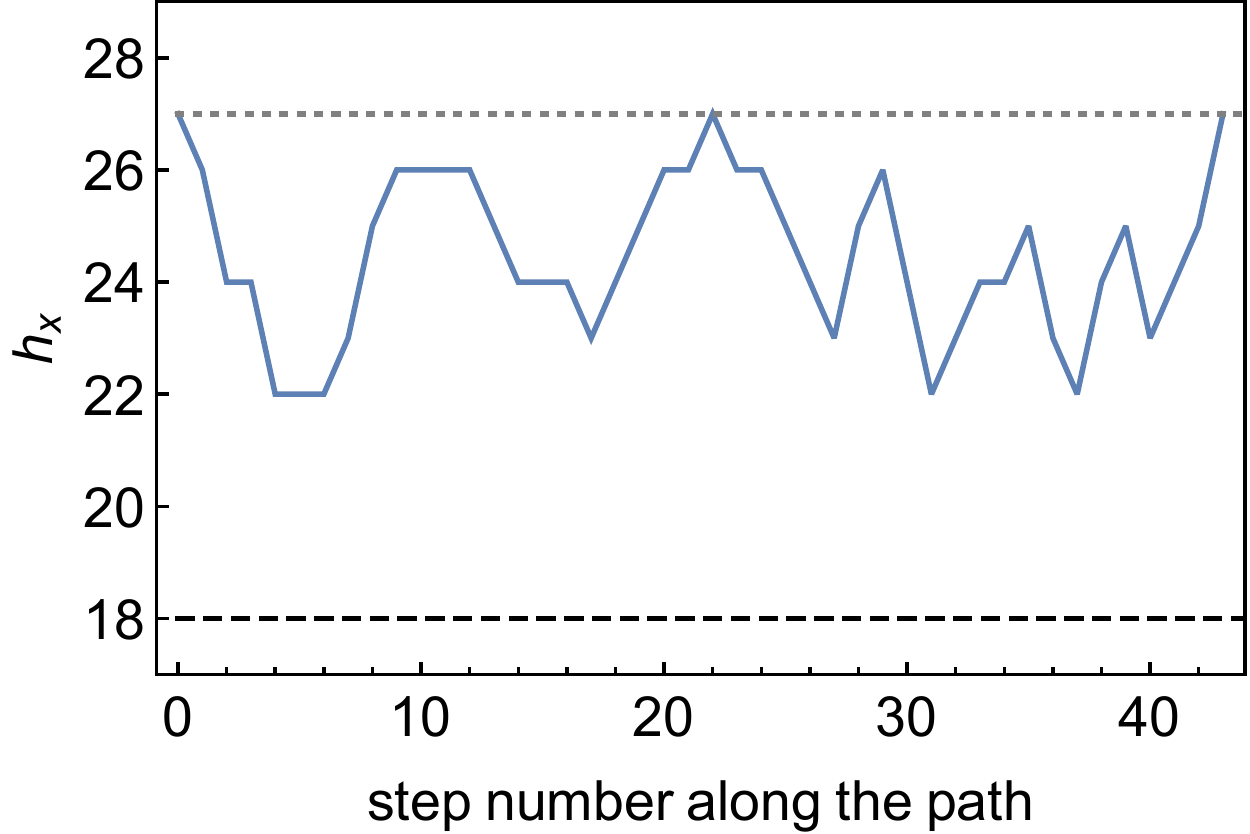} 
\includegraphics[width= 0.5 \textwidth]{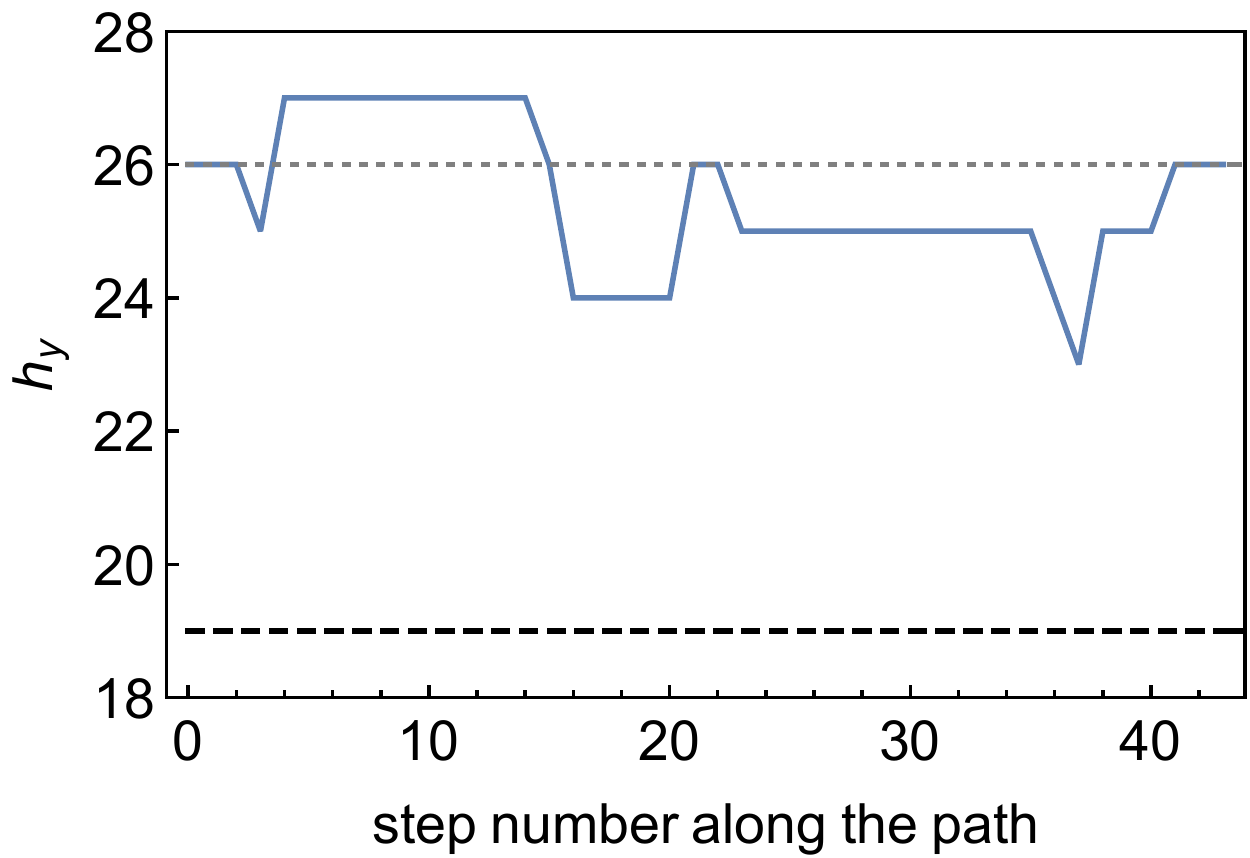}
\includegraphics[width= 0.5 \textwidth]{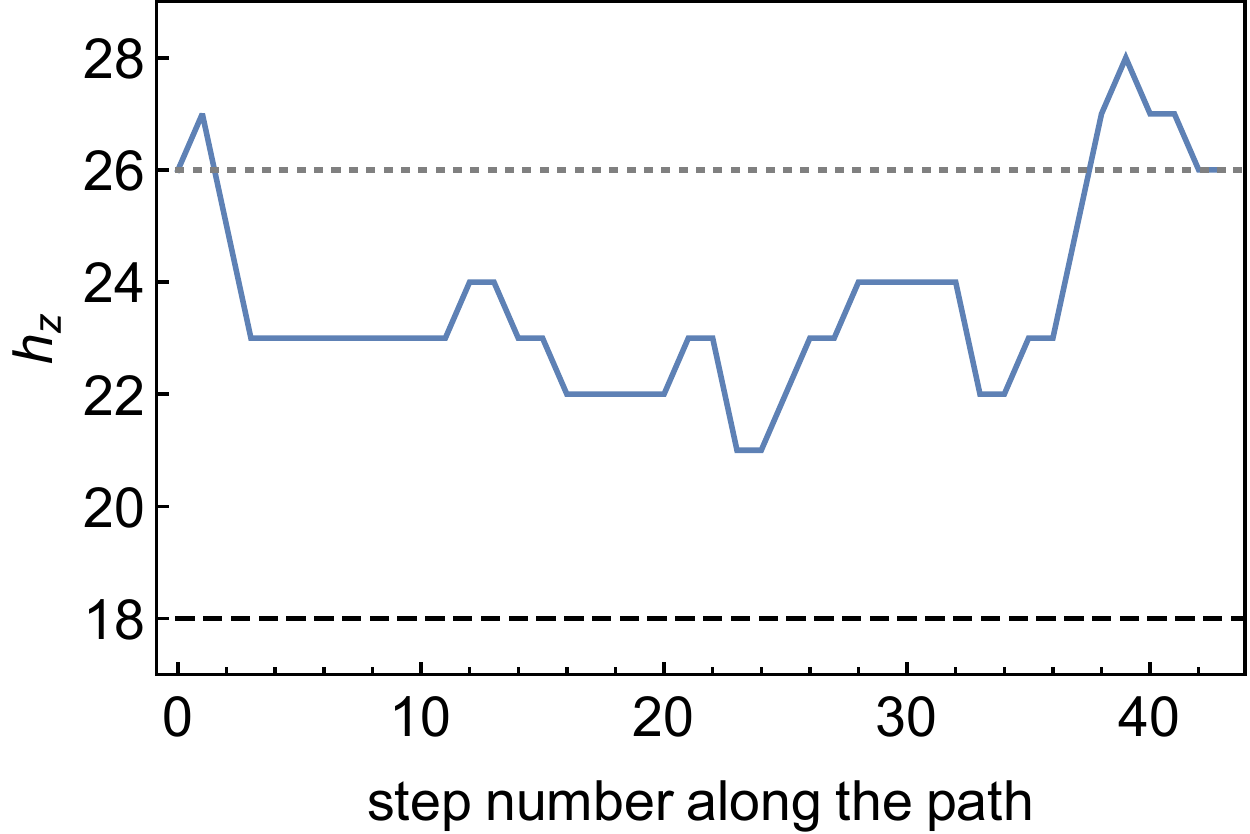} 
\includegraphics[width= 0.5 \textwidth]{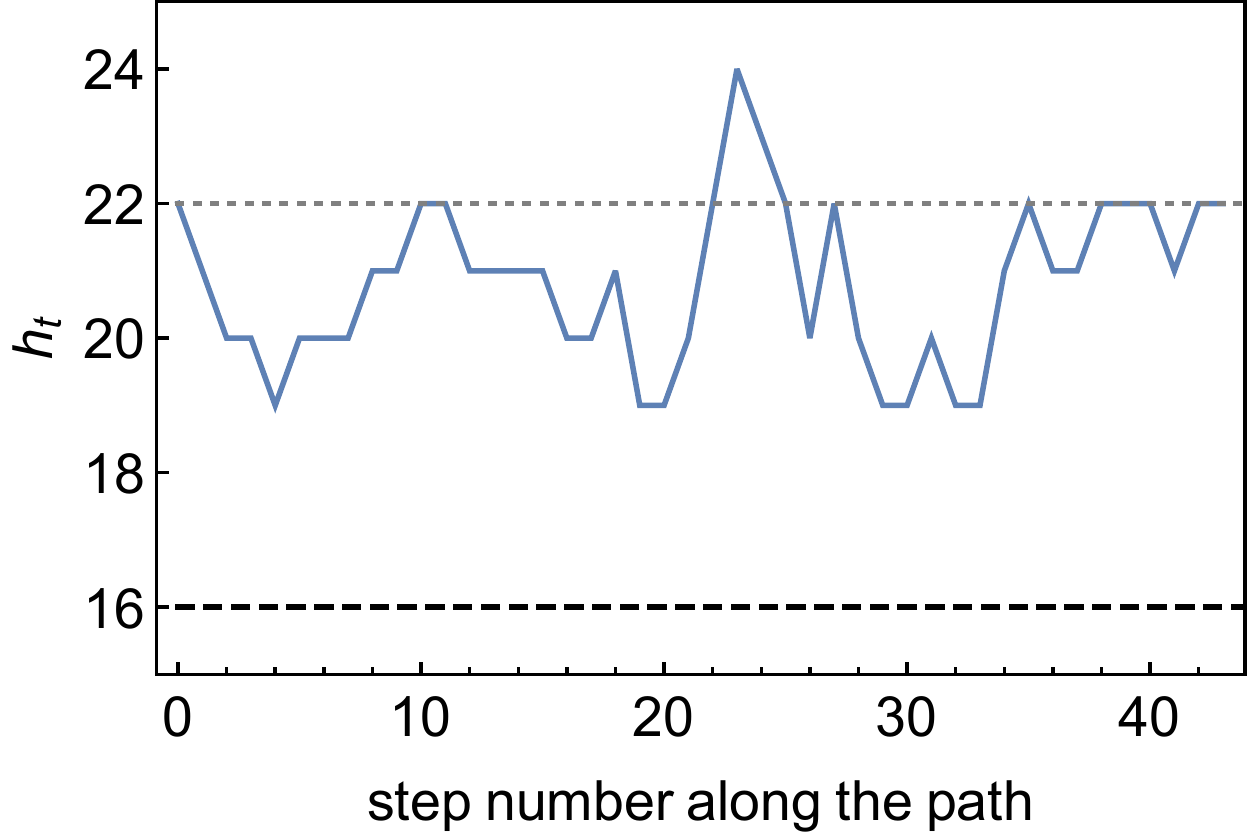}

\caption{The four basic heights of simplices belonging to one of the loops from Table \ref{tab:loops}: a loop of length 43 with winding numbers  $\{-2,3,-1,1\}$. }
\label{fig:b}
\end{figure}

\begin{figure}

\includegraphics[width= 0.5 \textwidth]{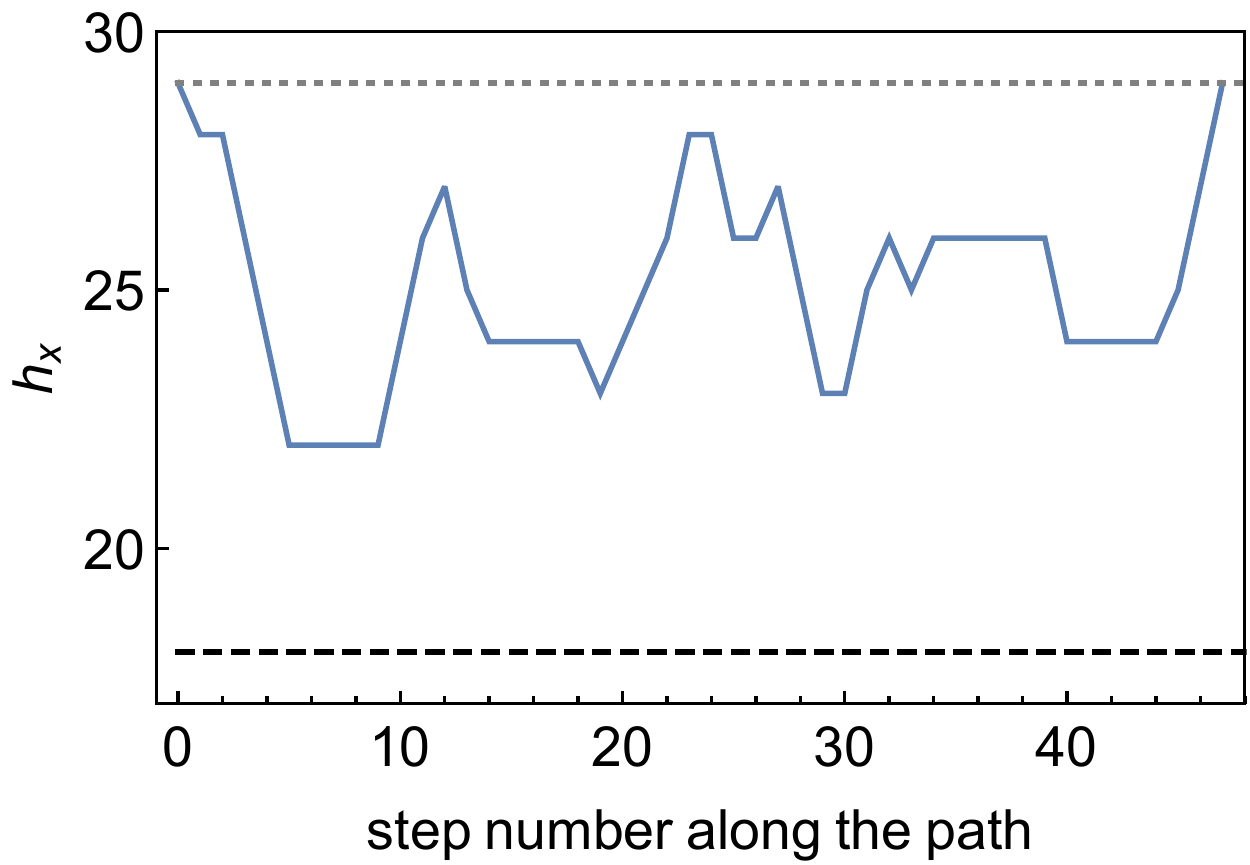} 
\includegraphics[width= 0.5 \textwidth]{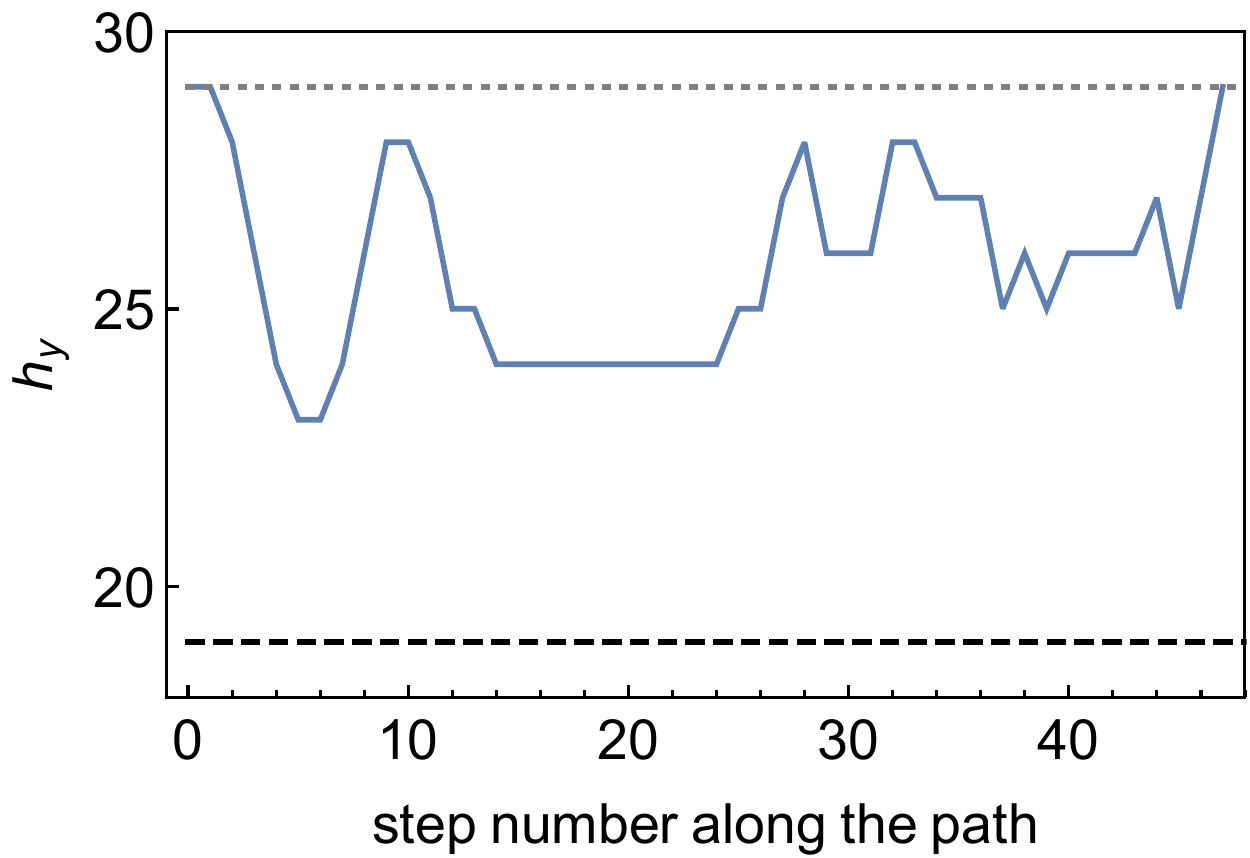}
\includegraphics[width= 0.5 \textwidth]{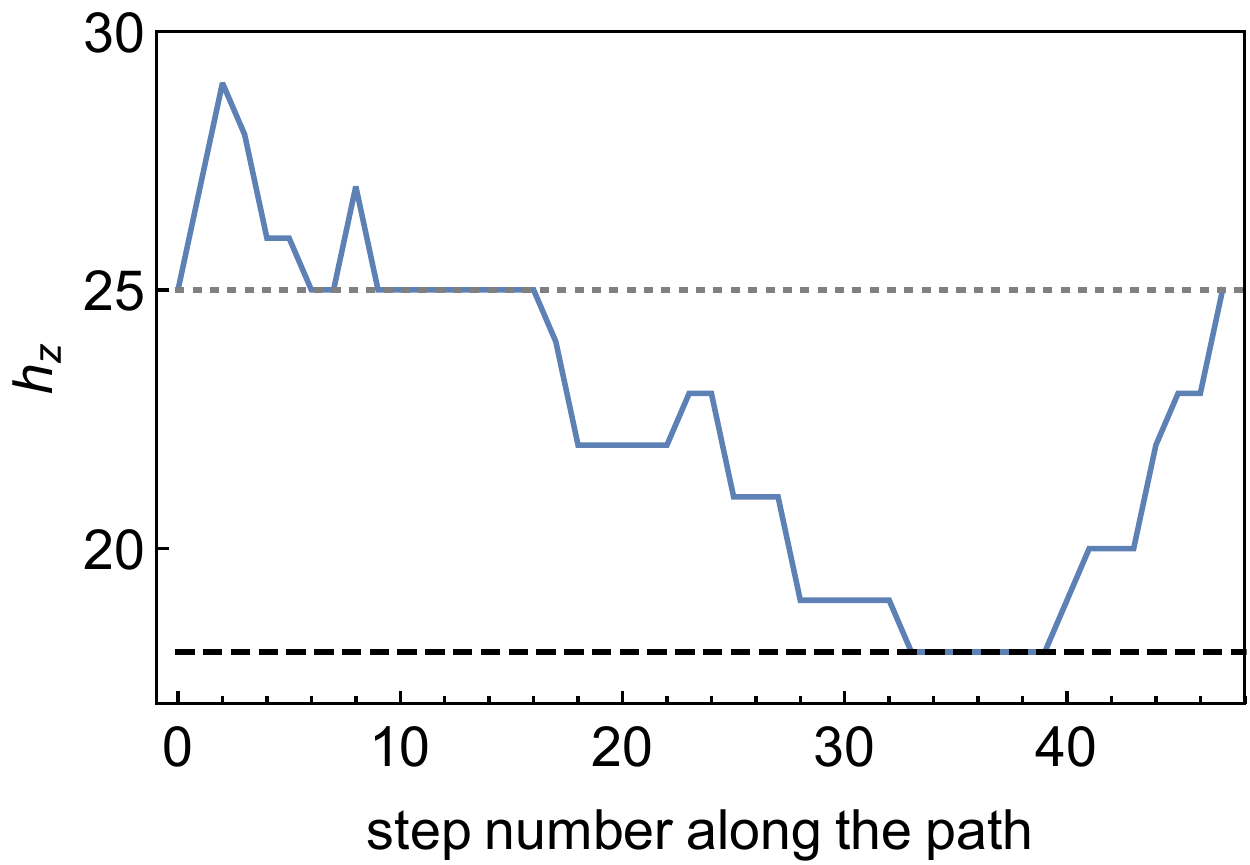} 
\includegraphics[width= 0.5 \textwidth]{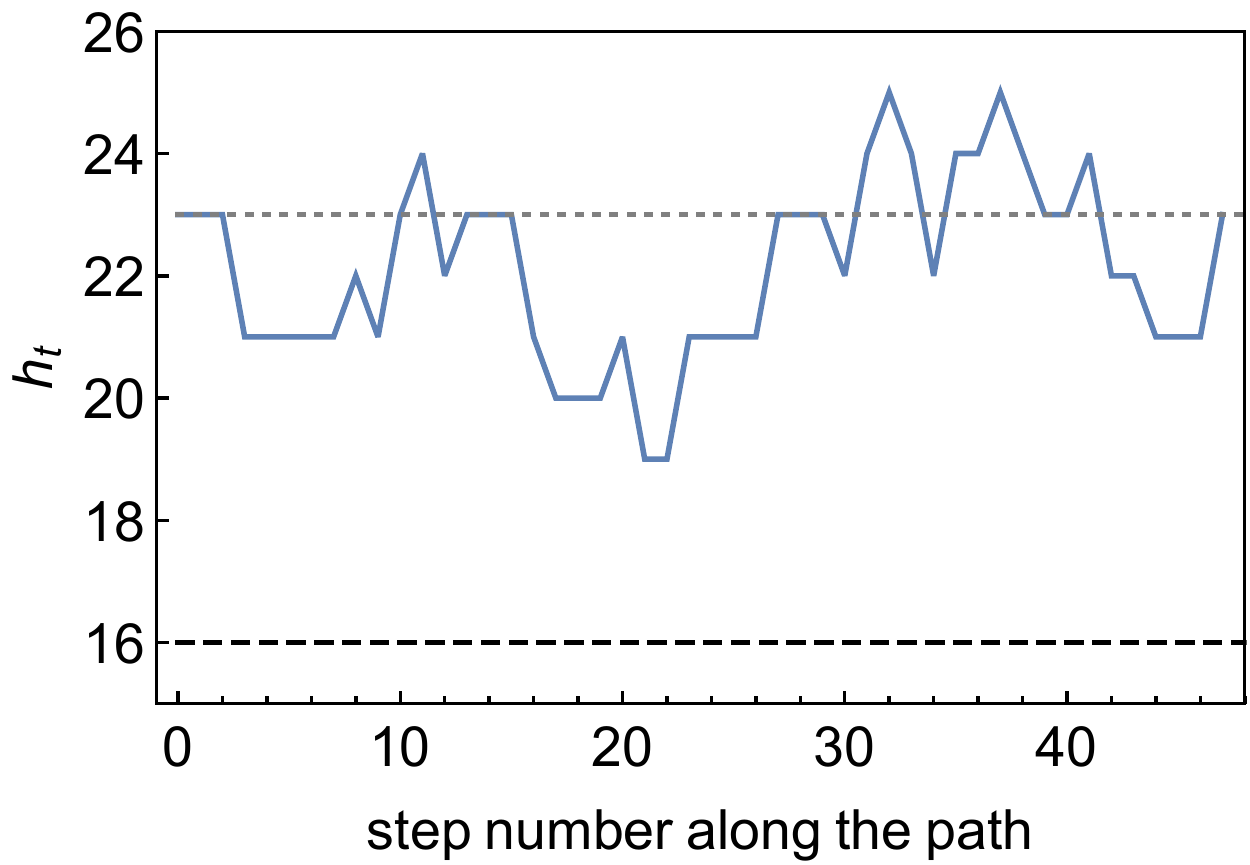}

\caption{The four basic heights of simplices belonging to one of the loops from Table \ref{tab:loops}: a loop of length 47 with winding numbers $\{2,-1,3,0\}$.}
\label{fig:c}
\end{figure}

Figs.~\ref{fig:a}-\ref{fig:c} show the heights in the four basic directions of the consecutive simplices along
the shortest loops described above. The heights, while not minimal,
are quite low compared to the average (which is, as mentioned before, above 30). This signifies
that these paths too belong to the base (``bulk'') region of the torus
rather than to the more common outgrowth regions.

\begin{figure}
\includegraphics[width= \textwidth]{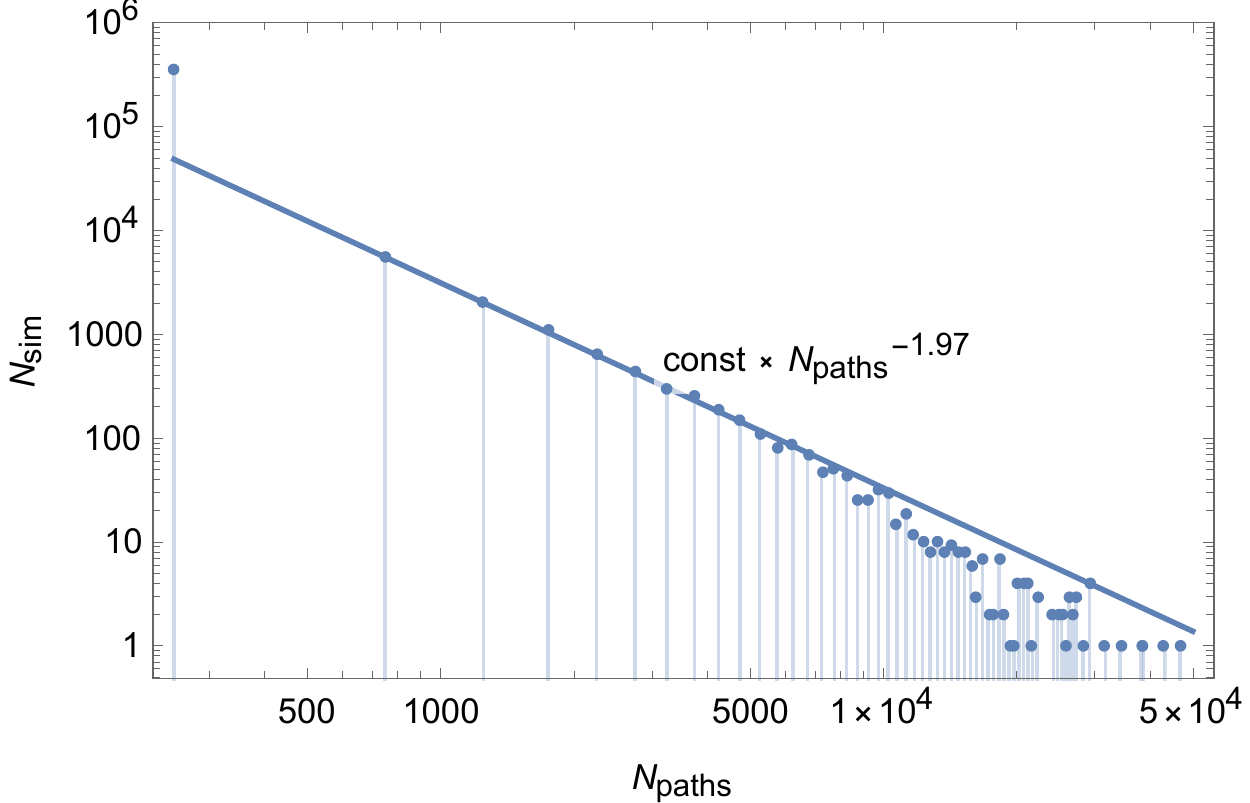}
\caption{A log-log scale histogram of the number of simplices
crossed by a given number of shortest loops.}
\label{fig:histogfit1}
\end{figure}

The algorithm that was used stores
for each simplex reached at step $R+1$ of the diffusion wave
the label of a neighboring simplex that had been reached at step $R$.
After a loop with a desired set of winding numbers
is found, tracing those simplex labels back to step $1$ and to the starting simplex
yields an ordered list of simplices belonging to the loop.
This was used to obtain the the lists of simplices
belonging to the shortest loops of winding numbers $\{1,0,0,0\}$,
one loop for each of the 370724 simplices of the configurations.
The same was repeated for loops with winding numbers $\{0,1,0,0\}$ and $\{0,0,1,0\}$.
Next, the initial (and at the same time final) simplex was removed from each list,
and the number of appearances of each simplex in the lists was counted. 
A log-log scale histogram is shown in Fig.~\ref{fig:histogfit1}.
The maximum value was more than 40000, which corresponded to one of the bulk simplices,
and the minimal value was zero, which occurred numerous times and corresponded to simplices at the far ends
of the outgrowths. The latter simplices are not a part of any geodesics apart from those that start within them.
A good fit to the histogram was obtained using a power law curve with
the exponent very close to $-2$, which seems to bear a certain 
significance. This functional relationship is different than in the cases of, e.g., a branched polymer or a regular lattice.
We have not yet found an algorithmic method of constructing a toroidal graph with behavior described 
by the same exponent. Perhaps 
this point will be investigated in future research.

\section{The structure of a typical quantum geometry}

\begin{figure}
\includegraphics[width= \textwidth]{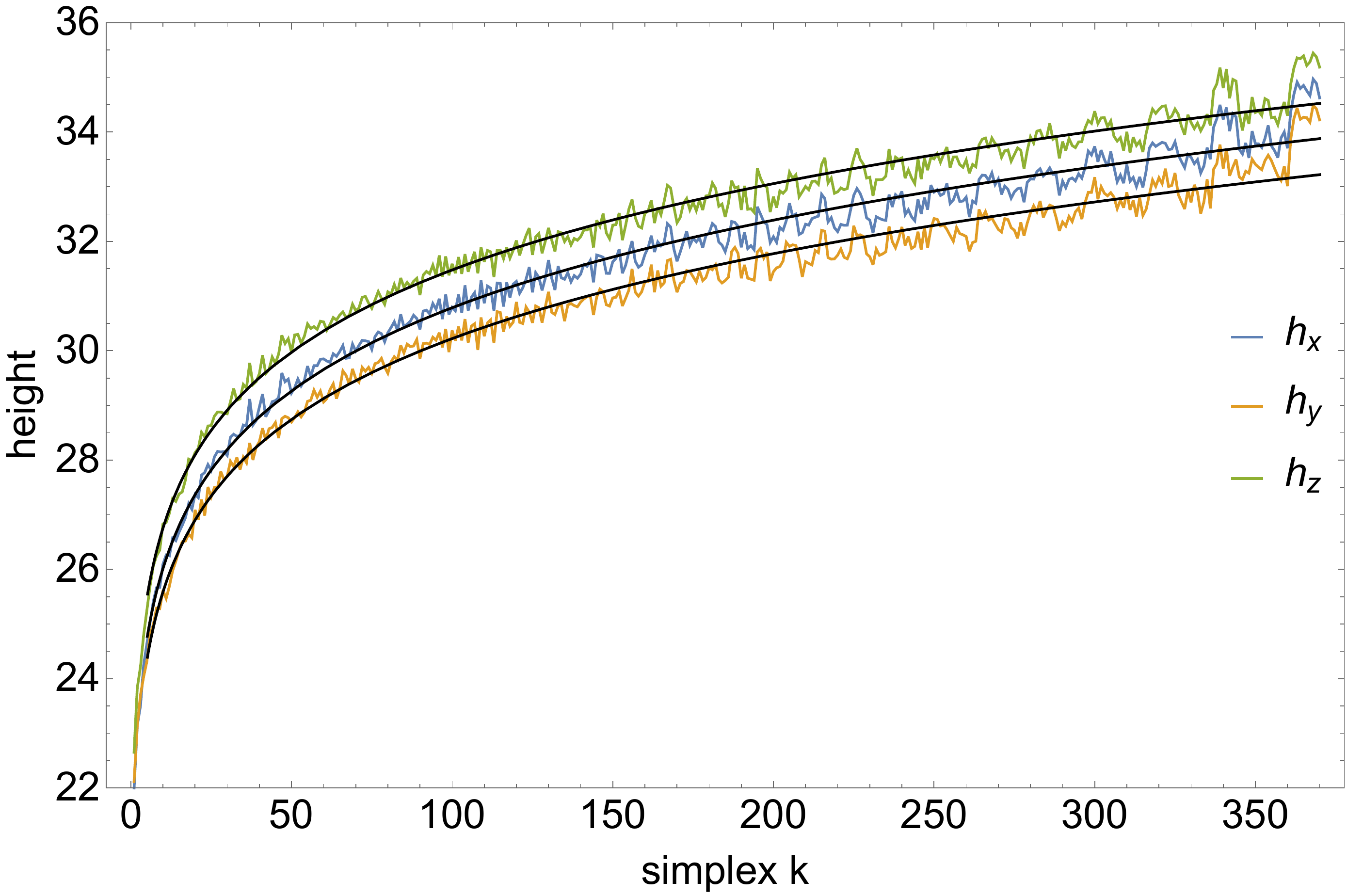}
\caption{The $x$-, $y$-, and $z$-heights of simplices. The simplices were sorted in the order of
descending number of loops passing through them, and then the heights were averaged over blocks
containing 1000 simplices each. The fitted functions are $h_x=21.98 k^{0.0732}$, $h_y=21.66 k^{0.0723}$ and $h_z=22.74 k^{0.0706}$.}
\label{fig:histogfit2}
\end{figure}

The following analysis showed that 
the heights of the simplices and the number of shortest loops passing through them
are strongly correlated.
All simplices of the configuration were sorted 
in the order of descending number of loops passing through them, 
then the list was divided into blocks containing 1000 simplices each, and within each block
the heights in each of the three spatial directions were averaged.
With this ordering of simplices, the heights
turned out to be increasing functions of the ordinal number of simplices in the list, modulo statistical
fluctuations (see Fig.~\ref{fig:histogfit2}). The fluctuations in all three directions were strongly correlated.
The shape of the curves was described well by a power law fit.
The exponent is probably the same for all the directions, 
and the constant factor depends on the shape of the torus: it is higher for the directions in which
the torus is more elongated (and so the average height of the simplices is greater).

This shows that the number of loops passing through a given simplex can serve as another indicator
of its position within the torus. Most of the geodesics between distant points
pass through the bulk simplices in the semiclassical region
and do not enter the outgrowths, which are the regions of quantum fluctuations.
If a geodesic passes through a simplex in the outgrowth, 
it usually means that it had its beginning even
deeper in the same outgrowth.

As observed, geodesics between distant simplices tend to pass through simplices of low heights
and avoid simplices of middle and greater heights. It is tempting
to interpret the former as belonging to a semiclassical bulk,
and the latter to outgrowths, which are a result of quantum fluctuation.
If so, then those regions should differ also in other properties apart from such 
non-local ones as height and
the number of loops passing through the simplices.
This is in fact the case.
The neighborhoods of bulk simplices and outgrowth simplices 
look considerably different, which allows for a construction
of local quantities distinguishing between them.

\begin{figure}
\includegraphics[width= \textwidth]{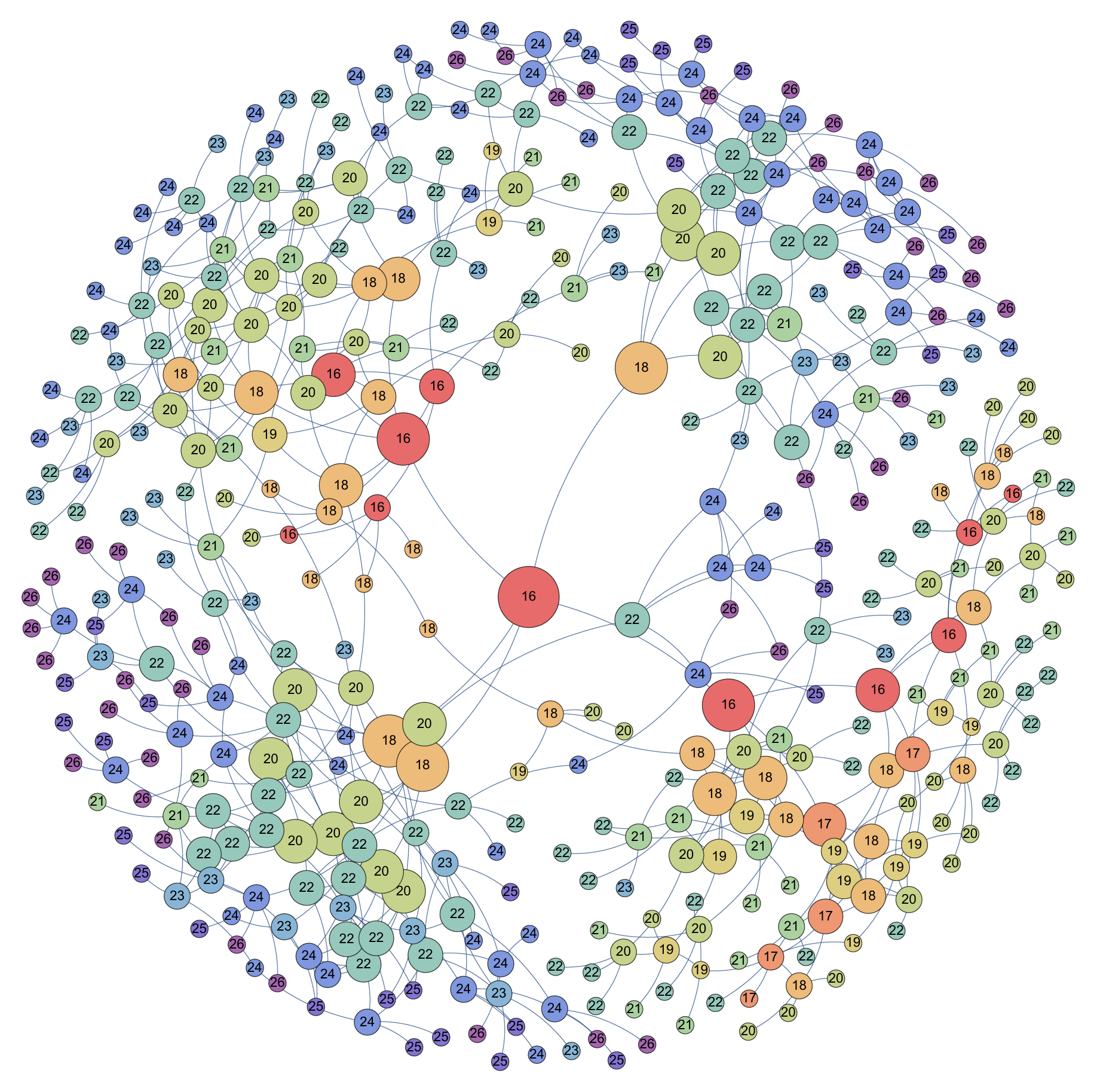}
\caption{Six concentric shells around a simplex of height equal to 16.  The colors and the numbers indicate the height of a simplex, and the size of a vertex its distance from the starting simplex. The central simplex lies in the bulk region.}
\label{fig:83621}
\end{figure}

\begin{figure}
\includegraphics[width= \textwidth]{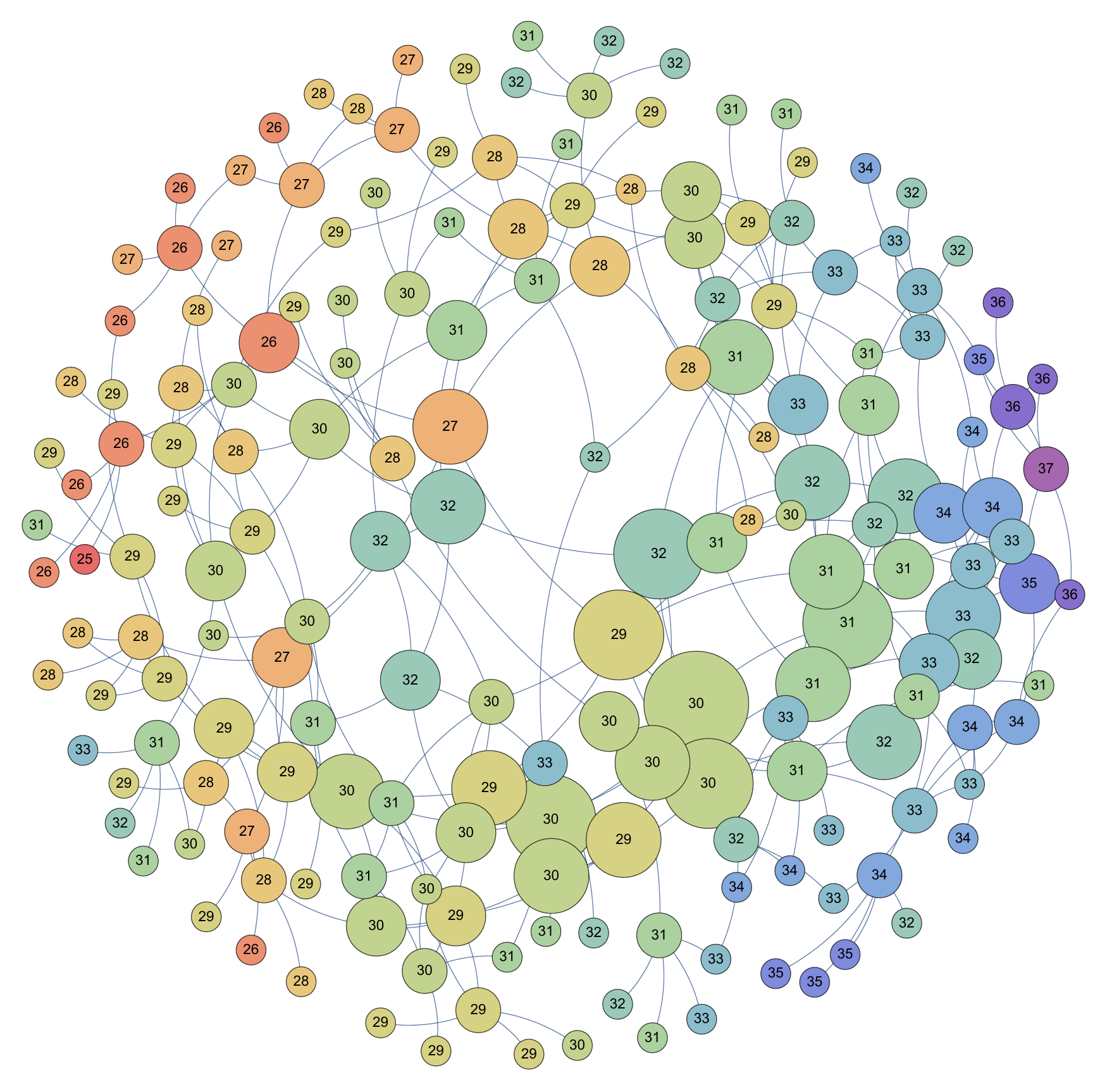}
\caption{Six concentric shells around a simplex of height equal to 30.  The colors and the numbers indicate the height of a simplex, and the size of a vertex its distance from the starting simplex. The central simplex lies in the middle of an outgrowth.}
\label{fig:25}
\end{figure}

\begin{figure}
\includegraphics[width= \textwidth]{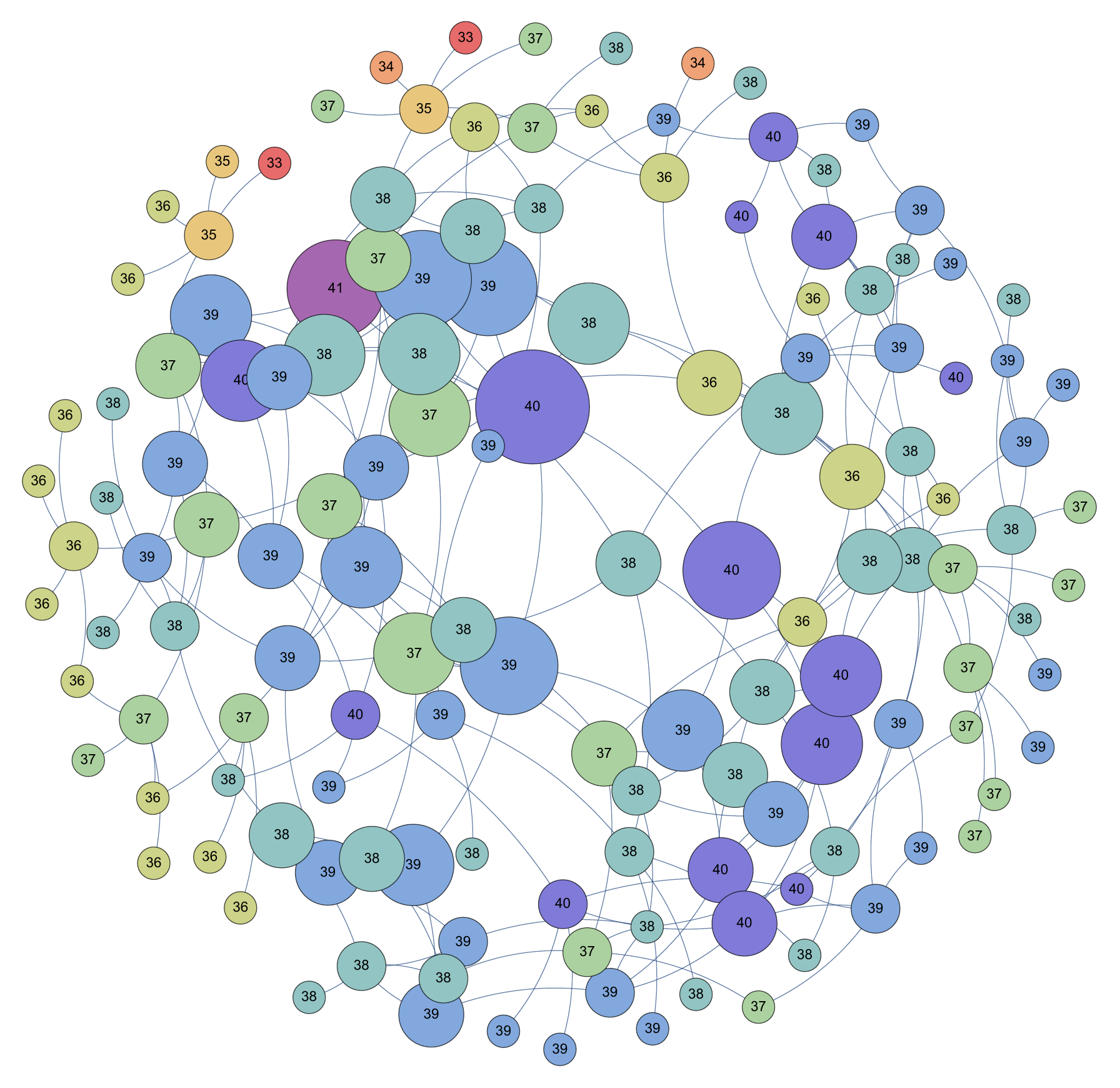}
\caption{Six concentric shells around a simplex of height equal to 40.  The colors and the numbers indicate the height of a simplex, and the size of a vertex its distance from the starting simplex. The central simplex lies near the deep end of an outgrowth.}
\label{fig:113}
\end{figure}

Figs.~\ref{fig:83621} - \ref{fig:113} show subgraphs of the dual lattice, depicting all the simplices
of distance up to 6 from a starting simplex, 
together with the connections between them.
The simplices – vertices of the dual lattice – are represented by circles whose size is a decreasing 
function of distance from the starting simplex. Colors of the circles indicate heights, with red corresponding to the shortest and violet to the longest loops in a given figure. 
The heights are also noted as numbers in the circles. It should be noted that figures in this section are based on data obtained using the version of the algorithm
with periodic boundary conditions in directions $y$, $z$ and $t$. 
Heights defined in this alternative way have similar values and interpretation to $x$-heights
defined previously, though they are not directly equivalent,
e.g., here the minimal value is 16,
which occurs in simplices that form the shortest loop with winding numbers $\{1,-1,1,0\}$ (see Table \ref{tab:loops}).

In Fig.~\ref{fig:83621} the starting simplex had the
height equal to 16. 
As that is the minimal value, we can see that in the neighborhood of the simplex the height tends
gradually to increase together
with distance from it. By the same token, loop length decreases with increasing distance from a simplex
with height equal to 40, which is among largest in the configuration – Fig.~\ref{fig:113}. 
Even a cursory glance at the graphs, moreover, suffices to note
a strong dependence of the total number of simplices
of distance up to 6 from the starting simplex on its loop-length.
As expected, outgrowth regions have an elongated shape and 
a lower Hausdorff dimension than the bulk region, which
is a sign of their fractality.

\begin{figure}
\includegraphics[width= \textwidth]{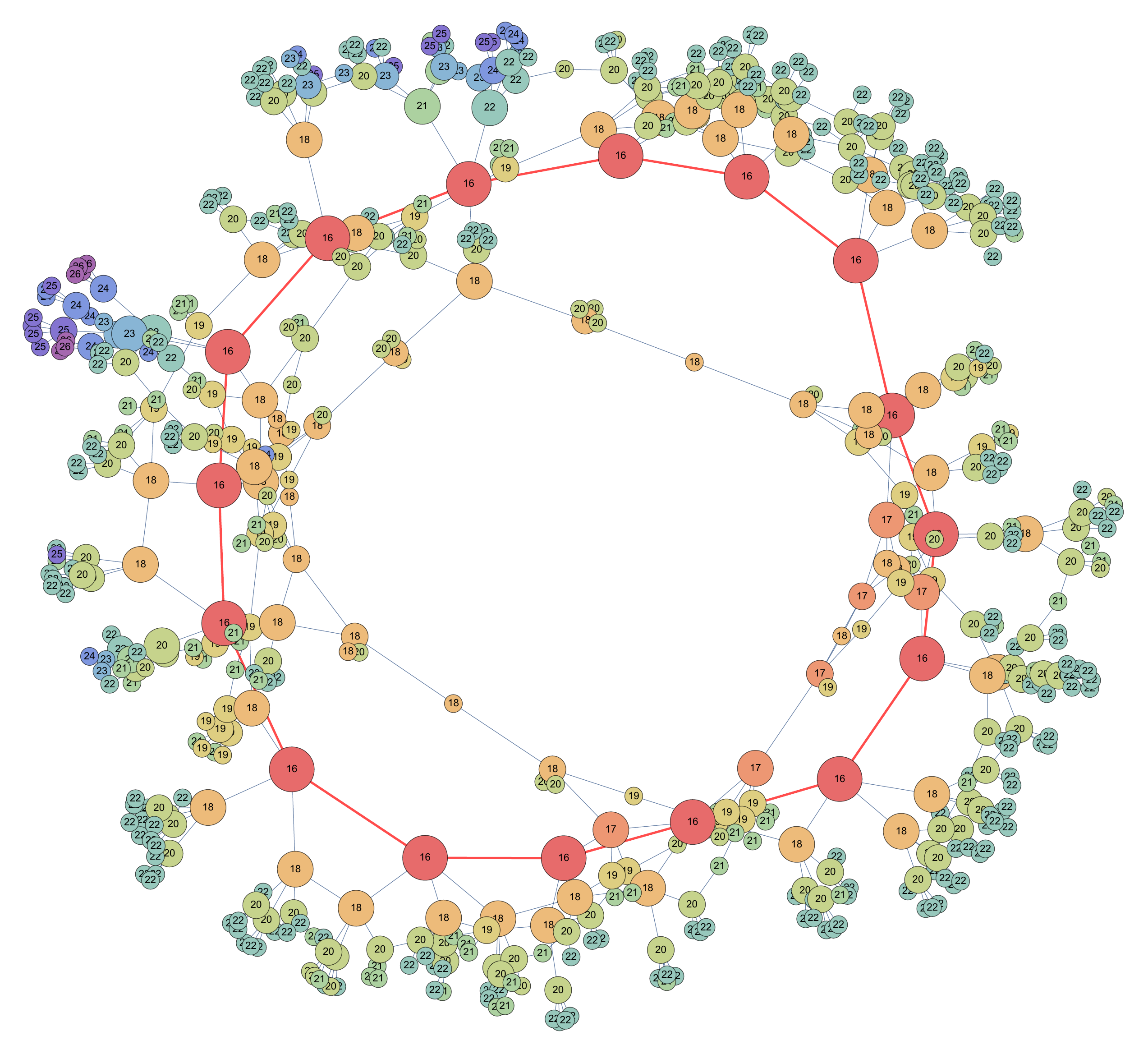}
\caption{The shortest \{1,-1,1,0\} loop together with its neighborhood. The colors and the numbers indicate the height of a simplex, and the size of a vertex its distance from the loop.}
\label{fig:loopgraph}
\end{figure}

Fig.~\ref{fig:loopgraph} shows the shortest $\{1,-1,1,0\}$ loop and its neighborhood.
This loop is among the very shortest loops in the configuration, as
noted in Table \ref{tab:loops}.
It is the only loop with that set of winding numbers and length 16. It is readily seen
that the number of loops with the same winding numbers and increasing length $17, 18, \dots$ in 
the vicinity of the marked loop grows approximately exponentially.

Summarizing, the detailed measurements performed on a typical toroidal configuration which appears in the CDT path integral in the $C$ phase shows the following: the spatial $T^3$ part 
consists of a relatively small 
bulk region, {\it the toroidal center}, which may be called semiclassical, and numerous
fractal outgrowths of an almost spherical topology (with a small boundary)
which contain most of the simplices. Introducing
the lengths of the shortest non-contractible loops in 
the coordinate directions as the heights associated with 
a given simplex made it possible to classify the simplex as belonging
to the toroidal center or to an outgrowth. Furthermore, the number 
of simplices in the outgrowths where the height is a local 
maximum is not small. The interpretation of this is that the 
outgrowths are quite fractal. An important feature
of the length distributions of non-contractible loops associated
to the simplices is that they scale as $N_4^{1/4}$, where 
$N_4$ denotes the size of the triangulation, i.e., the number of 
four-simplices, as shown in Fig.~3.4. The most likely
consequence of such a ``canonical'' scaling is that the
volume of the toroidal center, although small compared to the
volume of the outgrowths,  will also scale canonically with $N_4$. It might 
thus be justified to think of it as semiclassical.

 A similar approach had been successfully utilized in an analysis of 2D Euclidean quantum gravity \cite{ab2}.
In the two-dimensional case the valleys are
not semiclassical, i.e., they do not scale canonically but are a quantum phenomenon.
The situation is different in the case of 
four-dimensional toroidal CDT, where the Hausdorff
dimension is four,
the same as the canonical dimension of the spacetime. The valleys of $T^3$ thus constitute a semiclassical configuration which can act as a starting point for a description of a semiclassical spatial geometry.

\chapter{Quantum Ricci curvature}\label{chap:ricci}
\markboth{Quantum Ricci curvature}{Quantum Ricci curvature}

Chapter based on an article in preparation.

This chapter describes the measurement of a 
recently introduced observable, the quantum Ricci curvature, and shows its connection with 
topological observables such as the height (length of minimal noncontractible loop
passing through a given simplex) defined in the previous chapter.
The understanding of geometry of CDT spacetimes with toroidal spatial topology,
gained through the analysis of noncontractible loops,
led to the discovery of a class of simplices with properties that are
interesting in the point of view of the quantum Ricci curvature.
Such a class of simplices was not found in spacetimes with spherical spatial topology.

\section{Definitions}

The quantum Ricci curvature was introduced by Klitgaard and Loll in \cite{loll1}. It was
calculated by the same authors for smooth model spaces and regular lattices in the same paper, and then tested
on 2D DT in \cite{loll2} and on 4D CDT with spherical spatial topology in \cite{loll3}.
This section will summarize the motivation and the definitions; a complete discussion of the results
is contained in the articles cited.

In general relativity on piecewise linear manifolds,
described by Regge calculus, the scalar curvature is defined
in terms of deficit angles. Such is the definition used in 
the calculation of the bare action of CDT, i.e., the Regge action, 
the analogue of the Einstein-Hilbert action, 
as described in Chapter \ref{chap:introduction}. However, 
that curvature is not a good choice as an observable used to 
understand the structure of the triangulations and 
draw comparisons to semiclassical geometries, as 
deficit angles exist only at the cutoff scale, the edge length $a$ of the simplices.
Attempts to integrate such curvature over larger regions lead in the continuum limit to divergences rather than to averaging out to a
classical expression, since with decreasing $a$ the deficit angles persist,
while their number in any region increases,
and the series to be calculated in general does not happen to be convergent.

The quantum Ricci curvature has the advantage that it can be 
defined at any scale, and the results obtained for a triangulation
can be compared with an analogous quantity calculated for
regular lattices and smooth manifolds.
It is based on the average distance between two spheres centered at nearby
points, which on a Riemannian manifold depends on the (classical) Ricci
curvature.

More concretely, as defined in \cite{loll1}, 
the quantum Ricci curvature $K_q(p,p')$ for a pair of points $(p, p')$ is defined
by the formula
\beq\label{eq:avsph}
\fr{\bar{d}\left(S_p^\delta, S_{p'}^\delta \right)}{\delta}=c_q \left(1-K_q(p,p')\right),
\eeq
where $\delta=d(p,p')$ is the distance between the points, $\bar{d}\left(S_p^\delta, S_{p'}^\delta \right)$ is the distance of the sphere of radius $\delta$ with the center at $p$
to the sphere of the same radius centered at $p'$ (by which is meant
the distance from a point on one sphere to a point on the other sphere, averaged
over the whole spheres), and $c_q$ is a positive constant.
In the case of CDT, the points $p$ and $p'$ are two vertices on a dual lattice (see Chapter 2),
and the definition of distance used for these calculations
is the graph distance, i.e., the number of links on the dual lattice constituting the shortest path from $p$ to $p'$.
The graph distance is also used if this quantity is to be calculated on lattices of other
kinds; in the case of a standard manifold, the distance following from the metric can be used.

It is shown in the same article that on a $D$-dimensional Riemannian manifold
the average distance of spheres of radius $\epsilon$ centered at points $p$ and $p'$
that are at a distance $\delta$ from each other along a unit vector $v$
can be expressed as
\beq
\bar d \left(S_p^\epsilon,S_{p'}^\epsilon \right)=\delta \left(1-\fr{\epsilon^2}{2D}\mathrm{Ric}(v,v)+\mathcal{O}(\epsilon^3+\delta \epsilon^2) \right),
\eeq
where $\mathrm{Ric(v,v)}$ is the (classical) Ricci curvature associated with $v$.
This explains the connection between the Ricci curvature and the quantity that is the 
subject of this chapter (the word ``quantum'' in the name is derived 
not from the character of its definition, which in fact requires no quantum physics,
but from the area of its main prospective usage, which is quantum gravity \cite{loll1}).

The normalized average sphere distance $\bar{d} / \delta$ from eq.~(\ref{eq:avsph})
was in \cite{loll1} calculated and plotted as a function of $\delta$ for 
smooth hyperbolic, flat and spherical model spaces, regular square, hexagonal and
honeycomb lattices, and for so-called Delauney triangulations of a plane.
In the next article, \cite{loll2}, the same was performed for two-dimensional DT, 
and, finally, in the third article of the same authors, \cite{loll3},
the quantity was plotted for configurations in the C phase of four-dimensional CDT
with the spatial topology of a 3-sphere.

\section{Results in the toroidal CDT}

Because the spatial structure of a toroidal CDT is better understood and, moreover, apparently richer than that of spherical CDT,
it is of interest to check if analyzing quantum Ricci curvature 
can add something to this understanding or at least confirm the previous discoveries,
and if the results differ from those reported in \cite{loll3} for the case 
of spherical spatial topology.

\begin{figure}
\centering
\begin{tabular}{c}
\includegraphics[width=0.67 \textwidth]{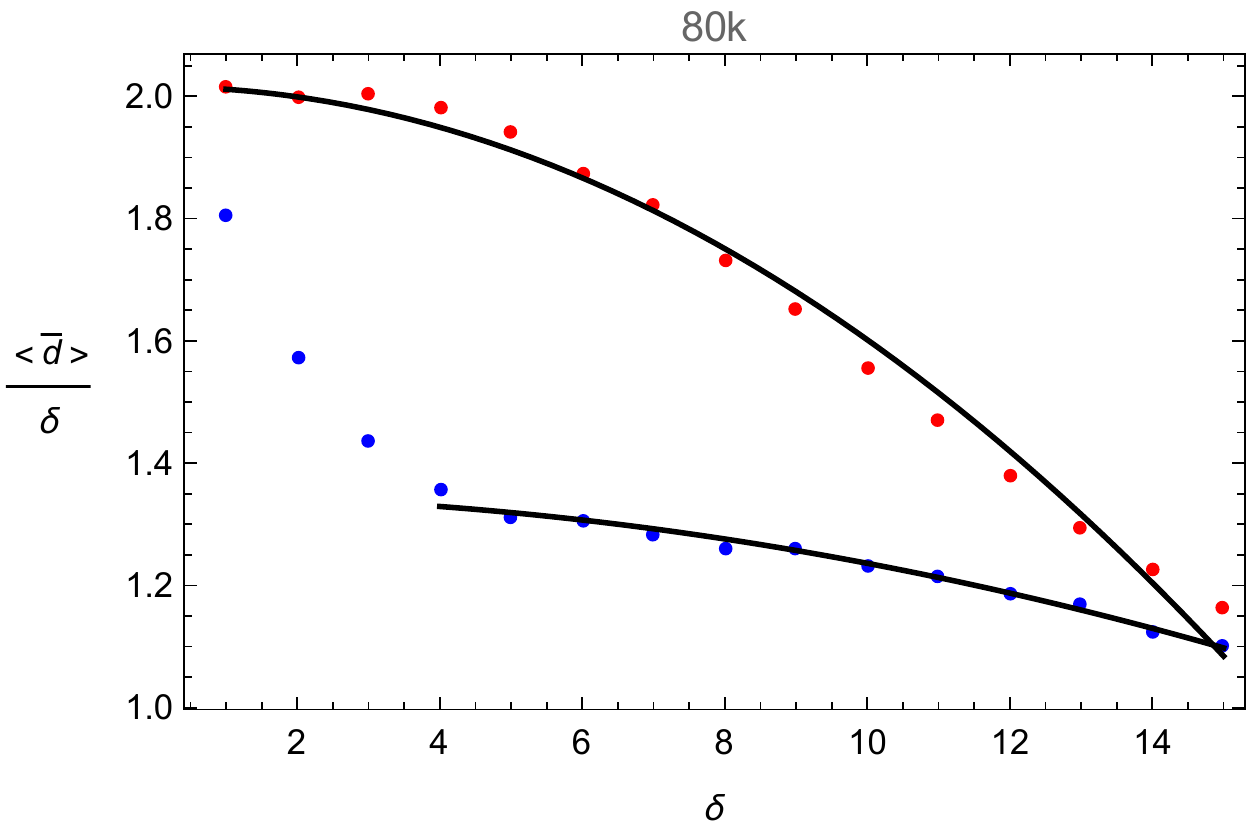}\\
\includegraphics[width=0.67 \textwidth]{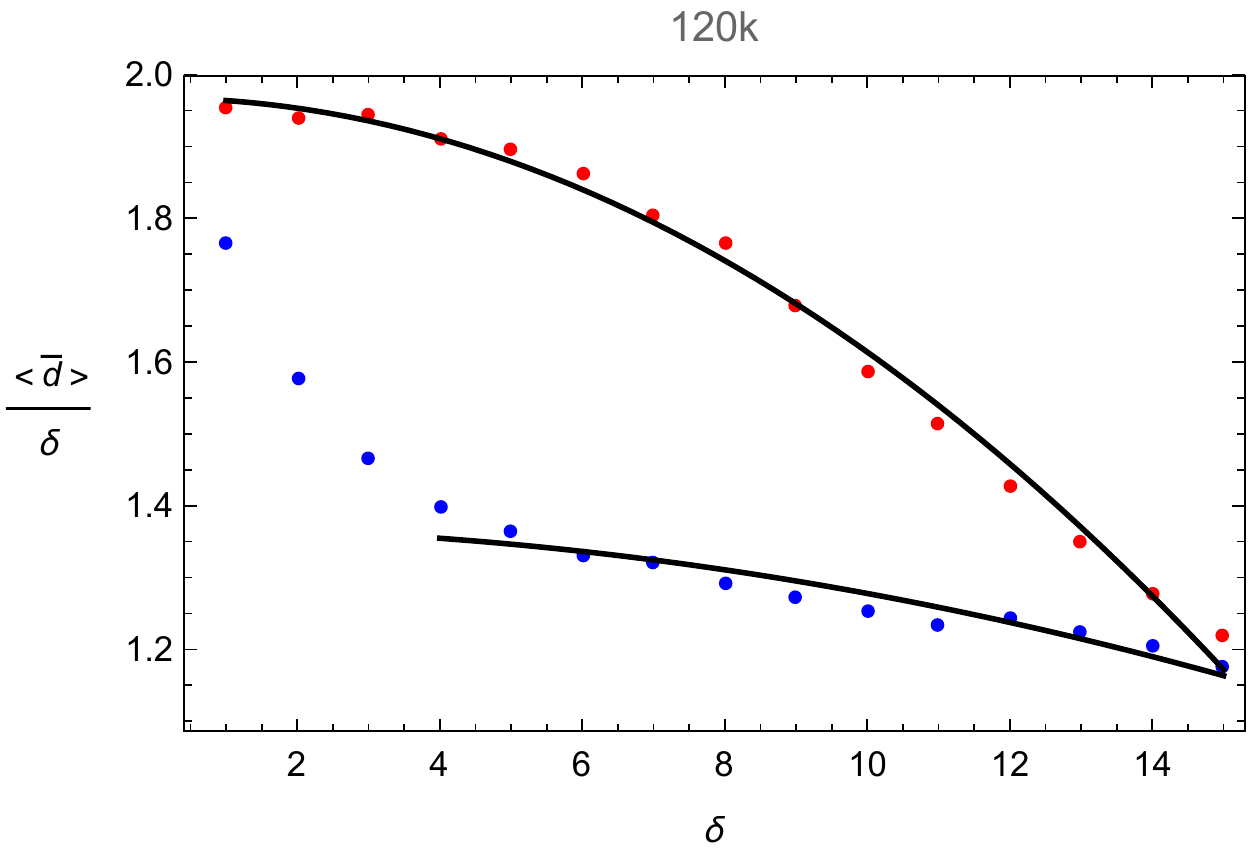}\\
\includegraphics[width=0.67 \textwidth]{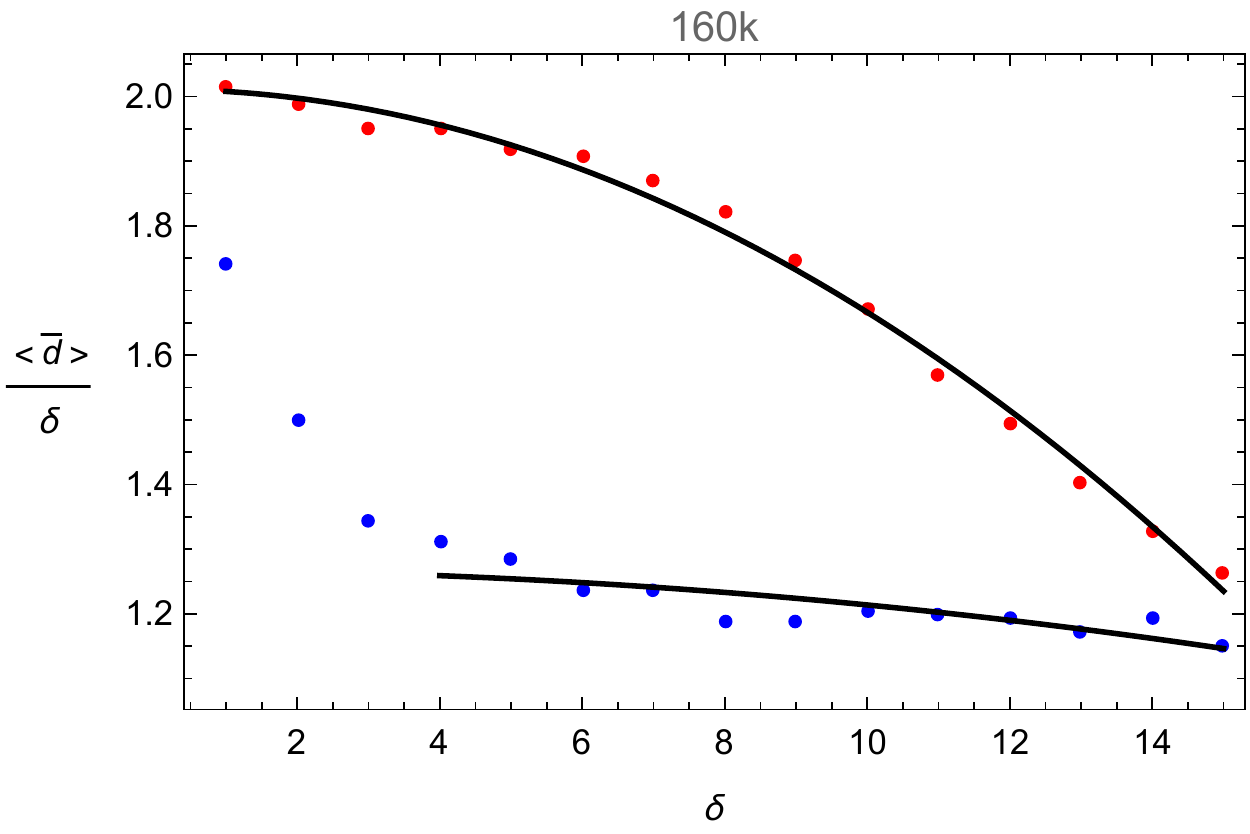}\\
\end{tabular}
\caption{Normalized average sphere distance as a function of $\delta$, measured for several hundred simplices with the longest (blue) and the shortest (red) loops with nontrivial winding numbers in configurations of various sizes ($N_{4,1}=80000, 120000, 160000$). Shown in black are the best fits to four-dimensional spheres.}\label{fig:d80-160k}
\end{figure}

Therefore, the present author measured the normalized average sphere distance (eq. \ref{eq:avsph})
as a function of $\delta$ for simplices in various regions of several configurations
in the C phase of four-dimensional toroidal CDT.
The results of the measurements are shown in Fig. \ref{fig:d80-160k}.
The plots are the result of averaging over from 100 to 300 simplices of the smallest (red) and largest (blue) loop length (height) in
configuration of sizes ranging from $N_{4,1}=80000$ (denoted 80k) to $N_{4,1}=160000$ (160k).
These two groups of simplices were the most interesting ones.
The fits correspond to the averaged sphere distance measured starting at any point on a four-dimensional sphere of radius $r$,
multiplied by a constant $c$.

As numerically calculated in \cite{loll2}, for a four-dimensional manifold 

\beq
\fr{\left< \overline{d} \right>}{\delta} = 1.6524 -  \delta^2 \left(-0.0469 \mathrm{Ric}(v,v) - 0.0067 R + \mathcal{O}(\delta) \right).
\label{eq:4d}
\eeq
On an n-sphere of radius $r$, $Ric(v,v) =\fr{n-1}{r^2}$, and $R = \fr{n(n-1)}{r^2}$. For a 
4-sphere this leads to 
\beq
\fr{\left< \overline{d} \right>}{\delta} = c \left (1.6524 - \frac{0.2211 \delta^2}{r^2} \right).
\label{eq:fit}
\eeq
This is the function fitted to the plots.
In each of the fits for large-height simplices the data points for the three lowest $\delta$
were discarded.

One can readily see that the behavior is markedly different depending on the choice of the simplex at which the quantum Ricci curvature 
is measured. In the small-height case (red dots), the classical solution for a 4-sphere agrees almost perfectly for all the data points,
while in the large-height case (blue dots), the agreement is found only for $\delta$ greater than 3 or 4.
Moreover, the fitted values of the sphere radius $r$ also differ (Table  \ref{tab:1}), being much larger for the 
large-height simplices. This seems to suggest that the central ``valley'' part behaves similarly to a small classical 4-sphere, while
the outgrowths show on the shortest scales a different behavior, perhaps a result of significant quantum fluctuations,
and on the larger scales behave like a 4-sphere with a larger radius of curvature.

As seems to follow from \cite{loll3}, the results for a 4D spherical configuration
regardless of the choice of simplex at which the average sphere distance is measured
correspond to the outgrowth simplices in the toroidal case.
It suggests that a 4D spherical configuration can be visualized as built out only of outgrowths, with no classical part.
On larger scales, the quantum Ricci curvature of the outgrowths behaves similar to a classical 4-sphere, but the
short-scale behavior is very different.

\begin{table}
\begin{center}
\begin{tabular}{ |c|c|c|  }
\hline
 data set&c &r\\
 \hline
80 S  & 1.22    &8.07\\
100 S & 1.20  & 8.39  \\
120 S & 1.19 & 8.63 \\
140 S & 1.19 & 8.75 \\
160 S & 1.22 & 8.83\\ 
 \hline
 80 L & 0.815 & 12.76  \\
 100 L & 0.840 & 14.34  \\
 120 L & 0.829 & 14.15 \\
 140 L & 0.780 & 20.54  \\
 160 L & 0.767 & 17.73  \\
 \hline
\end{tabular}
\end{center}
\caption{The values of best fits of the function (\ref{eq:fit}) to data from configurations of various sizes. S denotes small height simplices, L denotes large height simplices.}\label{tab:1}
\end{table}

\begin{figure}
\centering
\includegraphics[width=0.9 \textwidth]{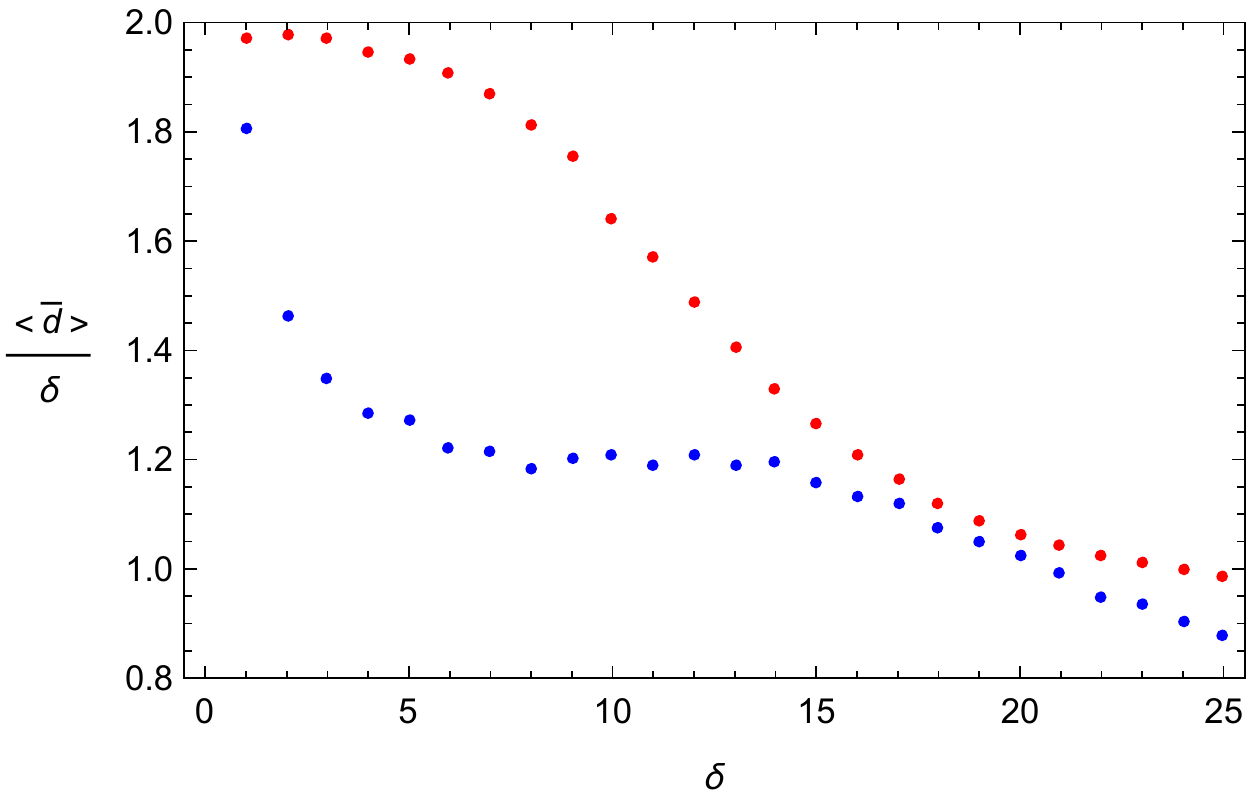}
\caption{Normalized average sphere distance as a function of $\delta$, measured for a number of simplices with the longest (blue) and the shortest (red) loops with nontrivial winding numbers in configurations of size $N_{4,1}=$ 160000. For $\delta>15$, the results become similar for both sets of simplices.}\label{fig:dist25}
\end{figure}

For larger $\delta$, the method becomes unreliable, partly because of the finite size of the systems, as also mentioned in \cite{loll3}.
For $\delta>15$ the extent of
$3 \delta$ becomes close to the size of the system.
Moreover, results for all simplices become similar, because in those cases mostly simplices in outgrowths (which are globally more numerous)
are probed by the spheres no matter what the starting point was (see Fig. \ref{fig:dist25}).
This shows again that the behavior of the large height simplices is the generic one.

It could be argued that curves corresponding to some other manifolds
rather than four-dimensional spheres should also be fitted to the data points,
as the neighborhood of simplices in CDT may conceivably be better approximated
by another manifold,
and, especially in the case of simplices
in the outgrowths, the Hausdorff dimension may differ from 4.
There are, however, technical difficulties in performing such an analysis since for
other manifolds, e.g., a four-torus, the (classical) Ricci curvature appearing
in eq.~(\ref{eq:4d}) is not uniform, and therefore finding the fitting function
would be much more difficult. As for spheres of dimension other than four, 
the fitting function obtained up to the term quadratic in $\delta$ would be 
equivalent to that in eq.~(\ref{eq:fit}). The only change would be in the values of the constants, 
which are unimportant because they can be absorbed into parameters $c$ and $r$. This method 
would thus not answer the question whether a sphere of dimension other than four can provide
a better fit. To answer that question, it would be necessary either to calculate
another term in the expression (\ref{eq:4d}) and in analogical expressions for manifolds of other
dimension or to find a numerical fit in other way. This should be possible and might be 
an interesting subject for a future article.

In conclusion, considering in CDT a universe with toroidal spatial topology makes it
possible to measure some additional quantities such as the height, defined in the previous
chapter, from which we can get insight into the shape of the universe, and which we can compare with the quantum Ricci curvature, presented here.
The class of simplices (corresponding to red points in figure \ref{fig:d80-160k}) whose quantum Ricci curvature behaves very similarly
to the case of a classical four-sphere 
exists in toroidal 4D CDT but apparently not in spherical 4D CDT.
Those simplices are the small-height simplices located in the semiclassical
part of the configuration, which seems to be absent in the spherical case.

\chapter{Scalar fields in toroidal four-dimensional CDT}\label{chap:scalar}
\markboth{Scalar fields in toroidal four-dimensional CDT}{Scalar fields in toroidal four-dimensional CDT}

\noindent  Some parts of this chapter are copied from the following publications, on which the chapter is based:
\begin{enumerate}[start=3,label=\lbrack \arabic*\rbrack]
\item J.~Ambjorn J, Z.~Drogosz, J.~Gizbert-Studnicki, A.~Görlich, J.~Jurkiewicz, D.~Németh, 
\textit{CDT Quantum Toroidal Spacetimes: An Overview}, Universe  \textbf{7}(4) (2021) 79
\item J.~Ambjorn, Z.~Drogosz, J.~Gizbert-Studnicki, A.~Görlich, J.~Jurkiewicz, D.~Németh, 
\textit{Cosmic Voids and Filaments from Quantum Gravity}, arXiv:2101.08617 
\item J.~Ambjorn, Z.~Drogosz, J.~Gizbert-Studnicki, A.~Görlich, J.~Jurkiewicz, D.~Németh, 
\textit{Matter-driven phase transition in lattice quantum gravity}, arXiv:2103.00198
\item J.~Ambjorn, Z.~Drogosz, J.~Gizbert-Studnicki, A.~Görlich, J.~Jurkiewicz, D.~Németh, 
\textit{Scalar Fields in Causal Dynamical Triangulations}, arXiv:2105.10086.
\end{enumerate}

\section{Scalar fields defining coordinates}

As the previous chapters make clear, there exists a well-defined geometric structure
underlying a typical configuration in the C-phase of toroidal
four-dimensional CDT. 
The discovery that most of the simplices
are contained in outgrowths of spherical topology, each of which is
connected to the rest of the triangulation by a three-dimensional surface with 
a relatively small area, suggests that a coordinate system defined via a classical scalar field
satisfying the Laplace equation, a procedure well-known in classical general relativity,
might be well-suited to these geometries.
The boundary conditions of the fields are defined by jumps at the boundaries of the elementary cell.
Because of the averaging property of the Laplace equation, 
the scalar field is expected to vary mostly within the toroidal center
and to remain close to constant in an outgrowth, thereby putting greater emphasis on
the semiclassical part of the triangulation.

The nontrivial fractal structure of the geometries makes 
the pseudo-Cartesian coordinates described in Chapter \ref{chap:cart} nonoptimal.
Hypersurfaces of constant coordinates behave much better in coordinates defined
with the help of scalar fields and may be used to define an analogue of foliations
in all space-time directions, making it possible to
visualize and measure multidimensional correlations in all directions. 

The scalar fields in question map a four-dimensional 
Riemannian manifold $\mathcal{M}$ with the topology of $T^4=(S^1)^4$ and an
arbitrary metric $g_{\mu\nu}$ to a Riemannian manifold $\mathcal{N}$ with the same
topology and the trivial flat metric $h_{\alpha\beta}=\delta_{\alpha\beta}$. Each of the fields 
$\phi^\alpha (x)$, $\alpha =1,2,3, 4$, is a map ${\cal M} \to S^1$. 
Together they minimize the action
\beq
S_M[\phi,{\cal M}]	= \frac{1}{2} \int \dd^{4}x \sqrt{g(x)}  g^{\mu \nu} (x)  h_{\rho \sigma}(\phi^\gamma(x)) \partial_\mu \phi^\rho(x) \partial_\nu \phi^\sigma(x)).
\label{classical_field_eq}
\eeq
Since the metric $h_{\alpha\beta}$ is diagonal, the minimization of this action 
is equivalent to solving four independent Laplace equations: 
\begin{eqnarray}
\Delta_x \phi^\sigma(x) = 0,\quad 
\Delta_x = \frac{1}{\sqrt{g(x)}}\partial_\mu \sqrt{g(x)}g^{\mu\nu}(x)\partial_\nu.
\label{laplace_eq}
 \end{eqnarray}
 
 \begin{figure}
\includegraphics[width=\textwidth]{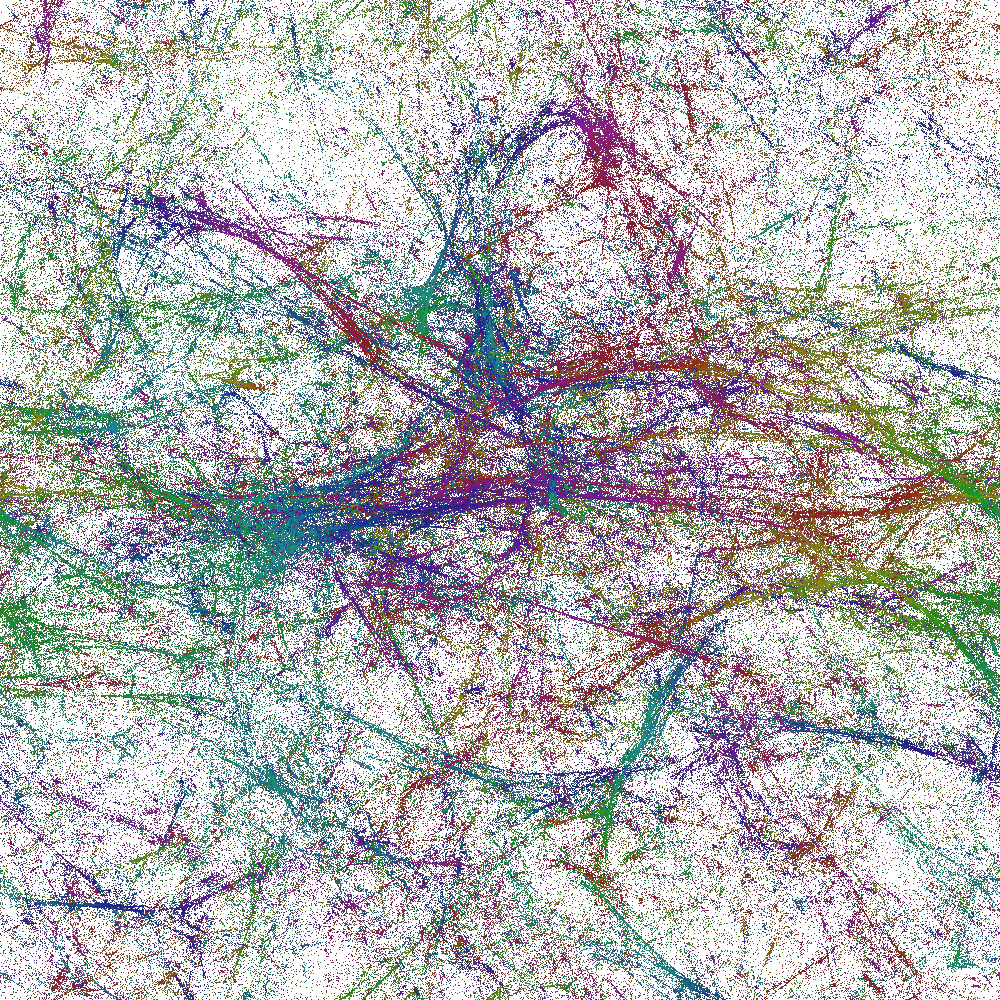} 
\caption{The projection of four-volume on the $xy$-plane, as defined by (\ref{hypers_vol_eq_psi}) for a CDT configuration in phase $C$. Different colors correspond to different times $t$ of the original $t$-foliation. 
}\label{phi-xy_plot}
\end{figure}
\begin{figure}
\includegraphics[width=\textwidth]{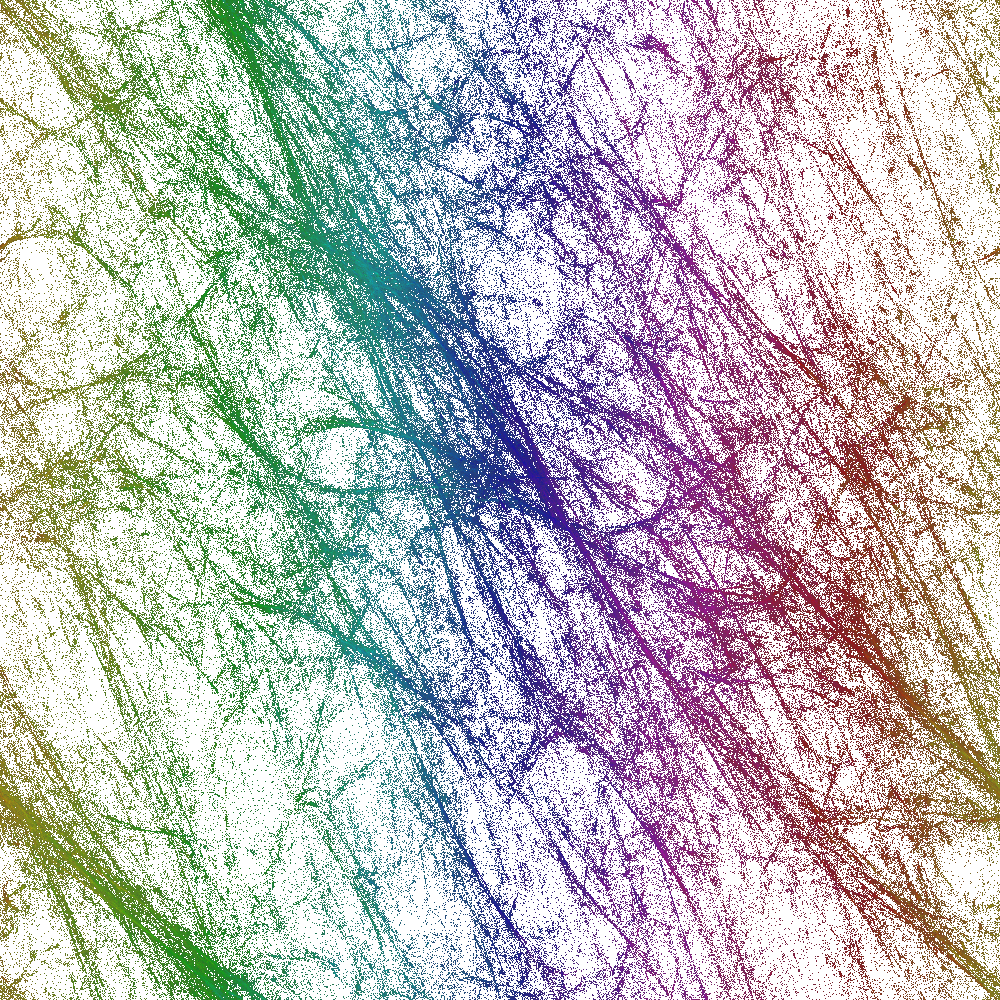}
\caption{The projection of four-volume on the $tx$-plane for a CDT configuration in phase $C$. There is a strong correlation between the original $t$-foliation (color) and new time coordinate $\psi^t$ (horizontal axis).} 
\label{phi-xt_plot}
\end{figure}

To facilitate the following discussion, it is convenient to also consider as an example a 
one-dimensional analogue of this problem.
Let $\mathcal{M}$ be $S^1$ with a unit circumference and a positive $\sqrt{g(x)}$.
Then, $x\to \phi(x)$ is a map $S^1\to S^1$ that can be used to define
a new coordinate, replacing $x$. 
If the codomain of $\phi(x)$ is a circle with a circumference $\delta$,
then there is a solution to the Laplace equation (\ref{laplace_eq}) that satisfies
$\phi(x+n)=\phi(x)+n \delta$ and $ \dd \phi(x) = \delta \cdot \sqrt{g(x)}  \dd x$.
With no loss of generality, it is possible to rescale the field and set $\delta=1$.
The solution $\phi(x)$ is fixed by picking $x_0$ where $\phi(x_0)=0$. 
The map $x\to \phi(x)$ becomes a monotonically increasing invertible map in the whole domain $\mathbb{R}$ with the shape of the simplicial manifold repeated (modulo a constant)
with the period $\delta$.
The parametrization of $\cal M$ in terms of $\phi$ instead of $x$
leads to the volume density in the range $\phi$ to $\phi+\dd \phi$ proportional 
to $\sqrt{g(x)}\dd x$, so that effectively $g(\phi) = 1$.
One can also perform the modulo operation and define a function
$\psi(x) = \textrm{mod}(\phi(x)-\phi(x_1),\delta)$. This function satisfies the Laplace equation in the range between $x_1$ (where $\psi(x)=0$) and $x_1+1$ (where $\psi(x)=\delta$). The equation satisfied by $\psi(x)$ becomes a Poisson equation with the extra 
inhomogeneous local term producing jumps at boundary points $x=x_1$ and $x=x_1+1$. It can also
be considered a Laplace equation with non-trivial boundary conditions defined 
by the jumps. Back in the case of ${\cal M}$ with the topology of $T^4$, 
this would mean solutions to eq.\ (\ref{laplace_eq}) that wrap once around $S^1$ 
in direction $\sigma$. The points $x$
in ${\cal M}$ that satisfy $\phi^\sigma(x) =c$ form hypersurfaces $H^\sigma(c)$ whose union for $c$ varying in a range of length 1 covers the whole ${\cal M}$.

The example described above is modified in the following way to be suitable for piecewise linear manifolds of CDT. The action in eq.\ (\ref{classical_field_eq}) is replaced
by its discrete analogue, which can be split into contributions from each 
field $\phi_i^\sigma$:
\begin{eqnarray}
\label{CDT_laplace_eq}
S_M^{CDT}[\phi^\sigma,{T_E}]= \frac{1}{2} 
\sum_{i \leftrightarrow j} (\phi_i^\sigma - \phi_j^\sigma -\delta \mathbf{B}_{ij}^\sigma)^2,
\end{eqnarray}
where $B$ is the matrix defining the boundaries (see Chapter \ref{chap:implementation}).
The summation is performed over all pairs of neighboring four-simplices in the triangulation $T_E$ representing the manifold $\mathcal{M}(g_{\mu\nu})$ in eq.\ (\ref{classical_field_eq}).

The parameter $\delta$ plays the same role as in the one-dimensional example, and here too it can be set to $1$ by a rescaling of the field. The action (\ref{CDT_laplace_eq}) is invariant under a constant shift of the scalar field (the Laplacian zero mode). More importantly, it is
also locally invariant under a modification of the boundary $\mathbf{B}_{ij}^\sigma$ 
by moving a simplex $i$ to its other side (possibly changing the total number of faces belonging to the boundary) and compensating for the change of the field in the center of the simplex 
with a shift by $\pm\delta$ (depending on the side of the boundary) of the field value.

The field $\phi_i^\sigma$
minimizes the action (\ref{CDT_laplace_eq}), and thus has to satisfy the non-homogeneous Poisson-like equation
\begin{equation}
\label{classical}
    \mathbf{L}\phi^\sigma =  b^\sigma,
\end{equation} 
where $\mathbf{L}=5 \mathbf{1} - \mathbf{A}$ is the $N_4\times N_4$ Laplacian matrix, and $\mathbf{A}_{ij}$ is the adjacency matrix with entries of value 1 if simplices $i$ and $j$ are neighbors and 0 otherwise. The Laplacian matrix $\mathbf{L}$ has a constant zero mode but 
can be inverted after fixing the value of the field $\phi_{i_0}^\sigma=0$ for an arbitrary simplex $i_0$. The inversion is computationally demanding, as the matrix $\mathbf{L}$,
albeit sparse, is typically of size $N_4 \approx 10^6$. Then, the multi-dimensional analogue of the one-dimensional function $\psi(x)$ is given by
$\psi_i^\sigma$:
\begin{equation}
    \psi_i^\sigma  = \textrm{mod}(\phi_i^\sigma, 1),
\end{equation}
where we assume $\delta=1$.
New boundaries of the elementary cell, implied by the jump of the scalar field $\psi_i^\sigma$ by $\delta=1$, are defined by $\bar{b}^\sigma = \mathbf{L}\psi^\sigma$.
Equivalently, new boundaries can be defined by hypersurfaces from a family $H(\alpha^\sigma)$
defined by equations
\begin{equation}
\label{alpha}
    \psi_i^\sigma(\alpha^\sigma) = \textrm{mod}(\phi_i^\sigma-\alpha^\sigma, 1),\quad
      \bar{b}^\sigma(\alpha^\sigma) = \mathbf{L}\psi^\sigma(\alpha^\sigma).
\end{equation}
The union of hypersurfaces for $\alpha^\sigma \in [0,1)$ covers the whole elementary cell.
This defines a foliation in the direction $\sigma$, and $\psi_i^\sigma= \psi_i^\sigma(0)$ 
can serve as a coordinate.
The same procedure can be performed in all four directions
$\sigma\in \{ x,y,z,t\}$ for any configuration obtained in the numerical simulations, and in this way every simplex $i$ will be assigned a unique set of coordinates $\{\psi^x_i,\psi^y_i,\psi^z_i,\psi^t_i\}$, all in the range between 0 and 1. 
The coordinate $\psi^t_i$ is not the same as the one coming from the original foliation of the CDT model.

These coordinates preserve the triangulation structure thanks to the 
averaging property of solutions to the Laplace equation: the coordinates of each simplex are equal to the mean value of the coordinates of its neighbors (up to the shift of the field at the boundary).
They permit to analyze the distribution of the four-volume (the number of simplices) contained in hypercubic blocks with sizes $\{\Delta\psi^x_i,\Delta\psi^y_i,\Delta\psi^z_i,\Delta\psi^t_i\}$, 
which is equivalent to measuring the integrated $\sqrt{g(\psi)}$:
\begin{eqnarray}
\Delta N(\psi) = \sqrt{g(\psi)} \prod_\sigma \Delta \psi^\sigma = N(\psi) \prod_\sigma \Delta \psi^\sigma.
\label{hypers_vol_eq_psi}
\end{eqnarray}

Measuring the four-dimensional volume distribution $N(\psi)$
is a significant upgrade compared to the volume profile (the volume distribution 
in the time dimension parametrized by CDT's inherent foliation), which
was described in some of the previous studies.
Figures \ref{phi-xy_plot} and \ref{phi-xt_plot} show projections of the four-volume density distribution of a typical configuration in phase C on two-dimensional parameter subspaces,
the $xy$-plane and the $tx$-plane, respectively,
integrating over the remaining two directions.
The patterns of voids and filaments are visually strikingly similar to
voids and filaments observed in our real Universe (see e.g.\ \cite{compare} and the plots in \cite{galaxies,figurefrom}).

The higher-density domains tend to attract each other, forming denser clouds, which survive in the imaginary time evolution in a quantum trajectory
(see Fig. \ref{phi-xt_plot}). There seems to appear a sequence of scales, characterizing a gradual condensation of gravitationally interacting ``objects''.
However, it should be noted that the system described in this section,
contrary to the next one, contains no matter. It is quantum fluctuations
of geometry that appear as if there were massive interacting objects, somewhat analogous to dark matter. This indicates that quantum geometric
fluctuations are highly correlated, an effect which could not be easily analyzed without introducing a global set of coordinates. 
Of course, these are quantum effects of Planckian size,
but if a more extended model exhibited inflation, one could imagine that aspects of these objects would be frozen when entering the horizon, like the standard Gaussian fluctuations in simple inflationary models, and then would re-enter the horizon at a later stage, after reheating, as classical densities.

\section{Dynamical scalar fields}

The previous section described scalar fields that were fictitious in the sense that they 
were added to fixed, well-thermalized configurations without matter,
and no further simulations were performed after adding them. 
The geometry was taken as given.
The fields served only as a tool for analysing existing configurations, using the fact that 
a classical solution to the discrete Laplace equation  
has in such toroidal geometries properties desirable for defining a set of coordinates, being a close analogue of the harmonic coordinate condition used in the context of GR.

This section, on the other hand, describes a model of CDT with dynamical matter fields, in which
the scalar field action has a form similar as previously but 
is added to the Ricci action of CDT, and the coupling is
preserved throughout the Monte Carlo simulations, influencing the choice of moves 
and the geometry of the triangulations.
The scalar fields considered in this section assume values on a torus with
circumference $\delta$ in each spatial direction. The parameter $\delta$ 
can also be interpreted as the size of the jump of the field values
on the boundaries of an elementary cell.
The formulation is topological, i.e., the matter action does not depend on the specific position of the boundary but just on the value of the jump.
After a thermalization phase, such a configuration may look considerably different than 
a configuration without dynamical scalar fields, described in the previous section
and in the previous chapters.
Results indicate that for values of $\delta$ above a certain threshold,
there is a new phase transition, previously unobserved in four-dimensional CDT.

The results can be contrasted with EDT, where 
matter does not seem to influence the geometry in a significant way \cite{matterEDT1, matterEDT2, matterEDT3, matterEDT4, matterEDT5}.
In two-dimensional CDT models, adding a coupling to scalar 
\cite{2dscalar1, 2dscalar2, 2dscalar3, 2dscalar4, 2dscalar5} and gauge \cite{2dgauge} fields
did result in a notable change of the geometry, as described in the articles cited.
However, no substantial impact of scalar field on spacetime geometry or
the phase diagram was shown in four-dimensional CDT 
in studies previous to the articles on which this chapter is based.

A $d$-component massless scalar field $\phi$ is the simplest quantum matter that can be added to the quantum geometry of CDT.
Adding it to the model follows several steps described in the previous
section. 
With the definition of the boundary jump vector 
\beql{def:jump_vector}
b_i^\sigma = \sum_j B_{ij}^\sigma.
\eeq
and the three-volume (i.e., the number of tetrahedra) of the boundary
\beql{eq:bvol} 
V^\sigma = \frac{1}{2} \sum_{ij} {B_{ij}^\sigma}^2 = \frac{1}{2}\sum_i | b_i^\sigma |, 
\eeq
the discrete matter action (\ref{CDT_laplace_eq}) can be written as 
\beq
 S_M^{CDT}[\phi,T] 
 =\sum_{\sigma=1}^d \left(\sum_{i, j} \phi_i^\sigma \mathbf{L}_{ij}(T) \phi_j^\sigma -2\delta\sum_i \phi_i^\sigma b_i^\sigma + \delta^2\cdot V^\sigma \right)
 \label{actionb}
\eeq
As previously, 
the $d$-component scalar field $\phi$ takes values on a torus 
$\mathcal{N}=(S^1)^d$ with circumference $\delta$ in each direction. 
Each component of the field $\phi^\sigma(x) \in S^1$ 
is required to wind around the circle once 
as $x$ goes around any non-contractible loop in $T$ that crosses a boundary in direction $\sigma$.

Simulations for $\phi^\sigma(x) \in \mathbb{R}$ without the jump boundary condition imposed
were also performed, but in that case the scalar field did not lead to a noteworthy
change of the geometry, as for those boundary conditions a constant function
is a solution to the Laplace equation, and it does not contribute to the action.
The case $\phi^\sigma(x) \in S^1$ with the winding requirement
is much more interesting, as a constant function has winding number zero and therefore
is not allowed as a classical solution; the allowed classical solutions do contribute to the matter action, 
as the following discussion will make clear.

The scalar field can be decomposed into the classical solution part $\bar{\phi}^\sigma$ 
and the quantum part $\xi^\sigma$ describing quantum fluctuations,
$\phi^\sigma  = \bar{\phi}^\sigma + \xi^\sigma$:
\begin{eqnarray}
 S_M^{CDT}[\phi,T] = \sum_{\sigma,i,j}  \xi_i^\sigma \mathbf{L}_{ij}(T) \xi_j^\sigma 
 +\sum_{\sigma=1}^d \left(\sum_{i, j}\bar \phi_i^\sigma \mathbf{L}_{ij}(T) \bar\phi_j^\sigma -2\delta\sum_i \bar\phi_i^\sigma b_i^\sigma + \delta^2\cdot V^\sigma \right) = \nonumber  \\
 = \sum_{\sigma,i,j}  \xi_i^\sigma \mathbf{L}_{ij}(T) \xi_j^\sigma + S_M^{CDT}[\bar \phi,T]. \quad\mbox{~~~~~~~~}
 \label{actionb2}
\end{eqnarray}

\begin{figure}
\centering
\includegraphics[scale=0.13]{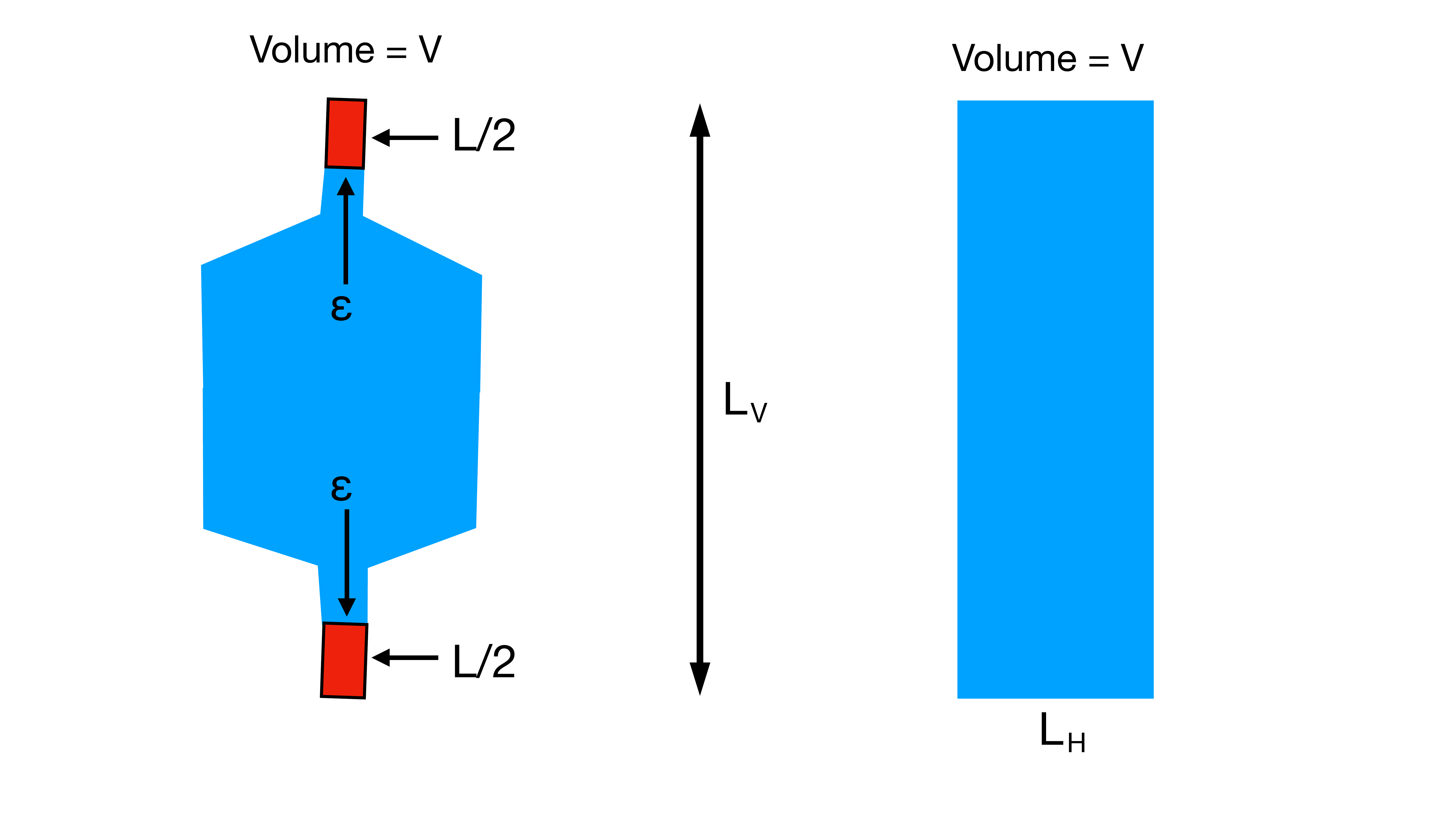}
\caption{Left: a torus (opposite sides identified) with a pinch. The region in red is the region where $\phi$ changes from 0 to $\delta$ from bottom to top. In the lower red part it changes from 0 to $\delta/2$, in the blue part it stays constant and equal to $\delta/2$ and in the upper red part it changes from $\delta/2$ to $\delta$. $\phi$ is constant in horizontal direction. The volume of the red region is $L \cdot \varepsilon$. Right:  a torus where $\phi$ is constant in horizontal direction and uniformly increases from 0 to $\delta$ from bottom to top. The two tori are assumed to have the same vertical length $L_v$ and the same volume $V$ (which for the right figure can be written as  $V=L_V L_H$).}\label{fig-pinch}
\end{figure}

The equation for the classical field $\bar\phi^\sigma$
\begin{equation}\label{laplace_eq3} 
    \mathbf{L}\bar\phi^\sigma =  \delta \cdot b^\sigma,
\end{equation}
has a boundary term and thus admits nontrivial solutions. 
Since $\phi^\sigma$ and  $\bar \phi^\sigma$ have winding number one, the fluctuation field $\xi^\sigma$ is a scalar field with winding number zero, i.e., an ordinary scalar field taking values in $\mathbb{R}$.
Thus the difference between the impact of the scalar field in $(S^1)^d$ and that of the ordinary field in $\mathbb{R}^d$ is the dependence of the effective matter action on the non-trivial classical solution $\bar\phi$; note that the size of the jump $\delta$ fixes the scale of the classical field.
Setting $\delta=0$ recovers the case $\mathbb{R}^d$ where $S_M^{CDT}[\bar\phi,T]$ is zero (the absolute minimum), as $\bar\phi=const$ for any quantum geometry $T$. 
For $\delta>0$ the constant solution is not allowed, and the action $S_M^{CDT}[\bar\phi,T]$ depends on the specific geometry $T$. 
However, the matter action can be reduced almost to zero by a process named
``pinching'', illustrated in Fig.\ \ref{fig-pinch} 
for the simple case of a two-dimensional torus with a one-dimensional field $\phi$ changing in the vertical direction.
The argument is clearly valid in higher dimensions and only depends on one direction being periodic.  
The left-hand side picture is a torus with volume $V$ and vertical length $L_V$,
which is pinched to a cylinder of circumference $\varepsilon$ and length $L$.
The field $\phi(x)$ changes uniformly from $0$ to $\frac{1}{2}\delta$ over a distance $\frac{1}{2}L$ in the lower red part.
In the blue part, the field $\phi(x)$ is constant and equal $\frac{1}{2}\delta$,
and in the upper red region the field changes uniformly from $\frac{1}{2}\delta$ to $\delta$.
The red and blue regions are smoothly joined.
The total matter action of this field configuration is then 
\begin{equation}\label{jan50}
S_{M}^{CDT}[\phi, T_L] = \Big( \frac{\delta}{L}\Big)^2 L \, \varepsilon = \delta^2 \frac{\varepsilon}{L},
\end{equation}
and thus the minimal action for a classical field configuration $S_{M}^{CDT}[\bar\phi,T_L]$ for this geometry is even lower. This can clearly be made arbitrarily small when $\varepsilon \to 0$, and this is even more true in higher dimensions. 
The right-hand side picture is a torus with volume $V$ and vertical length $L_V$. For this geometry, the action is minimal for a field changing uniformly from 0 to $\delta$, and equal to
\begin{equation}\label{jan51}
S_{M}^{CDT}[\bar\phi,T_R] = \Big( \frac{\delta}{L_V}\Big)^2 L_V L_H  = \delta^2 
\frac{V}{L_V^2},~ V= L_H L_V,
\end{equation}
which is bounded from below when $V$ and $L_V$ are fixed. 

The consequence of this possibility for restructuring the triangulation
to decrease the contribution to the action from
the classical solution of the scalar field is the competition between 
the scalar field action and the gravitational Regge action.
The ``pinched'' triangulations have a small matter action but a larger Regge action
than triangulations with a shape similar to those with pure gravity and no matter.
A simple minisuperspace model, like the Hartle-Hawking model, suggests that for small jumps $\delta$ the geometric part of the action dominates and the generic configurations in the path integral are quite similar to the ones which dominate when no matter field with a jump is present. However, for large $\delta$ the total action will be the lowest for pinched configurations and the system will instead fluctuate around pinched configurations. Thus, the system 
can be supposed to undergo a phase transition as a function of the jump magnitude $\delta$.

\begin{figure}
\includegraphics[width=\textwidth]{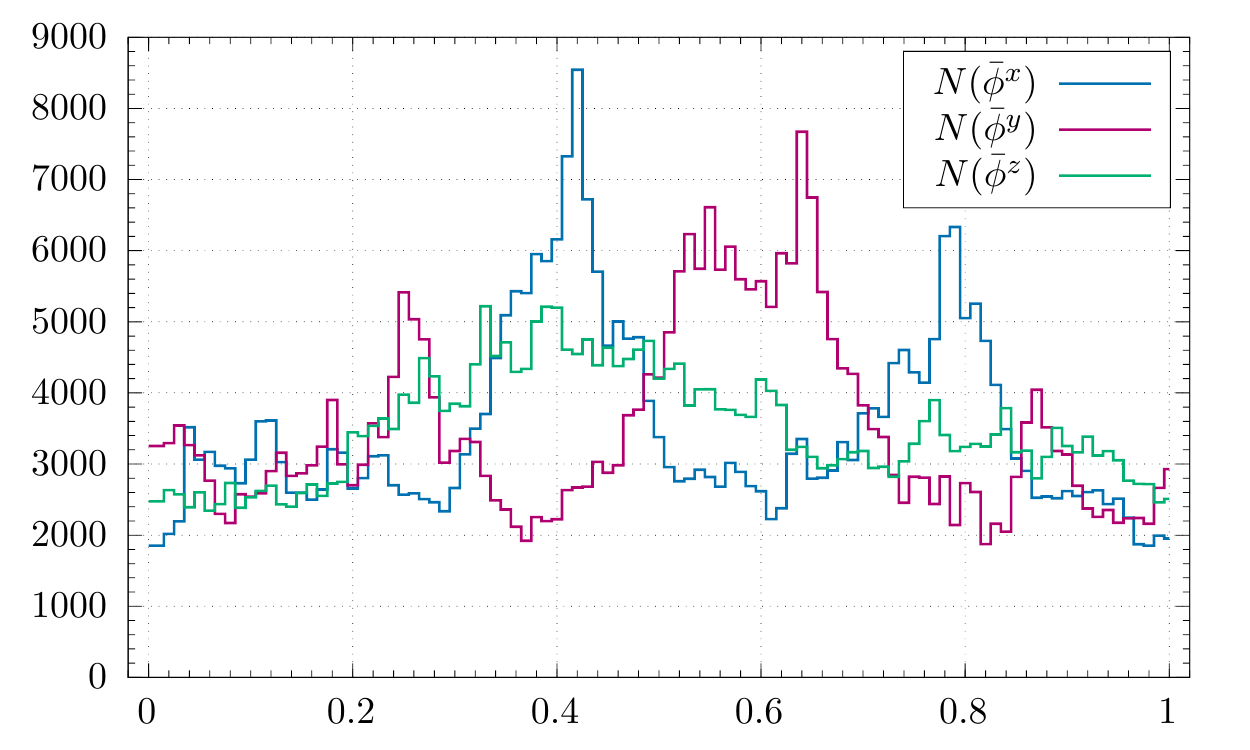} 
\caption{The projection of four-volume, as defined by (\ref{hypers_vol_eq_psi}),  on one spatial direction ($x$, $y$ or $z$) for a typical  CDT configuration in phase C with  small jump magnitude ($\delta=0.1$).
{The horizontal axis is $\bar{\phi}/\delta$.}}\label{1dsmall}
\end{figure}

\begin{figure}
\includegraphics[width=\textwidth]{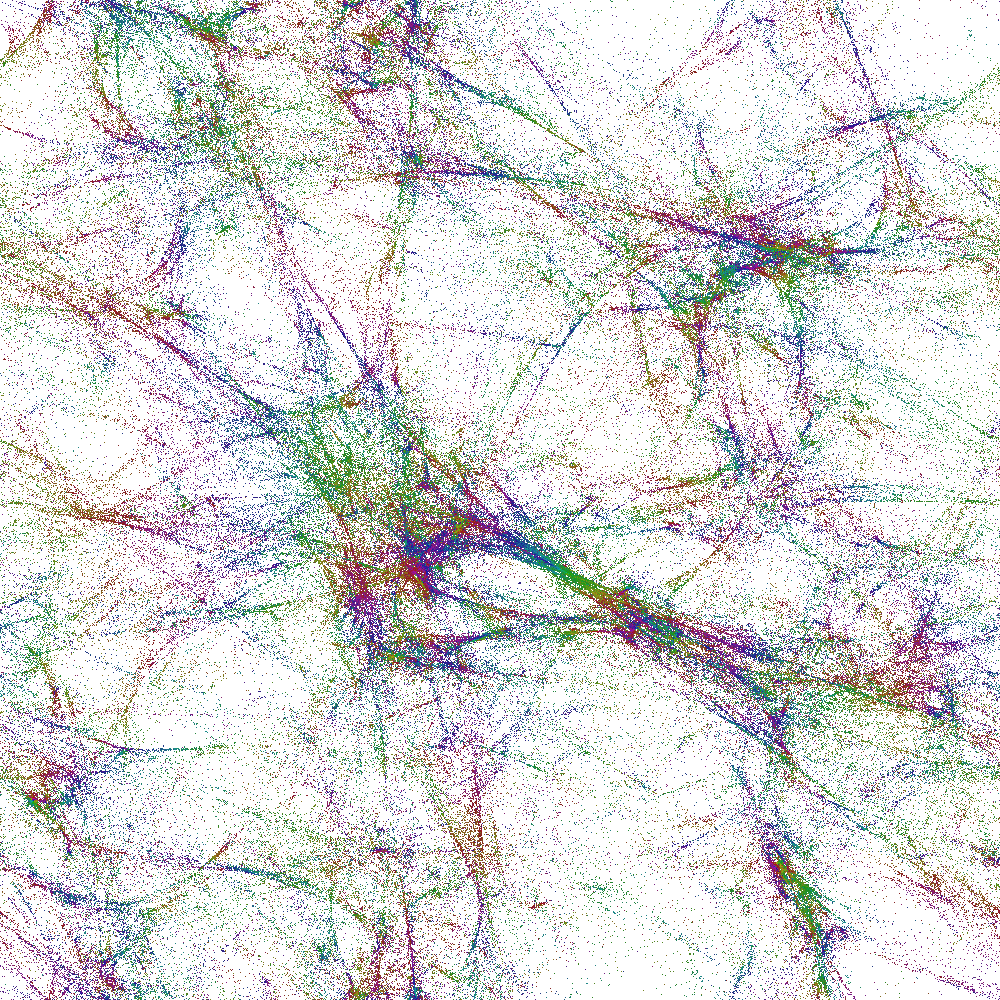} 
\caption{The projection of four-volume, as defined by (\ref{hypers_vol_eq_psi}), on the $xy$-plane for a typical CDT configuration in phase C with small jump magnitude ($\delta=0.1$).
Different colors correspond to different times $t$ of the original (lattice) time-foliation.}\label{xysmall}
\end{figure}

\begin{figure}
\includegraphics[width=\textwidth]{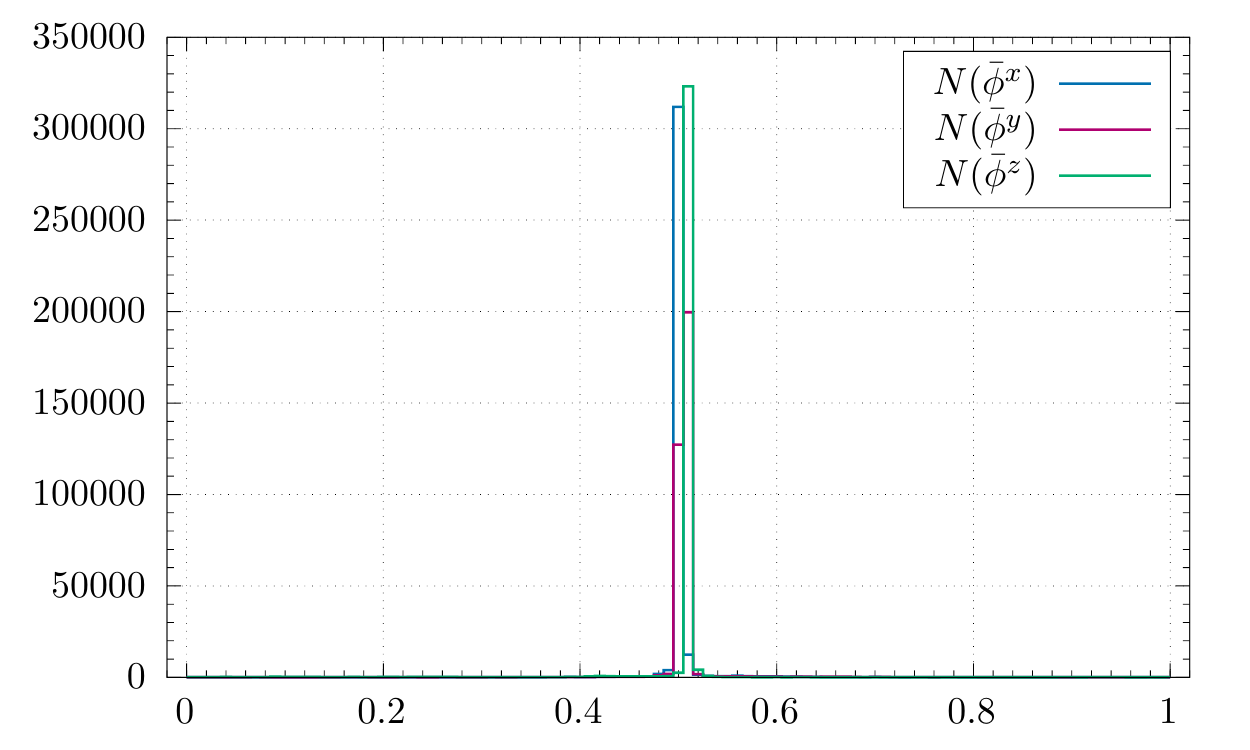}
\caption{The projection of four-volume, as defined by (\ref{hypers_vol_eq_psi}),  on one spatial direction ($x$, $y$ or $z$) for a typical  CDT configuration in phase C with large jump magnitude ($\delta=1.0$). {The horizontal axis is $\bar{\phi}/\delta$}.}
\label{1dlarge}
\end{figure}

\begin{figure}
\includegraphics[width=\textwidth]{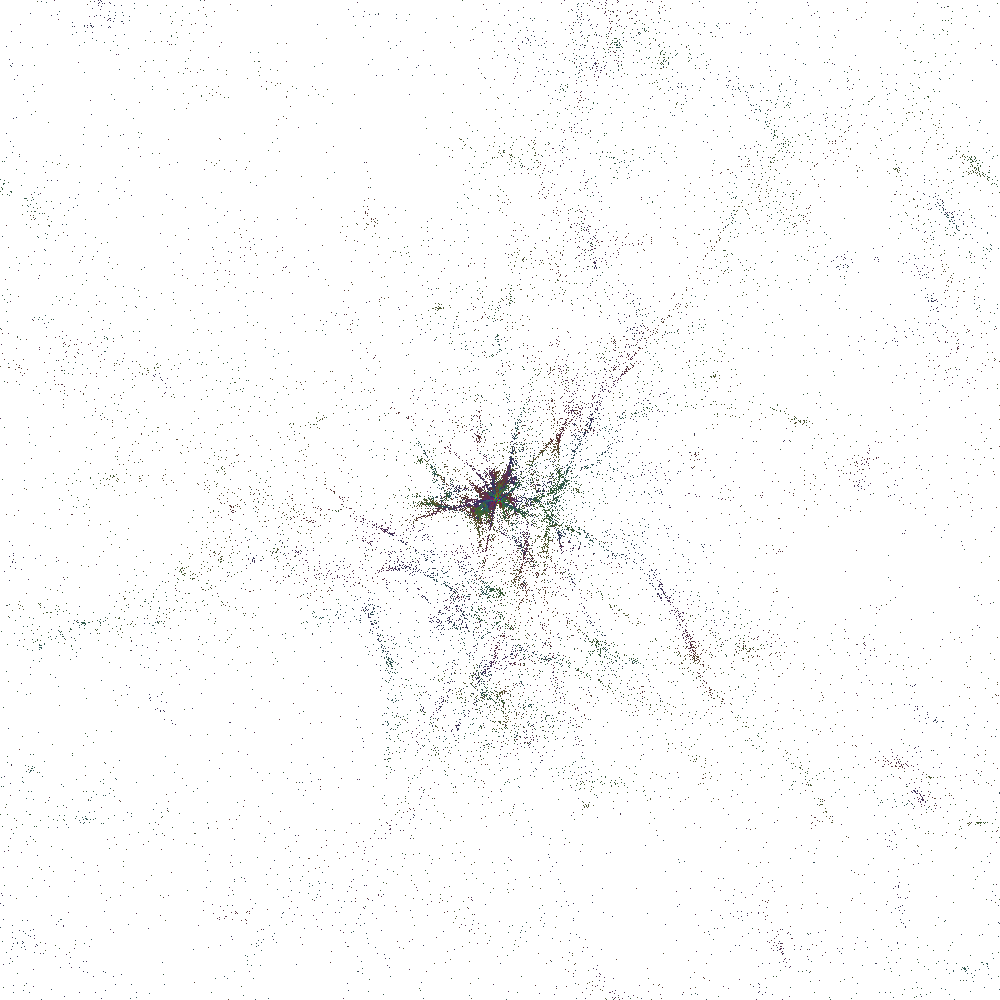} 
\caption{The projection of four-volume, as defined by (\ref{hypers_vol_eq_psi}), on the $xy$-plane for a typical CDT configuration in phase C with  large jump magnitude ($\delta=1.0$).
Different colors correspond to different times $t$ of the original (lattice) time-foliation. 
}\label{xylarge}
\end{figure}

Numerical Monte Carlo simulations confirm this hypothesis.
The analyzed systems were all in the same point in the $C$ phase 
on the phase diagram, and in each case we used $d=3$ scalar fields winding around spatial directions, but the parameter $\delta$, the jump magnitude, varied between
simulations.

For the purpose of visualization, the spatial coordinates defined by the
classical scalar field were used, in the way described in the previous section.

Figures \ref{1dsmall} and \ref{1dlarge} show projections of the volume density distribution $N(\bar\phi)$ (eq. \ref{hypers_vol_eq_psi}):
\begin{eqnarray}
\Delta N(\bar\phi) = \sqrt{g(\bar\phi)} \prod_\sigma \Delta \bar\phi^\sigma = N(\bar\phi) \prod_\sigma \Delta \bar\phi^\sigma.
\label{hypers_vol_eq_psi2}
\end{eqnarray}
in a typical toroidal CDT configuration on one spatial direction ($x$, $y$ or $z$).
Figures \ref{xysmall} and \ref{xylarge} show projections on two-dimensional ($x-y$ directions) parameter subspace, integrating over the remaining directions.
The parameter $\delta$ was small ($\delta=0.1$) in systems depicted in
Figures \ref{1dsmall} and \ref{xysmall} and large ($\delta=1.0$) in systems depicted in Figures \ref{1dlarge} and \ref{xylarge}.
The change of geometry caused by the magnitude of the jump is clearly visible.
The pictures for a small $\delta$  resemble those without dynamical scalar fields,
as in the previous section, whereas for a large $\delta$ the geometry is ``pinched'' in all spatial directions (which manifests itself as the small-volume region in Fig. \ref{1dlarge} and the low-density region in Fig. \ref{xylarge}).

The results imply that for a globally hyperbolic spacetime with the toroidal $(S^1)^3$
spatial topology 
an addition of a three-component scalar field with matching topological boundary conditions
can have a dramatic effect on the geometries that dominate the CDT path integral.

This new kind  of coupling between the topology of the matter fields and the topology of spacetime is likely to result in a phase transition for sufficiently strong coupling (sufficiently large $\delta$ in the presented model), a transition where the universe breaks up into several universes with no non-contractible loops: this is a natural extrapolation of what is schematically shown in Fig.\ \ref{fig-pinch} and for actual configurations in the path integral in Figs.~\ref{1dlarge} and \ref{xylarge}, 
using as coordinates in the non-contractible ``directions'' the classical scalar field with non-trivial boundary conditions
in these directions. 
For a spacetime with the spatial topology of a three-sphere $S^3$, a scalar field
with such boundary conditions cannot be defined and thus this phase transition
is not possible.

\chapter{Conclusions}
\markboth{Conclusions}{Conclusions}

Causal Dynamical Triangulations is a model of quantum gravity, which is 
background independent and coordinate free in the spatial directions.
These properties fit the spirit of general relativity and are reminiscent of 
diffeomorphism invariance. However, the lack of coordinates 
makes it difficult to describe the geometry and measure observables other
than the global ones. Conversely, it was precisely the existence of a proper time coordinate
built-in into the model that allowed
for calculating the mini-superspace action – one of the main results in CDT.
That makes it apparent that there exists a need for coordinates and other observables with
which to probe the geometries of the triangulations in order to understand
their structure better and ultimately to extend the effective action
to include also the spatial directions.

One could argue that a single configuration
appearing in the CDT path integral is not physical and therefore uninteresting,
since physical observables are defined as averages over all configurations.
However, some observables could equivalently be estimated by a measurement
on a single configuration in a given point of the phase space, 
if we had at our disposal a representative configuration with a sufficiently large volume.
Most of the results in CDT could in principle be replicated that way, although it would not be as efficient from the practical point of view because
the computer Monte Carlo simulations of very large systems become very 
resource-consuming. 
This theoretical possibility makes it clear, however, that the understanding of
the geometry of a generic configuration is interesting and useful.

In four-dimensional CDT with spherical $S^3$ spatial topology,
the problem of introducing coordinates has not yet been resolved, and the only
notable observable usable to analyze the spatial structure of configurations
is the quantum Ricci curvature (see articles cited in Chapter \ref{chap:ricci}).
The situation is more interesting in the case of models with toroidal $T^3$
spatial topology, where there are several ways of studying the geometry of triangulations,
as described in this thesis. That is one of the reasons for which recently
the $T^3$ spatial topology has been the main focus of studies in four-dimensional CDT.

In the toroidal case, 
the existence of nonequivalent non-contractible loops helps to determine three spatial directions, 
and the boundaries of the elementary cell, 
which are in a sense objects dual to the minimal loops, can serve as a reference
frame for introducing coordinates. 
The simplest idea of using the distance (on the dual lattice) from a simplex
to the boundaries as coordinates, called here the pseudo-Cartesian coordinates,
was the first attempt at a more quantitative description of spatial geometries 
of four-dimensional CDT. These coordinates have some shortcomings, caused partly by their dependence on the position and shape of the fictitious boundaries, but
they provided a first glimpse into the structure of 
the semiclassical geometries of the $C$ phase.
The following developments of using the shortest loops themselves and of using scalar fields satisfying
Laplace's equation with a jump on the boundaries brought some improvement and corroborated the understanding of the irregular and complicated geometries of CDT.
The measurements of quantum Ricci curvature showed a connection between
topological observables such as the lengths of shortest loops
with nonzero winding numbers passing through a given simplex with an observable
defined in a more local way and on shorter length scale.
They also showed once more that configurations with a toroidal spatial topology
seem to have a more interesting structure than those with a spherical one.
It is hoped that measurements of spacetime
correlations in suitable coordinates will make it possible to determine 
the effective continuum action in all four directions, which might help to understand 
if CDT is an UV-complete quantum field theory of gravity,
as hoped for in the asymptotic safety scenario.

In Chapter \ref{chap:scalar}, the scalar fields $\phi_i^\mu$ were first treated
as classical field used to define coordinates and did not influence the simulations
and therefore the geometry of the triangulations.
The field configurations revealed structures bearing striking similarity to the 
void-and-filament patterns known from astronomical observations. 
As CDT offers a model of what could be generic fluctuations of geometry at the Planck scale,
it is conceivable that those structures magnified during a period of inflation
could be the origin of the voids and filaments observed in the present-day Universe.

Finally, the second section of Chapter \ref{chap:scalar} broadens somewhat the scope
of the thesis by discussing new observables not in a system with pure gravity
and no matter content but in a system with matter in the form of 
dynamical scalar fields. 
Contrarily to the scalar fields taking values in $\mathbb{R}$, which had been considered
previously in the context of CDT and did not have a large impact on the geometry,
the fields considered here were taking values on $S^1$
and were forced to wind around it when one moves around a non-contractible loop on the manifold.
Considering such dynamical fields was a new development, which however logically 
followed the non-dynamical usage of analogous fields, described in the previous section of that chapter.
Interestingly, it seems that when the diameter $\delta$ of the circle $S^1$ is sufficiently
large, the system undergoes a phase transition of a kind not observed previously 
in CDT. It is caused by an interplay of the matter action, which is minimized 
if the geometry of the triangulation is squeezed at some section of the torus
(and most of the change of the 
scalar field occurs in the squeezed region), 
and of the Regge action, which is minimized for configurations analogous to those
with pure gravity.
For large $\delta$, the matter action dominates, and the squeezing of the geometry
leads to configurations with very few simplices located
in the toroidal part (i.e., in the region whose simplices would have the smallest loop lengths)
and the great majority of simplices located in 
a single large outgrowth of almost spherical topology.
Thus, effectively there appears to be a change from a toroidal to a spherical topology.
This phase transition needs further investigation, but it is already evident that 
it is an interesting one, as it is the first phase transition in 
higher-dimensional CDT caused by matter.
The phase transition is clearly visible on plots made using the new coordinates
based on non-dynamical scalar fields, which is a further proof of their usefulness.

In the future, it would be interesting to determine the effective action
of four-dimensional CDT in the C phase and to analyze more deeply also the structure
of geometries in the other phases.

\end{document}